\documentclass[]{pasj01}
\usepackage{lscape}
\usepackage{pdflscape}

\begin{document} 
\Received{2018/12/17}
\Accepted{2019/09/18}

\title{CO Multi-line Imaging of Nearby Galaxies (COMING) IV. Overview of the Project}

\author{Kazuo \textsc{Sorai}\altaffilmark{1, 2, 3, 4, 5}%
}
\altaffiltext{1}{Department of Physics, Faculty of Science, Hokkaido University, Kita 10 Nishi 8, Kita-ku, Sapporo 060-0810, Japan}
\email{sorai@astro1.sci.hokudai.ac.jp}

\author{Nario \textsc{Kuno}\altaffilmark{4, 5}}
\author{Kazuyuki \textsc{Muraoka}\altaffilmark{6}}
\author{Yusuke \textsc{Miyamoto}\altaffilmark{7, 8}}
\author{Hiroyuki \textsc{Kaneko}\altaffilmark{7}}
\author{Hiroyuki \textsc{Nakanishi}\altaffilmark{9}}
\author{Naomasa \textsc{Nakai}\altaffilmark{4, 5, 10}}
\author{Kazuki \textsc{Yanagitani}\altaffilmark{6}}
\author{Takahiro \textsc{Tanaka}\altaffilmark{4}}
\author{Yuya \textsc{Sato}\altaffilmark{4}}
\author{Dragan \textsc{Salak}\altaffilmark{10}}
\author{Michiko \textsc{Umei}\altaffilmark{2}}
\author{Kana \textsc{Morokuma-Matsui}\altaffilmark{7, 8, 11, 12}}
\author{Naoko \textsc{Matsumoto}\altaffilmark{13, 14}}
\author{Saeko \textsc{Ueno}\altaffilmark{9}}
\author{Hsi-An \textsc{Pan}\altaffilmark{15}}
\author{Yuto \textsc{Noma}\altaffilmark{10}}
\author{Tsutomu, T. \textsc{Takeuchi}\altaffilmark{16}}
\author{Moe \textsc{Yoda}\altaffilmark{16}}
\author{Mayu \textsc{Kuroda}\altaffilmark{6}}
\author{Atsushi \textsc{Yasuda}\altaffilmark{4}}
\author{Yoshiyuki \textsc{Yajima}\altaffilmark{2}}
\author{Nagisa \textsc{Oi}\altaffilmark{17}}
\author{Shugo \textsc{Shibata}\altaffilmark{2}}
\author{Masumichi \textsc{Seta}\altaffilmark{10}}
\author{Yoshimasa \textsc{Watanabe}\altaffilmark{4, 5, 18}}
\author{Shoichiro \textsc{Kita}\altaffilmark{4}}
\author{Ryusei \textsc{Komatsuzaki}\altaffilmark{4}}
\author{Ayumi \textsc{Kajikawa}\altaffilmark{2, 3}}
\author{Yu \textsc{Yashima}\altaffilmark{2, 3}}
\author{Suchetha \textsc{Cooray}\altaffilmark{16}}
\author{Hiroyuki \textsc{Baji}\altaffilmark{6}}
\author{Yoko \textsc{Segawa}\altaffilmark{2}}
\author{Takami \textsc{Tashiro}\altaffilmark{2}}
\author{Miho \textsc{Takeda}\altaffilmark{6}}
\author{Nozomi \textsc{Kishida}\altaffilmark{2}}
\author{Takuya \textsc{Hatakeyama}\altaffilmark{4}}
\author{Yuto \textsc{Tomiyasu}\altaffilmark{4}}
\author{Chey \textsc{Saita}\altaffilmark{9}}


\altaffiltext{2}{Department of Cosmosciences, Graduate School of Science, Hokkaido University, Kita 10 Nishi 8, Kita-ku, Sapporo 060-0810, Japan}
\altaffiltext{3}{Department of Physics, School of Science, Hokkaido University, Kita 10 Nishi 8, Kita-ku, Sapporo 060-0810, Japan}
\altaffiltext{4}{Division of Physics, Faculty of Pure and Applied Sciences, University of Tsukuba, 1-1-1 Tennodai, Tsukuba, Ibaraki 305-8571, Japan}
\altaffiltext{5}{Tomonaga Center for the History of the Universe (TCHoU), University of Tsukuba, 1-1-1 Tennodai, Tsukuba, Ibaraki 305-8571, Japan}
\altaffiltext{6}{Department of Physical Science, Osaka Prefecture University, Gakuen 1-1, Sakai, Osaka 599-8531, Japan}
\altaffiltext{7}{Nobeyama Radio Observatory, Minamimaki, Minamisaku, Nagano 384-1305, Japan}
\altaffiltext{8}{Chile Observatory, 2-21-1 Osawa, Mitaka, Tokyo 181-8588, Japan}
\altaffiltext{9}{Graduate School of Science and Engineering, Kagoshima University, 1-21-35 Korimoto, Kagoshima, Kagoshima 890-0065, Japan}
\altaffiltext{10}{Department of Physics, School of Science and Technology, Kwansei Gakuin University, Gakuen 2-1, Sanda, Hyogo 669-1337, Japan}
\altaffiltext{11}{Institute of Space and Astronautical Science, Japan Aerospace Exploration Agency, 3-1-1 Yoshinodai, Chuo-ku, Sagamihara, Kanagawa 252-5210, Japan}
\altaffiltext{12}{Institute of Astronomy, Graduate School of Science, The University of Tokyo, 2-21-1 Osawa, Mitaka, Tokyo 181-0015, Japan}
\altaffiltext{13}{The Research Institute for Time Studies, Yamaguchi University, Yoshida 1677-1, Yamaguchi, Yamaguchi 753-8511, Japan}
\altaffiltext{14}{Mizusawa VLBI Observatory, National Astronomical Observatory of Japan, 2-21-1 Osawa, Mitaka, Tokyo 181-8588, Japan}
\altaffiltext{15}{Institute of Astronomy and Astrophysics, Academia Sinica, 11F of AS/NTU Astronomy-Mathematics Building, No.1, Sec. 4, Roosevelt Rd, Taipei 10617, Taiwan}
\altaffiltext{16}{Division of Particle and Astrophysical Science, Nagoya University, Furo-cho, Chikusa-ku, Nagoya, Aichi 464-8602, Japan}
\altaffiltext{17}{Tokyo University of Science, Faculty of Science Division II, Liberal Arts, 1-3, Kagurazaka Shinjuku-ku Tokyo 162-8601 Japan}
\altaffiltext{18}{College of Engineering, Nihon University, 1 Nakagawara, Tokusada, Tamuramachi, Koriyama, Fukushima 963-8642, Japan}


\maketitle

\begin{abstract}
Observations of the molecular gas in galaxies are vital to understanding the evolution and star-forming histories of galaxies. 
However, galaxies with molecular gas maps of their whole discs having sufficient resolution to distinguish galactic structures are severely lacking. 
Millimeter wavelength studies at a high angular resolution across multiple lines and transitions are particularly needed, severely limiting our ability to infer the universal properties of molecular gas in galaxies. 
Hence, we conducted a legacy project with the 45 m telescope of the Nobeyama Radio Observatory, called the CO Multi-line Imaging of Nearby Galaxies (COMING), which simultaneously observed 147 galaxies with high far-infrared flux in \atom{C}{}{12}\atom{O}{}{}, \atom{C}{}{13}\atom{O}{}{}, and \atom{C}{}{}\atom{O}{}{18} $J=1-0$ lines. 
The total molecular gas mass was derived using the standard \atom{C}{}{}\atom{O}{}{}--to--\atom{H}{}{}$_2$ conversion factor and found to be positively correlated with the total stellar mass derived from the WISE $3.4\,\micron$ band data. 
The fraction of the total molecular gas mass to the total stellar mass in galaxies does not depend on their Hubble types nor the existence of a galactic bar, although when galaxies in individual morphological types are investigated separately, the fraction seems to decrease with the total stellar mass in early-type galaxies and vice versa in late-type galaxies. 
No differences in the distribution of the total molecular gas mass, stellar mass, and the total molecular gas to stellar mass ratio was observed between barred and non-barred galaxies, which is likely the result of our sample selection criteria, in that we prioritized observing FIR bright (and thus molecular gas-rich) galaxies.
\end{abstract}

\section{Introduction}
\label{section:introduction}



How and where stars form in galaxies are clues to understanding galaxy evolution, and require information about the distribution, dynamics, and physical properties of their molecular gas content. 
H\,\emissiontype{II} regions and massive stars are found in spiral arms (\cite{Lynds1980}; \cite{GarciaGomez+1993}; \cite{Thilker+2002}; \cite{Oey+2003}; \cite{Bresolin+2005}), while only a few are found in the bar of some barred spiral galaxies (\cite{Koopmann+2001}; \cite{James+2004}; \cite{Hernandez+2005}; \cite{Erroz-Ferrer+2015}). 
Interacting and merging galaxies often display an abundance of star-forming regions in both their interface regions, especially compared to their spiral arms (\cite{Koopmann+2001}; \cite{Wang+2004}; \cite{Torres-Flores+2014}), while little new stars form even in the spiral arms of some galaxies (\cite{vanDenBergh1976}; \cite{Kennicutt+1986}; \cite{Masters+2010}; \cite{Fraser-McKelvie+2016}). 
These observational results indicate that star formation is not uniform both within and between different galaxies. 
Some questions must be answered for us to understand the causes of a variety of star formations within a galaxy and among galaxies.

Many studies have observed the distribution and dynamics of molecular gas in galaxies. 
Molecular gas in spiral galaxy M\,51 is primarily concentrated along the two grand-design spiral arms, but also detected in the interarm regions (\cite{Garcia-Burillo+1993}; \cite{Nakai+1994}). 
The velocity of molecular gas qualitatively changes at the spiral arm in accordance with density wave theory, and the estimated elliptical motion can explain the surface density contrast of the molecular gas between the spiral arms and the interarm regions (\cite{Kuno+1997a}). 
Flocculent galaxies also display molecular gas concentrations along their spiral arms, such as in NGC\,5055 (\cite{Kuno+1997b}). 
On-the-fly (OTF) observations of the barred spiral galaxy M\,83 showed that the CO disc has a sharp edge, while the H\,\emissiontype{I} disc more gradually extends to larger radii (\cite{Crosthwaite+2002}).

In the recent years, CO observations with high spatial resolution have resolved giant molecular clouds (GMCs) in galaxies. 
Giant molecular cloud associations (GMAs) are dominant in the spiral arms and broken up into GMCs in the interarm regions in M\,51 (\cite{Koda+2009}). 
In the barred spiral galaxy NGC\,4303, the molecular gas in the bar has a lower star formation efficiency (SFE) than that in the spiral arms, where the SFE is the star formation rate (SFR) divided by the molecular gas mass (\cite{Momose+2010}). 
The SFE depends on the environment at sub-kpc scales, and increases with the surface density of the molecular gas (\cite{Momose+2013}). 
Meanwhile, in the local spiral galaxy M\,33, the molecular gas fractions are loosely correlated with the neutral gas fraction observed at the GMC scales, with particular variations in the inner disc (\cite{Tosaki+2011}). 
A CARMA (Combined Array for Research in Millimeter Astronomy interferometer) and Nobeyama Nearby galaxies (CANON) survey resolved approximately 200 GMCs in the inner discs of five galaxies and revealed that they are similar to those in the Milky Way (\cite{DonovanMeyer+2013}). 
PAWS (Plateau de Bure Interferometer Arcsecond Whirlpool Survey, \cite{Schinnerer+2013}) observed M\,51 at $\sim\,40\,\mathrm{pc}$ resolution and found that the dynamical environment of the GMCs significantly influences their star-forming capability (\cite{Meidt+2013}), and that feedback from massive stars affects the dependency of the GMC properties on the environment (\cite{Colombo+2014}). 
 Observations of M\,100 with Atacama Large Millimeter/Submillimeter Array (ALMA) revealed that the GMA properties depend on the environment: GMAs are compact in the circumnuclear region, but diffuse in interarm regions, and their velocity dispersions are higher in the circumnuclear region and the bar than the other regions (\cite{Pan+2017}).

A few notable systematic surveys have mapped gas across the entire surface of galaxies at a sub-kpc resolution. 
Berkeley-Illinois-Maryland Association millimeter interferometer Survey of Nearby Galaxies, also known as BIMA SONG, imaged 44 nearby galaxies (\cite{Helfer+2003}) via interferometry and single-dish observations. 
Meanwhile, the Nobeyama CO Atlas of nearby galaxies (\cite{Kuno+2007}) mapped 40 galaxies with a single-dish telescope. 
These observations revealed higher molecular gas concentrations toward the galactic center in barred spiral galaxies compared to unbarred spirals (\cite{Sheth+2005}; \cite{Kuno+2007}). 
Some galaxies located near the center of the Virgo cluster have revealed a higher fraction of molecular gas to the total neutral gas, including H\,\emissiontype{I} gas, which is interpreted as ram pressure stripping of H\,\emissiontype{I} gas or induced molecular gas formation caused by a higher external pressure in cluster environments (\cite{Nakanishi+2006}). 
A total of 28 Virgo cluster spirals were also mapped with the Five College Radio Astronomy Observatory (FCRAO) 14\,m telescope (\cite{Chung+2009b}); however, some galaxies overlap with one another. 
The total number of mapped galaxies in these three surveys was 74.

Many mapping observations of molecular gas, whose sample size numbers were 10 or fewer, and surveys with interferometers covering only the central regions have also been made [\cite{Sakamoto+1999}; \cite{Sofue+2003}; CARMA STING (Survey Toward Infrared-bright Nearby Galaxies), \cite{Rahman+2012}]. 
However, combining such data is not necessarily suitable for comparing many galaxies because spatial resolutions and instrument sensitivity can wildly differ between surveys. 
In addition, observations with interferometers alone miss extended emission (i.e., are ``resolved out''); hence, there are concerns that such observations underestimate the total molecular gas mass of the target galaxies. 
If mapping does not extend across the entirety of the galactic disc, then correct information on the molecular gas and star formation in outer regions, particularly in interacting galaxies, are impossible to obtain.

Surveys targeting higher-$J$ transitions have also been conducted, although they carry added caveats for estimating the total molecular gas masses of the target galaxies. 
HERACLES (HEterodyne Receiver Array CO Line Extragalactic Survey, \cite{Leroy+2009}) provided sensitive images of 48 nearby galaxies in \atom{C}{}{12}\atom{O}{}{} $(J=2-1)$. 
The relation between the surface density and the velocity dispersion of GMCs in nearby galaxies was reported (\cite{Sun+2018}) based on the recent very high-resolution observations with ALMA in \atom{C}{}{12}\atom{O}{}{} $(J=2-1)$ [PHANGS-ALMA (Physics at High Angular resolution in Nearby Galaxies with ALMA), A.~K.,~Leroy, et~al. (in preparation)]. 
JCMT Nearby Galaxies Legacy Survey (NGLS, \cite{Wilson+2012}) mapped 155 galaxies in \atom{C}{}{12}\atom{O}{}{} $(J=3-2)$. 
Such high transition data are particularly useful for excitation analysis combined with $J=1-0$. 
However, the estimation of the molecular gas mass assuming a constant intensity ratio has a considerable uncertainty (e.g., $J=2-1/J=1-0$) because the ratio is not constant within a galaxy (\cite{Sakamoto+1997}; \cite{Koda+2012}; \cite{Leroy+2013}).

Several single-point observations have provided key insights into the relation between molecular gas content and galaxy morphology and evolution. 
The FCRAO survey of 300 galaxies reported that the molecular gas distribution against the optical galaxy size depends on morphology (\cite{Young+1995}). 
\citet{Komugi+2008} observed 68 galaxies. 
They showed larger central concentrations of molecular gas in earlier-type galaxies and the impact of the inner bulge on the gas concentrations. 
Large CO surveys have recently provided clues of galaxy evolution by comparison with stellar mass information. 
COLD GASS [CO Legacy Data base for the GASS (GALEX Arecibo SDSS Survey)] observed $\sim\,350$ galaxies and showed that the relation between molecular gas fraction and stellar mass strongly depends on $\mathrm{NUV} - r$ color (\cite{Saintonge+2011}). 
The Herschel Reference Survey observed 59 galaxies (\cite{Boselli+2014a}) and illustrated that the molecular gas mass fraction only slightly depends on morphology, but strongly depends on the stellar mass and the specific SFR (\cite{Boselli+2014b}). 
The fraction increases with the redshift in the range of $0 \lesssim z \lesssim 3$ (\cite{Daddi+2010}; \cite{Popping+2012}; \cite{Saintonge+2013}; \cite{Popping+2015}; \cite{Dessauges-Zavadsky+2017}), and this evolution depends on the stellar mass (\cite{Dessauges-Zavadsky+2015}; \cite{Morokuma-Matsui+2015b}).

We have limited spatially resolved information on the physical conditions of molecular gas (e.g., whether the density and the temperature of molecular gas in GMCs differ between the arm, interarm, and bar regions of disc galaxies). 
Although the most general tracer of molecular gas in galaxies is \atom{C}{}{12}\atom{O}{}{} $(J=1-0)$, we cannot estimate the molecular gas density and the temperature from a single transition. 
Molecular lines are excited under various physical conditions; hence, we have to observe multiple lines to constrain the physical conditions of molecular gas. 
In the case of multiple \atom{C}{}{12}\atom{O}{}{} line observations, we have to carefully compare the data because the line frequencies are very different from each other, and as a result, different lines are measured with different telescopes or taken at different spatial resolutions. 
Although we can observe \atom{C}{}{13}\atom{O}{}{} or \atom{C}{}{}\atom{O}{}{18} lines with the same telescope at nearly the same spatial resolution, these lines are very weak, and mapping the whole disc is a time-consuming task [\cite{Watanabe+2011}; CARMA STING, \cite{Cao+2017}; EMPIRE (EMIR Multiline Probe of the ISM Regulating Galaxy Evolution) survey, \cite{Cormier+2018}]. 
Such efforts have revealed the properties of molecular gas in several local galaxies. 
In the bar ends of NGC\,3627, the \atom{C}{}{12}\atom{O}{}{} and \atom{C}{}{13}\atom{O}{}{} measurements suggest a very high molecular gas density, which results in a very high SFE (\cite{Watanabe+2011}). 
In contrast, the \atom{C}{}{12}\atom{O}{}{} / \atom{C}{}{13}\atom{O}{}{} intensity ratios do not clearly correlate with the SFR (\cite{Cao+2017}). 
Despite such efforts, a great deal of progress must still be made on understanding how the properties of molecular gas varies both within and between galaxies.

High-resolution and high-sensitivity mapping capabilities have recently expanded targets from local spiral galaxies to early-type or low-$z$ galaxies. 
The CARMA ATLAS$^{3\mathrm{D}}$ molecular gas imaging survey observed 30 early-type galaxies and showed various \atom{C}{}{}\atom{O}{}{} morphologies and a wide distribution of \atom{C}{}{13}\atom{O}{}{} / \atom{C}{}{12}\atom{O}{}{} ratios (Alatalo et al.\ \yearcite{Alatalo+2013}, \yearcite{Alatalo+2015}). 
Meanwhile, the Evolution of Molecular Gas in Normal Galaxies (EGNoG) survey imaged 31 star-forming galaxies from $z = 0.05$ to $z = 0.5$ and illustrated molecular gas depletion times and fractions (\cite{Bauermeister+2013}). 
The Extragalactic Database for Galaxy Evolution (EDGE) -- Calar Alto Legacy Integral Field Area (CALIFA) survey (\cite{Bolatto+2017}) observed 126 relatively distant galaxies and presented a fairly constant molecular-to-stellar mass ratio across spiral galaxies and an approximately linear relation between the resolved surface densities of the SFR and molecular gas. 
Meanwhile, the Valpara\'{i}so ALMA Line Emission Survey (VALES) observed 67 galaxies up to $z = 0.35$ with ALMA and found that the molecular gas distribution is, on average, $\sim\,0.6$ times more compact than the optical size (\cite{Villanueva+2017}).

We conducted the project CO Multi-line Imaging of Nearby Galaxies (COMING), which is one of the Nobeyama Radio Observatory (NRO) legacy projects, using the 45 m telescope to quantitatively improve our understanding of the spatially resolved, galaxy-scale distribution of molecular gas. 
The OTF observations with the multi-beam receiver, FOur-beam REceiver System on the 45 m Telescope (FOREST) (\cite{Minamidani+2016}) enabled us to make efficient maps toward a large number of galaxies. 
We simultaneously observed the \atom{C}{}{12}\atom{O}{}{}, \atom{C}{}{13}\atom{O}{}{}, and \atom{C}{}{}\atom{O}{}{18} lines using the wide intermediate frequency (IF) band of FOREST. 
Some preliminary results for individual galaxies have already been published (\cite{Muraoka+2016}; \cite{Hatakeyama+2017}; \cite{Yajima+2019}), and this paper presents a project overview. 
Sections \ref{section:sample} and \ref{section:observation} present the sample selection and observations, respectively. 
Sections \ref{section:reduction} and \ref{section:analysis} show the data reduction, analysis software development, data analysis, and archival data, respectively. 
Section \ref{section:results} presents the results and discussion. 
Finally, section \ref{section:summary} summarizes the project overview.

\section{Sample}
\label{section:sample}
The initial sample selection consisted of 344 far-infrared (FIR) bright galaxies from the ``Nearby Galaxies Catalog'' (\cite{Tully1988}). 
The number of the CO images of galaxies was much smaller than that in optical or infrared regimes; thus, we gave priority to galaxies expected to be bright in CO, although the completeness of the samples is important in understanding the galaxy evolution. 
Accordingly, we selected candidates biased to the FIR flux that is known to be well correlated with the CO flux (\cite{Young+1991}) from the Nearby Galaxies Catalog. 
We used the selection criteria of $100\,\micron$ flux $S_{100\,\micron} \geq10\>\mathrm{Jy}$ in ``IRAS catalogue of Point Sources'' (\cite{Helou+1988}) or the $140\,\micron$ flux $S_{140\,\micron} \geq10\>\mathrm{Jy}$ in ``AKARI/FIS All-Sky Survey Point Source Catalogues'' (\cite{Yamamura+2010}). 
We checked that all sources common to both samples satisfied both criteria. 
Elliptical galaxies were removed from our sample candidates even if they satisfied the criteria because CO emission was not expected to be detected within a reasonable observing time. 
M\,31 and M\,33 were also eliminated in spite of satisfying the criteria because both galaxies would require excessively large maps that demanding a long observing time.

Next, we selected galaxies from the abovementioned parent sample, most of which have a large extent in optical images, and have not yet been observed using the 45 m telescope, thereby resulting in 238 galaxies. 
The sample included galaxies previously observed with the 45 m telescope because we intended to compare our new OTF maps with the previous ones (\cite{Muraoka+2016}). 
In addition, some previous observations were less sensitive. 
This selection was done to increase the number of galaxy CO maps available within a limited observation time. 
We resolved the galactic structure in detail by assigning a higher priority to galaxies with (1) a large optical diameter ($D_{25}$), (2) a lower inclination angle when $D_{25}$ is similar, and (3) with former observational works having a higher spatial resolution, such as CANON (\cite{DonovanMeyer+2013}) (we added Mrk\,33) or CARMA STING (\cite{Rahman+2012}). 
We also included pair galaxies, where only one of the pair satisfies the criteria because the molecular gas in the interacting systems may spread over a wide area, including intergalactic regions (\cite{Kaneko+2013}). 
We assigned the priority code (A, B, and C) to the 238 selected galaxies according to the apparent galaxy size with or without previous CO (interferometric) data and interest of individual members of the survey team. 
Practical constraints, such as available observing time and weather, limited the actual number of the observed galaxies to 147.

\begin{figure}
 \begin{center}
  \includegraphics[width=8cm]{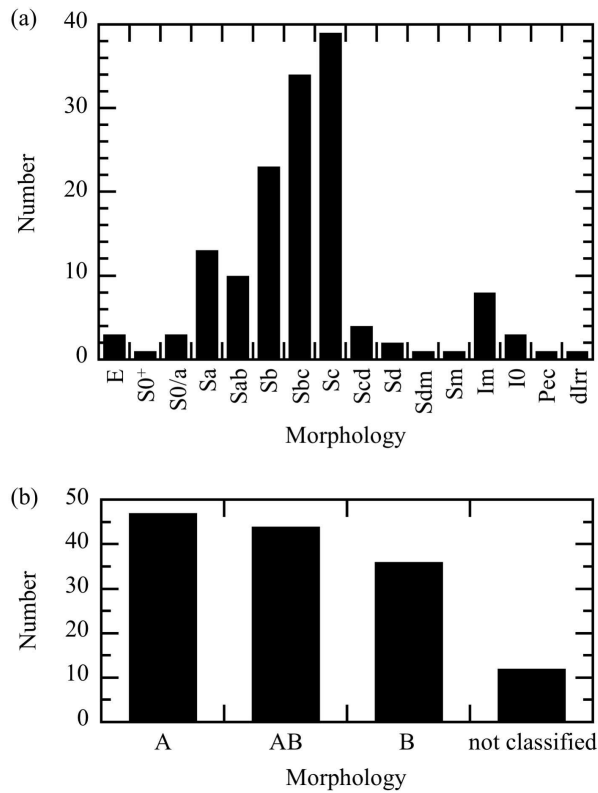}
 \end{center}
\caption{Morphological type of the sample galaxies: (a) Hubble types and (b) barred or non-barred spirals. 
The ``E'' and ``Pec'' types are not included in panel (b).}
\label{fig:morphology}
\end{figure}

\begin{figure}
 \begin{center}
  \includegraphics[width=8cm]{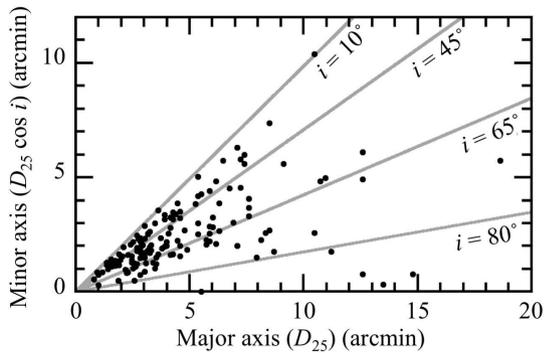}
 \end{center}
\caption{Apparent optical sizes of the sample galaxies. 
The gray lines indicate a specified inclination ($i$).}
\label{fig:sizes}
\end{figure}

\begin{figure}
 \begin{center}
  \includegraphics[width=8cm]{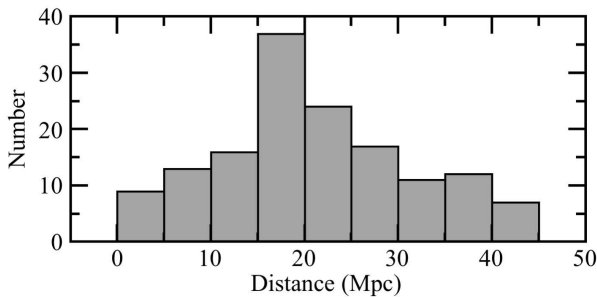}
 \end{center}
\caption{Distances of the sample galaxies.}
\label{fig:distance}
\end{figure}

\begin{figure}
 \begin{center}
  \includegraphics[width=8cm]{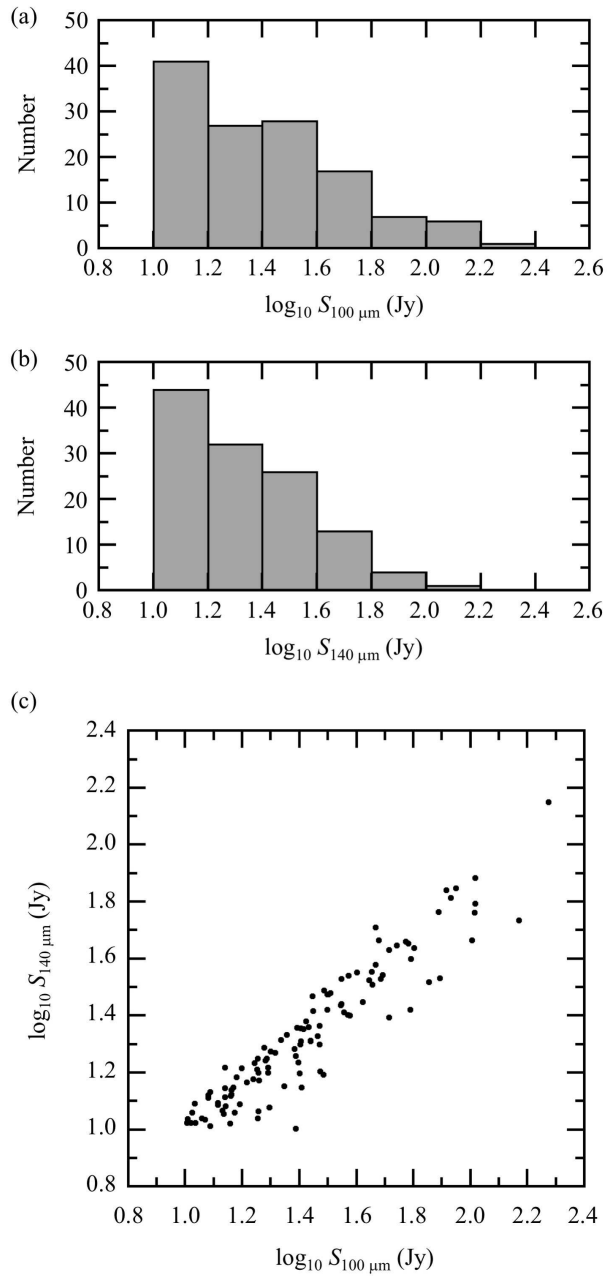}
 \end{center}
\caption{FIR flux of the sample galaxies. 
(a) IRAS $100\,\micron$. 
(b) AKARI $140\,\micron$. 
(c) Correlation plot of both fluxes. 
The histograms include galaxies without other wavelength data.}
\label{fig:sample}
\end{figure}

Table \ref{tab:Galpars1} and figures \ref{fig:morphology} -- \ref{fig:sample} show the morphology, optical diameter, distance, position angle (PA) of the major axis, inclination ({\it i}) of the galactic disc, and FIR fluxes of the 147 observed galaxies. 
We adopted the distance of each galaxy from the redshift-independent distances in the NASA/IPAC Extragalactic Database (NED)\footnote{$\langle$http://ned.ipac.caltech.edu$\rangle$.} prioritizing the following: (1) distance with the minimum error of the distance modulus taken after 2013; (2) the same as (1) if no matches, but the latest data taken after 2003; and (3) the same as (2) if no matches, but taken before 2003. 
The distances of the galaxies identified as members of the Virgo cluster are assumed to be the same value of $16.5\ \mathrm{Mpc}$ (\cite{Mei+2007}). 
The distances of the interacting galaxies are considered similar to each other. 
We adopted PA and {\it i} that were measured kinematically and determined at the same time, where possible. 
When no such data exist, we adopted the pair of PA and {\it i} measured by fitting brightness distribution by an ellipse or PA and {\it i} measured individually. 
PA is the receding side of the semi-major axis, unless no kinematical information exists. 
PA is expressed within $\pm\timeform{180D}$, where $\timeform{0D}$ corresponds to the north, and the angle is measured counterclockwise. 
Our kinematical analysis has shown that PA and {\it i} of NGC\,2967 are quite different from the data estimated in previous works (\cite{Salak+2019}); however, we do not adopt the latest values herein.

The observed galaxies have a bias toward Sb to Sc types with few early- and late-type galaxies because constraints on the observing conditions restricted our observation of the entire original sample. 
The presence or absence of a bar classified as SA, SAB, or SB is distributed throughout the sample. 
The observation ranking system described in subsection \ref{subsection:ObservationRankingSystem} particularly selected smaller galaxies because the weather conditions in the last two observation seasons were poor.



\section{Observations}
\label{section:observation}

The observations were made with the NRO 45 m telescope. 
Three \atom{C}{}{}\atom{O}{}{} isotopomers were simultaneously observed with the OTF mode. 
We developed an observation ranking system that objectively selects the optimal target to maximize the survey efficiency. 
This system was used in the last two seasons.

\subsection{System setting and OTF mapping}
\label{subsection:SystemSetting}
Simultaneous observations of \atom{C}{}{12}\atom{O}{}{} $(J=1-0)$ (rest frequency: $115.271202\,\mathrm{GHz}$), \atom{C}{}{13}\atom{O}{}{} $(J=1-0)$ (rest frequency: $110.201353\,\mathrm{GHz}$), and \atom{C}{}{}\atom{O}{}{18} $(J=1-0)$ (rest frequency: $109.782173\,\mathrm{GHz}$) were made with FOREST over four seasons: from 2015 April to May, from 2015 December to 2016 May, from 2016 December to 2017 May, and from 2017 December to 2018 April (table \ref{tab:obsterms}). 
FOREST has four beams, and each beam can receive dual-polarization $8\,\mathrm{GHz}$ bandwidth data for each sideband (upper and lower sidebands). 
The beam size of each beam was $\sim\,\timeform{14''}$ in $110\,\mathrm{GHz}$ and $115\,\mathrm{GHz}$ bands. 
SAM45 (\cite{Kuno+2011}; \cite{Kamazaki+2012}), which consists of 16 correlators, is available as a backend. 
We used SAM45 in a wide-band mode with $2\,\mathrm{GHz}$ bandwidth and 4096 channels corresponding to a frequency resolution of $488\,\mathrm{MHz}$.\footnote{Some observations of a standard source IRC\,+10216 were made in $1\,\mathrm{GHz}$ bandwidth per IF band, but this does not affect the calibration results.} 
Two correlators were assigned for each beam and polarization. 
The center frequencies were set to $115.271202\,\mathrm{GHz}$ for \atom{C}{}{12}\atom{O}{}{} $(J=1-0)$ and $109.991763\,\mathrm{GHz}$ for \atom{C}{}{13}\atom{O}{}{} $(J=1-0)$ and \atom{C}{}{}\atom{O}{}{18} $(J=1-0)$, respectively. 
We measured the image rejection ratio (IRR) of each side band of each beam every observation day. 
The typical system noise temperatures ($T_{\mathrm{sys}}$) in each observation season were $560\,\mathrm{K}$, $340\,\mathrm{K}$, $360\,\mathrm{K}$, and $390\,\mathrm{K}$ in \atom{C}{}{12}\atom{O}{}{} and $300\,\mathrm{K}$, $170\,\mathrm{K}$, $180\,\mathrm{K}$, and $180\,\mathrm{K}$ in \atom{C}{}{13}\atom{O}{}{} and \atom{C}{}{}\atom{O}{}{18}, respectively (table \ref{tab:obsterms}).

\begin{table}
  \tbl{Observation periods and system noise temperatures.}{
  \begin{tabular}{cccc}
    \hline              
    Season & Date & \multicolumn{2}{c}{Typical $T_{\mathrm{sys}}$ (K)} \\
     & & \atom{C}{}{12}\atom{O}{}{} & \atom{C}{}{13}\atom{O}{}{} \& \atom{C}{}{}\atom{O}{}{18} \\
    \hline
    1 & 2015 Apr. 4 -- 2015 May 6 & 560 & 300 \\
    2 & 2015 Dec. 21 -- 2016 May 24 & 340 & 170 \\
    3 & 2016 Nov. 28 -- 2017 March 17 & 360 & 180 \\
    4 & 2017 Dec. 24 -- 2018 Apr. 23 & 390 & 180 \\
    \hline
  \end{tabular}}
  \label{tab:obsterms}
\end{table}

\begin{figure}
 \begin{center}
  \includegraphics[width=8cm]{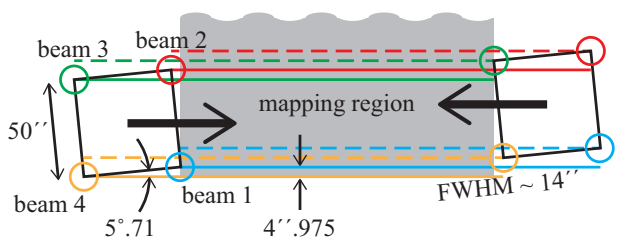}
 \end{center}
\caption{OTF scanning pattern. 
FOREST with $50''$ side was inclined $5^{\circ}.71$. 
The broad arrows indicate the scanning directions. 
The five-time round-trip scans with an offset of $4''.975$ uniformly cover the mapped regions.}
\label{fig:scan_pattern}
\end{figure}

We observed the OTF-mapping mode to achieve a typical sensitivity of $30\,{\rm mK}$ in the antenna temperature ($T_{\mathrm{A}}^{*}$) scale corresponding to $\sim\,70\,{\rm mK}$ in the main beam brightness temperature ($T_{\mathrm{MB}}$) scale. 
Each region in a given target was mapped with more than one beam to reduce the influence caused by the difference in the performances of the four beams. 
Figure \ref{fig:scan_pattern} shows that we rotated FOREST $\timeform{5D.71}$ against the scan direction and raster scanned in $\timeform{4''.975}$ separation, which was $\sim\,1.4$ times higher than the Nyquist sampling of the $\timeform{14''}$ beam to cover the mapped region with five round trips. 
This mapping method enabled us to minimize the peripheral regions with a high noise temperature. 
We adopted two orthogonal scan directions and made the noise temperatures of each direction as even as possible. 
Two points $\timeform{10'}$ offset from the map center were observed as the off-source positions. 
Although the off-source positions were kept far away from the Galactic disc, unfortunately, unknown Galactic molecular clouds may exist at the off-source points of UGCA\,86 and NGC\,1569 at the same receding velocity of the galaxies; thus, some profiles of these two galaxies were contaminated by ``absorption'' patterns.

We scanned each galaxy to cover $70\,\%$ of $D_{25}$ in diameter. 
Previous works (\cite{Nishiyama+2001b}; \cite{Kuno+2007}) have shown that $70\,\%$ of $D_{25}$ of a galaxy covers the main region, where the CO emission is concentrated within a reasonable observation time. 
The scan directions were along the galaxy major and minor axes to maximize efficiency, while those of the interacting systems were along the right ascension and declination covering the whole system. 
Observations were made referencing the equatorial coordinates, except for NGC\,2268 and NGC\,2276 with a declination of $> \timeform{80D}$. 
Both galaxies were observed in the galactic coordinates to avoid a large variation of the right ascension. 
The scan length along the minor axis of the galaxy was adopted by multiplying the length along the major axis by $\cos i$ listed in \citet{Tully1988}. 
The PA and {\it i} for setting an observing region were different from those listed in table \ref{tab:Galpars1}, which we used for the data analysis (section \ref{section:analysis}), because we observed referring the PAs mainly listed in the Third Reference Catalogue of Bright Galaxies (RC3, \cite{deVaucouleurs+1991}); however, it was convenient for the data analysis to refer to the kinematically derived PAs or use IR images. 
Table \ref{tab:Galpars2} lists the observation parameters.

\begin{longtable}{p{2cm}p{2cm}p{2cm}p{0.8cm}p{2cm}p{0.8cm}p{0.8cm}p{0.8cm}p{1cm}p{1.5cm}}
  \caption{Sample observational parameters.}\label{tab:Galpars2}
  \hline
 Galaxy & $\alpha_{\mathrm{J}2000.0}$ & $\delta_{\mathrm{J}2000.0}$ & & Mapping size & \multicolumn{2}{l}{PA of $+X$} & \multicolumn{2}{c}{$V_{\mathrm{LSR}} (\mathrm{km\>s}^{-1})$} & Season \\
 (1) & (2) & (3) & (4) & (5) & (6) & (7) & (8) & (9) & (10) \\
\endfirsthead
  \hline
 Galaxy & $\alpha_{\mathrm{J}2000.0}$ & $\delta_{\mathrm{J}2000.0}$ & & Mapping size & \multicolumn{2}{l}{PA of $+X$} & \multicolumn{2}{c}{$V_{\mathrm{LSR}} (\mathrm{km\>s}^{-1})$} & Season \\
 (1) & (2) & (3) & (4) & (5) & (6) & (7) & (8) & (9) & (10) \\
  \hline
\endhead
  \hline
\endfoot
  \hline
\multicolumn{10}{@{}l@{}}{\hbox to 0pt{\parbox{160mm}{\footnotesize
Notes. 
\par\noindent
Column (1) Galaxy name. 
Same as the column (1) in table \ref{tab:Galpars1}. 
\par\noindent
Columns (2) and (3): Adopted galaxy center coordinates. 
No remarks in column (4) indicates that the coordinates correspond to the reference position of the observations. 
\par\noindent
Column (4) Reference and notes of the galaxy center that is listed on the top of NED. 
No remarks: \cite{TwoMASS2003}. 
1: \cite{Anderson+2005}. 
2: \cite{Argyle+1990}. 
3: \cite{Becker+1995}. 
4: \cite{Clements1981}. 
5: \cite{Clements1983}. 
6: \cite{Condon+1991}. 
7: \cite{Condon+1998}. 
8: \cite{Evans+2010}. 
9: \cite{Herrnstein+2005}. 
10: \cite{Jackson+2007}. 
11: \cite{Kautsch+2006}. 
12: \cite{Krips2007}. 
13: \cite{Li+1998}. 
14: \cite{Liu+2005}. 
15: \cite{Minchin+2010}. 
16: \cite{Monet1998}. 
17: \cite{Nagar+2002}. 
18: \cite{NEDTEAM2011}. 
19: \cite{NEDTEAM2012}. 
20: \cite{Paturel+2000}. 
21: \cite{Petrov+2011}. 
22: \cite{SDSS2004a}. 
23: \cite{SDSS2004b}. 
24: \cite{SDSS2005}. 
25: \cite{SDSS2006}. 
26: \cite{SDSS2007}. 
27: \cite{Stone+1999}. 
28: \cite{TwoMASS2000}
29: \cite{Vigott+1989}. 
30: \cite{Villi+1998}. 
31: \cite{Whiting+2007}. 
32: \cite{XMM-Newton2010}. 
33: Averaged the coordinates of the two galaxies. 
\par\noindent
$*$: Observed in the Galactic coordinates, $(l, b) = (\timeform{127D40'11.939''}, \timeform{27D42'33.173''})$. 
\par\noindent
$\dag$: Observed in the Galactic coordinates, $(l, b) = (\timeform{129D14'20.697''}, \timeform{27D32'47.348''})$. 
\par\noindent
$\ddag$: We observed NGC\,2775 wrongly with the reference position of $(\alpha_{\mathrm{J}2000.0}, \delta_{\mathrm{J}2000.0}) = (\timeform{09h10m20.12s}, \timeform{+07D02'16.06''})$.
\par\noindent
$\S$: We observed NGC\,3079 with the reference position of $(\alpha_{\mathrm{J}2000.0}, \delta_{\mathrm{J}2000.0}) = (\timeform{10h01m57.80s}, \timeform{+55D40'47.2''})$.
\par\noindent
$\parallel$: We observed NGC\,3169 with the reference position of $(\alpha_{\mathrm{J}2000.0}, \delta_{\mathrm{J}2000.0}) = (\timeform{10h14m15.050s}, \timeform{+03D27'57.875''})$.
\par\noindent
$\sharp$: We observed NGC\,3147 with the reference position of $(\alpha_{\mathrm{J}2000.0}, \delta_{\mathrm{J}2000.0}) = (\timeform{10h16m53.65s}, \timeform{+73D24'02.695''})$.
\par\noindent
$**$: We observed NGC\,4527 wrongly with the reference position of $(\alpha_{\mathrm{J}2000.0}, \delta_{\mathrm{J}2000.0}) = (\timeform{12h34m08.427s}, \timeform{+02D39'13.19''})$. 
\par\noindent
$\dag \dag$: We adopted the declination different from the average of both galaxies to avoid observing a large blank sky. 
\par\noindent
$\ddag \ddag$: We observed Arp\,094 wrongly with the reference position of $(\alpha_{\mathrm{J}2000.0}, \delta_{\mathrm{J}2000.0}) = (\timeform{10h23m28.08s}, \timeform{+19D52'54''})$. 
\par\noindent
$\S \S$: The original reference is unidentified. 
\par\noindent
Column (5) Mapping size in the observation coordinates, $X$ (major axis) $\times$ $Y$ (minor axis). 
\par\noindent
Column (6) Position angle (PA) of the direction of $+X$. 
\par\noindent
Column (7) Reference and notes of PA. 
No remarks: \cite{deVaucouleurs+1991}. 
1: \cite{Fingerhut+2010}. 
2: \cite{Guhathakurta+1988}. 
3: \cite{Helfer+2003}. 
4: \cite{Jarrett+2003}. 
5: \cite{Koda+2002}. 
6: \cite{Koopmann+2001}. 
7: \cite{Kuno+2007}. 
8: \cite{Leroy+2009}. 
9: \cite{Moshir+1990}. 
10: \cite{Nishiyama+2001a}. 
11: \cite{Paturel+2003}. 
12: \cite{SDSS2007}. 
13: \cite{TwoMASS2003}. 
14: \cite{Vaduvescu+2008}. 
15: \cite{Wilson+2012}. 
\par\noindent
$\parallel \parallel$: Not kinematically determined but PA of observation. 
\par\noindent
Column (8) Tracking velocity with respect to the local standard of rest (LSR) in radio definition. In most cases, it corresponds to the receding velocity of galaxies. 
\par\noindent
Column (9) Reference and notes of $V_{\mathrm{LSR}}$. 
No remarks: \cite{deVaucouleurs+1991}. 
1: \cite{Boselli+2014a}. 
2: \cite{Bushouse+1998}
3: \cite{Huchra+1999}. 
4: \cite{Kuno+2007}. 
5: \cite{NEDTEAM1992}. 
6: \cite{Springob+2005}. 
7: \cite{Theureau+1998}. 
8: \cite{Toloba+2014}. 
9: \cite{Wilson+2012}. 
10: \cite{Yamauchi+2004}. 
$\sharp \sharp$: We adopted a wrong value. $V_{\mathrm{LSR}} = 1737\ \mathrm{km\ s}^{-1}$ is the correct value. 
$***$: Not referring NED. 
$\dag \dag \dag$: Averaged the two galaxies values. 
$\ddag \ddag \ddag$: We miscalculated the average velocity of the two galaxies, and the adopted value differed by $60\ \mathrm{km\ s}^{-1}$ from both galaxies. 
$\S \S \S$: We adopted a wrong value. $V_{\mathrm{LSR}} = 1452\ \mathrm{km\ s}^{-1}$ is the correct value. 
$\parallel \parallel \parallel$: The value listed in the reference differs by more than $200\ \mathrm{km\ s}^{-1}$ from both galaxies. 
\par\noindent
Column (10) Observation period in table \ref{tab:obsterms}.
}\hss}}
\endlastfoot
 \hline
\multicolumn{10}{c}{isolated galaxies}\\
 \hline
IC\,10	&	\timeform{00h20m17.34s}	&	\timeform{+59D18'13.6''}	&			& $\timeform{	366	''} \times \timeform{	330	''}$ & $\timeform{	125	D}$ &	14		&	-340	&	3	&	2	\\
\end{longtable}

Telescope pointing was checked roughly every hour by observing a bright source near the target galaxy, resulting in a pointing accuracy higher than $\timeform{5''}$. 
We observed Galactic SiO maser sources in SiO $(v=1, J=1-0)$ (rest frequency: $43.122090\,\mathrm{GHz}$) and $(v=2, J=1-0)$ (rest frequency: $42.820570\,\mathrm{GHz}$). 
We observed quasars in continuum at $43\,\mathrm{GHz}$ when no strong SiO maser sources were found near the target. 
We did not use the data taken between these two pointing observations if the offset of the telescope pointing between the two pointing observations was larger than $\timeform{5''}$. 
We also did not use the data taken under the bad condition of the radio seeing higher than $\timeform{5''}$ -- $\timeform{8''}$, except for a few instances where the seeing monitors malfunctioned.

The intensity calibration was made with the chopper wheel method. 
The hot load was observed for every approximately 10 min. 
Corrections among the four beams of FOREST were made through the observations of the standard sources every observation day. 
The standard sources were the W\,3\,core [$(\alpha, \delta)_{\mathrm{B}1950.0} = (\timeform{2h21m53s.2}, \timeform{+61D52'21''.0})$] or IRC\,+10216 [$(\alpha, \delta)_{\mathrm{B}1950.0} = (\timeform{09h45m15s.0}, \timeform{+13D30'45''.0})$]. 
We also tried observing TMC-1 [$(\alpha, \delta)_{\mathrm{B}1950.0} = (\timeform{04h38m38s.6}, \timeform{+25D35'45''})$], but the emission was very weak; thus, we did not use it as a standard source. 
We mapped a region with a side of approximately $\timeform{3'}$ around the standard source with all four beams of FOREST. 
The intensities of the IRC\,+10216 and W\,3\,core showed a daily variation of $\sim\,1 - 8\%$, which might be caused by the uncertainty of the IF attenuators. 
Subsection \ref{subsection:calibration} presents the details on the calibration method.

\subsection{Observation ranking system}
\label{subsection:ObservationRankingSystem}
We introduced an observation ranking system in the last two observation seasons, which typically reduced the total observing time to complete a map by 37\,\% compared to the first two seasons without this system. 
Selecting an optimum target from a sample of various right ascensions and declinations is a non-trivial exercise. 
Thus, we developed an observation system that ranked and selected the optimal target to minimize the : (1) observation time for each galaxy, (2) slewing time of the telescope, and (3) unusable data caused by pointing inaccuracy. 
We also put priority on finishing a galaxy instead of starting to map a new one, if possible, to promptly progress the data analysis.

We numerically express these factors as follows and compare the weights of the products of the factors, a priority index ``$R$'': 
\begin{equation}
R \equiv f_{\mathrm{obs, eff}}\ f_{\mathrm{slw}}\ f_{\mathrm{point}}\ f_{\mathrm{prop}}\ f_{\mathrm{comp}}, 
\label{eq:definition_R}
\end{equation}
where $f_{\mathrm{obs, eff}}$ is a factor representing the observation efficiency; $f_{\mathrm{slw}}$ represents the antenna slewing time; $f_{\mathrm{point}}$ denotes the pointing accuracy; $f_{\mathrm{prop}}$ expresses the scientific priority within our project team; and $f_{\mathrm{comp}}$ is the degree of completeness for a given target galaxy. 
A higher observation priority was given to the targets with a higher $R$ value. 
Quantity $R$ was immediately updated through a quick data reduction after each observation. 
Each factor and evaluation method were described in the subsections that follow, but note that this is somewhat a trial case for this system, and is, by no means, perfectly optimized. 
The goal was simply to select an optimal target under various observational conditions.

\subsubsection{Factor of observation efficiency}
\label{subsubsection:f_obs}
Factor $f_{\mathrm{obs,eff}}$ consists of two factors: 
\begin{equation}
f_{\mathrm{obs, eff}} \equiv f_{\mathrm{atm}}\ f_{\mathrm{map}},
\label{eq:definition_eta-eff}
\end{equation}
where $f_{\mathrm{atm}}$ represents an atmospheric factor, and $f_{\mathrm{map}}$ represents a mapping factor. 
We should observe a target with as low $T_{\mathrm{sys}}$ as possible, especially in an apparently large galaxy, to reduce the total survey time.

\begin{figure}
 \begin{center}
  \includegraphics[width=8cm]{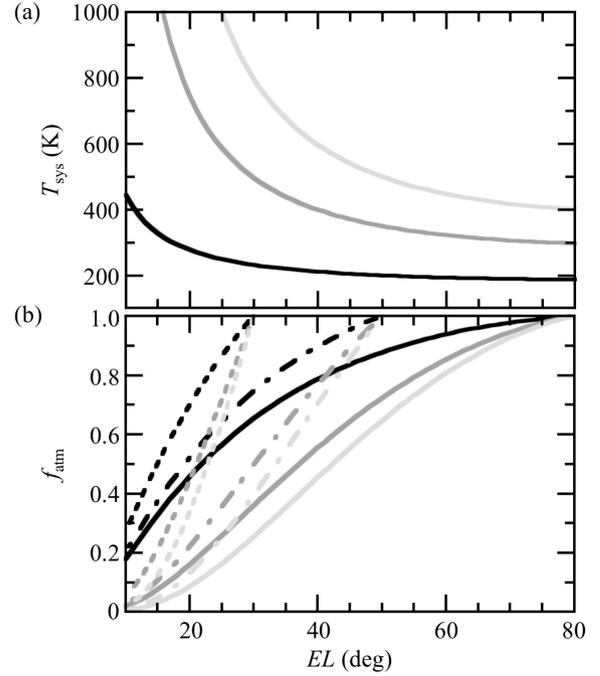}
 \end{center}
\caption{(a) System noise temperature ($T_{\mathrm{sys}}$) as a function of the elevation ($EL$). 
Lines for $T_{\mathrm{sys}} = 200\,\mathrm{K}$ ({\it black}), $350\,\mathrm{K}$ ({\it dark gray}), and $500\,\mathrm{K}$ ({\it light gray}) at $EL = \timeform{50D}$ are plotted. 
(b) Atmospheric factor ($f_{\mathrm{atm}}$) as a function of the elevation. 
The colors are the same as (a). 
The lines for the elevation at the culmination of $\timeform{80D}$ ({\it solid line}), $\timeform{50D}$ ({\it dash-dotted line}), and $\timeform{30D}$ ({\it dotted line}) are plotted.}
\label{fig:Tsys-fatm_EL}
\end{figure}

We formulated a $f_{\mathrm{atm}}$ weighted by $T_{\mathrm{sys}}$ because the observations at a lower $T_{\mathrm{sys}}$ [i.e., at a higher elevation ($EL$), figure \ref{fig:Tsys-fatm_EL}a] saved our limited observing time. 
The factor had a higher weight when a target was observed at a higher $EL$. 
Scaling is presented as follows: 
\begin{equation}
f_{\mathrm{atm}} \equiv \frac{t_{\mathrm{integ}} (EL_{\mathrm{max}})}{t_{\mathrm{integ}} (EL)} = \left [ \frac{T_{\mathrm{sys}}(EL_{\mathrm{max}})}{T_{\mathrm{sys}}(EL)} \right ]^2, 
\label{eq:definition_eta-atm}
\end{equation}
where $t_{\mathrm{integ}} (EL)$ is the necessary integration time for the observation at an elevation $EL$, and $EL_{\mathrm{max}}$ is the upper culmination $EL$ or $\timeform{80D}$, the latter being the $EL$ limit for the telescope operation. 
$T_{\mathrm{sys}}(EL)$ is $T_{\mathrm{sys}}$ at the $EL$ of the target. 
$T_{\mathrm{sys}}(EL)$ was calculated using the measured value before the observation and assuming that an optical depth only depends on the $EL$. 
That is, $f_{\mathrm{atm}}$ did not include effects caused by weather fluctuations. 
The $f_{\mathrm{atm}}$ factor was plotted in figure \ref{fig:Tsys-fatm_EL}b for various conditions.

\begin{figure}
 \begin{center}
  \includegraphics[width=8cm]{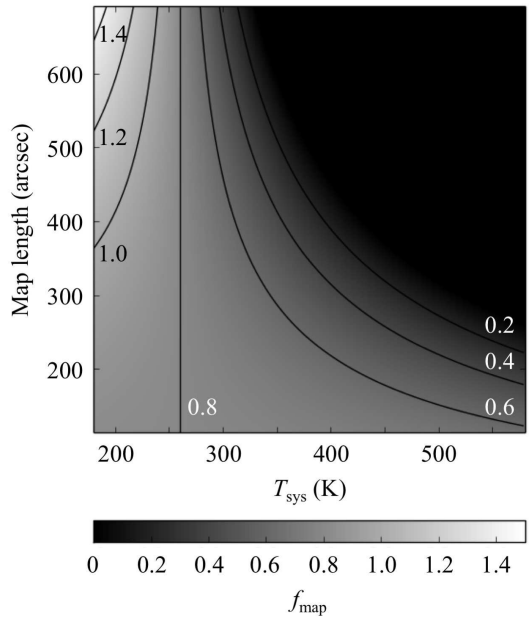}
 \end{center}
\caption{Mapping factor ($f_{\mathrm{map}}$) as a function of the system noise temperature and the map length. 
The target map is assumed to be square. 
Several contours are shown at the indicated levels spaced by 0.2.}
\label{fig:fct_map}
\end{figure}

We defined $f_{\mathrm{map}}$ to observe apparently larger galaxies under a lower $T_{\mathrm{sys}}$ and smaller galaxies under a higher $T_{\mathrm{sys}}$. 
We expressed this behavior as the relative integration time necessary for our survey sensitivity of $T_{\mathrm{A}}^{*} = 30\,\mathrm{mK}$. 
\begin{equation}
f_{\mathrm{map}} \equiv 0.8 + \frac{t_{\mathrm{integ}} (T_{\mathrm{sys}, 260}) - t_{\mathrm{integ}} (T_{\mathrm{sys}})}{t_{\mathrm{integ}} (T_{\mathrm{sys}, 260})} \times \frac{t_{\mathrm{dump}}}{t_{\mathrm{pix}}} \times 5, 
\label{eq:definition_eta-map}
\end{equation}
where $t_{\mathrm{integ}} (T_{\mathrm{sys}, 260})$ and $t_{\mathrm{integ}} (T_{\mathrm{sys}})$ are the necessary integration times under the condition of $T_{\mathrm{sys}} = 260\,\mathrm{K}$ of FOREST at $115\,\mathrm{GHz}$, which is typical of good observing conditions, and $T_{\mathrm{sys}}$ at the present instant, respectively. 
$t_{\mathrm{pix}}$ is the integration time for $1\>\mathrm{pixel}$ per observing script for OTF mapping. 
$t_{\mathrm{pix}}$ decreased with the size of the target galaxies because the scan speed was different from one galaxy to another. 
The dump time of the backend $t_{\mathrm{dump}}$ was introduced for normalization. 
Factors 0.8 and 5 were adopted as constants such that all factors have a similar weight. 
Figure \ref{fig:fct_map} shows the $f_{\mathrm{map}}$ behavior for a square-shaped observing map. 
In principle, efficiency has no maximum value for very low $T_{\mathrm{sys}}$; however, such conditions are unlikely in practice.

\subsubsection{Other factors}
\label{subsubsection:f_slw}
The speed of the telescope slew was $\sim\,\timeform{18D}\>\mathrm{min}^{-1}$; therefore, we defined an antenna-tracking factor to provide a higher weight for the closer target to the present telescope direction: 
\begin{equation}
f_{\mathrm{slw}} \equiv 0.5 + \frac{1}{1 + |AZ_{\mathrm{tel}} - AZ_{\mathrm{obj}}| / 90},
\label{eq:definition_eta-slw}
\end{equation}
where $AZ_{\mathrm{tel}}$ is the present azimuth of the telescope, and $AZ_{\mathrm{obj}}$ is the azimuth of the target object. 
Factor 0.5 was introduced such that $f_{\mathrm{slw}}$ became unity when the 45 m telescope caught up with a target in 5 min (i.e., the target was $\timeform{90D}$ away in azimuth).

Telescope pointing affects the usable data acquisition. 
The pointing factor $f_{\mathrm{point}}$ was affected by an angle between the telescope and the wind direction and the telescope elevation. 
A pointing source with a weak intensity needs longer integration time and iteration for accurate pointing. 
However, we simply set this factor to unity because no suitable relation was found between pointing and wind during the observational period.

We introduced factor $f_{\mathrm{prop}}$ for weighting by our scientific priority. 
$f_{\mathrm{prop}}$ is defined by the values of ``default'' priority in a decreasing order of: A, B, or C and the number of proposals by our project members containing the target, $N$, as: 
\begin{equation}
f_{\mathrm{prop}} \equiv R_{\mathrm{priority}} + \frac{N}{20}. 
\label{eq:definition_P-prop}
\end{equation}
Here, $R_{\mathrm{priority}}$ is a coefficient corresponding to the default priority, that is, 1.0 for priority A, 0.8 for B, and 0.0 for C.

The completeness of the observation is expressed as: 
\begin{equation}
f_{\mathrm{comp}} \equiv 1 + \frac{x}{1000}, 
\label{definition_f-comp}
\end{equation}
where $x$ is the percentile of the observation completeness, and $f_{\mathrm{comp}} = 0$ is set when the observations of a specific target have been completed. 
The factor took a value between 1.0 and 1.1 for the uncompleted targets.

\section{Data reduction}
\label{section:reduction}
Data reduction consisted of two parts: intensity scaling of the four beams of FOREST to a $T_{\mathrm{MB}}$ scale and data integration with baseline subtraction. 
We developed \textsc{python} scripts for the automatic data reduction: COMING Auto-Reduction Tools (COMING ART) for the latter step, which allowed the data reduction to be highly objective and reproducible.

\subsection{Calibration}
\label{subsection:calibration}
\begin{figure*}
 \begin{center}
  \includegraphics[width=16cm]{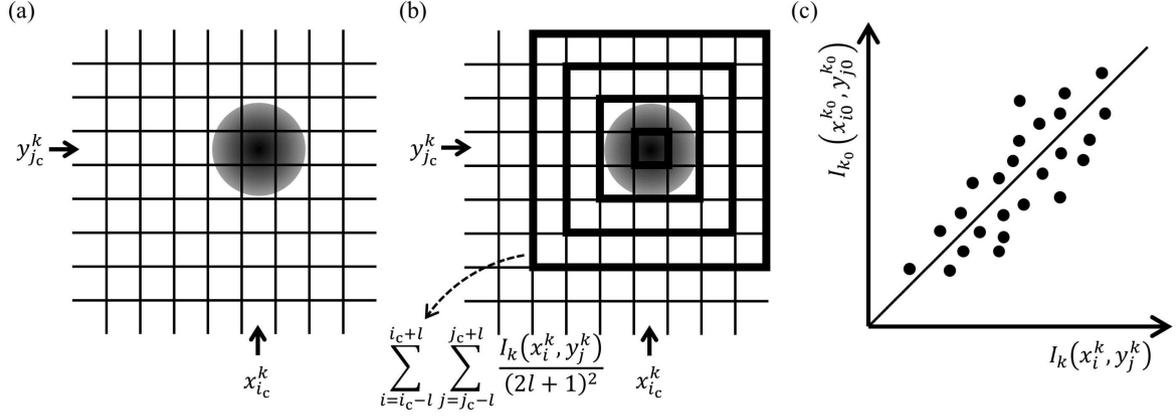}
 \end{center}
\caption{Illustration of our calibration methods. 
(a) First, the pixel corresponding to the peak intensity is identified. 
(b) An average is then taken over $l \times l\>\mathrm{pixels}$ ($l = 0, 1, 2, ...$) around the peak pixel $(x_{i_{\mathrm{c}}}^{k}, y_{j_{\mathrm{c}}}^{k})$. 
The averaged pixels are surrounded by the thick squares in the case of $l = 0, 1, 2, 3$. 
(c) Correlation among the $l \times l\>\mathrm{pixels}$ pixels around the peak position of the standard array and those of the other arrays was measured by the least squares fitting with a linear function.}
\label{fig:calib_methods}
\end{figure*}

We corrected the relative intensity among the observational instruments using the integrated intensity of the standard sources taken with each beam and the polarization of FOREST. 
First, we made the integrated intensity maps in the \atom{C}{}{12}\atom{O}{}{} and \atom{C}{}{13}\atom{O}{}{} lines observed with each beam in $25 \times 25\>\mathrm{pixels}$ in $\timeform{7''.5}$ spacing. 
The pixel spacing was different from the target maps of $\timeform{6''}$ (subsection \ref{subsection:COMING-ART}), but this did not affect our calibration results because 1) the spacing was near the Nyquist rates, and we compared the relative intensities among the calibration data and 2) we searched for the peak position in each \atom{C}{}{12}\atom{O}{}{} map (figure \ref{fig:calib_methods}a). 
We adopted the peak position of \atom{C}{}{12}\atom{O}{}{} with the same beam and polarization as the \atom{C}{}{13}\atom{O}{}{} peak because the \atom{C}{}{13}\atom{O}{}{} intensity was comparatively weak, making the definition of a precise peak position difficult. 
Some integrated intensity maps and the spectrum of the peak position appeared inconsistent when taken under poor pointing or seeing conditions. 
We did not use such data for the calibration.

One of the IF bands of beam 1, which seemed to be stable within each observation season, was adopted as the standard array. 
That is, the intensity scale taken with the other arrays was calibrated to the data taken with the standard array. 
Beam 1 was selected because it was on the optic axis of the telescope, and the beam efficiency was measured by the NRO.

Most of the integrated intensity maps of the calibration source obtained during the period of 2018 March to April contained unexpected artifacts in a band of pixels because the reference frequency signal shifted during the OTF scans, resulting in parts of the observed data block being separated in frequency. 
One source object block was constructed by combining the separated data blocks, but this resulted in an erroneous band of pixels at the separation boundary. 
Thus, this contaminated region was masked and reconstructed using the information from the unaffected areas. 
The reconstruction was done using a modified version of the Papoulis-Gerchberg algorithm, which extrapolates the masked region by iterative Fourier transformations with certain assumptions. 
The reconstruction method was tested on the calibration images without issue, and a successful reconstruction was observed with an error of $\sim\,1\%$. 
A detailed explanation about the obtained artifact, methodology, and reconstruction results for the affected images can be found in \citet{Cooray+2019}.

The precise calibration required measurements of the intrinsic intensity of the standard sources that was estimated by avoiding bad weather conditions as much as possible. 
The pointing errors induced by sudden wind or halation in an image because of bad seeing reduced the intensity calibration precision. 
Hence, we measured the peak intensity and the intensities averaged over $3 \times 3$, $5 \times 5$, $7 \times 7$, $9 \times 9$, and $11 \times 11\>\mathrm{pixels}$ around the peak of the integrated intensity map, and compared them with the corresponding data taken with the standard array.

We measured the temporal scaling factors in the two following ways: 
The first is defined as: 
\begin{eqnarray}
f_{1, l}^{k} &=& \sum_{i_{0} = i_{\mathrm{c}0} - l}^{i_{\mathrm{c}0} + l} \sum_{j_{0} = j_{\mathrm{c}0} - l}^{j_{\mathrm{c}0} + l} I_{k_0}(x_{i_0}^{k_0}, y_{j_0}^{k_0}) / \sum_{i = i_{\mathrm{c}} - l}^{i_{\mathrm{c}} + l} \sum_{j = j_{\mathrm{c}} - l}^{j_{\mathrm{c}} + l} I_{k}(x_{i}^{k}, y_{j}^{k}) \nonumber \\
&&(l = 0, 1, 2, 3, 4, 5). 
\label{eq:sclfct1}
\end{eqnarray}
Here, $l = 0, 1, 2, ...$ correspond to using only the peak, $3 \times 3\>\mathrm{pixels}$ around the peak, $5 \times 5\>\mathrm{pixels}$ around the peak, and so on (figure \ref{fig:calib_methods}b). 
We described each pixel in a $25 \times 25\>\mathrm{pixels}$ map taken with the array ``$k$'' as $(x_{i}^{k}, y_{j}^{k}) (i, j = 1, 2, 3, ..., 25; k = 1, 2, 3, ..., 16)$ and the found peak position as $(x_{i_{\mathrm{c}}}^{k}, y_{j_{\mathrm{c}}}^{k})$. 
Subscript ``$0$'' represents the standard array value. 
$I_{k}(x_{i}^{k}, y_{j}^{k})$ is the integrated intensity at the pixel $(x_{i}^{k}, y_{j}^{k})$ of the array ``$k$'' and $k_{0}$ represents the standard array. 
The other is defined as a gradient of the least squares fitting of the linear function (figure \ref{fig:calib_methods}c): 
\begin{eqnarray}
I_{k_0}(x_{i_0}^{k_0}, y_{j_0}^{k_0}) &=& f_{2, l}^{k} I_{k}(x_{i}^{k}, y_{j}^{k}) \nonumber \\
&&(i_0 = i_{\mathrm{c}0} - l, i_{\mathrm{c}0} - l + 1, ..., i_{\mathrm{c}0} + l; \nonumber \\
&& \ j_0 = j_{\mathrm{c}0} - l, j_{\mathrm{c}0} - l + 1, ..., j_{\mathrm{c}0} + l; \nonumber \\
&& \ i = i_{\mathrm{c}} - l, i_{\mathrm{c}} - l + 1, ..., i_{\mathrm{c}} + l; \nonumber \\
&& \ j = j_{\mathrm{c}} - l, j_{\mathrm{c}} - l + 1, ..., j_{\mathrm{c}} + l; \nonumber \\
&& \ l = 1, 2, 3, 4, 5).
\label{eq:sclfct2}
\end{eqnarray}

We compared these temporally scaling factors, $f_{1, l}^{k}\ (l = 0, 1, 2, 3, 4, 5)$, $f_{2, l}^{k}\ (l = 1, 2, 3, 4, 5)$, and found that the average of $11 \times 11\>\mathrm{pixels}$ around the peak ($f_{1, 5}^{k}$) seemed to be the most stable. 
However, the intensity showed a noticeable scatter between the observational days even when using $11 \times 11\>\mathrm{pixels}$. 
Thus, we did not calibrate the data daily, but calibrated them using the averaged scaling factor over some observational periods (3 -- 83 days, typically $\sim\,20$ days) divided by the maintenance of FOREST or the local oscillators that possibly affected the intensity scaling.

The data taken with the standard arrays were multiplied by the IRR correction factors as follows: 
\begin{equation}
f_{\mathrm{IRR}} = 1 + 10^{- \frac{IRR}{10}}
\label{eq:irr}
\end{equation}
where $IRR$ is the image rejection ratio in dB measured at the beginning of every observation day (subsection \ref{subsection:SystemSetting}). 
After this step, the intensity scale of the standard array became the $T_{\mathrm{A}}^{*}$ scale. 
This correction was not done for the data taken with the other arrays because we directly scaled the intensity taken with those arrays with the intensity of the standard array in the $T_{\mathrm{A}}^{*}$ scale.

The data of the standard sources taken in the second observation season were divided by the main beam efficiency ($\eta_{\mathrm{MB}}$) of the 45 m telescope, while those taken in the other seasons were scaled to be the same intensity as the former data. 
The main beam efficiencies of the telescope of $\eta_{\mathrm{MB}}(115\,\mathrm{GHz}) = 45\,\pm\,2\%$ and $\eta_{\mathrm{MB}}(110\,\mathrm{GHz}) = 43\,\pm\,2\%$ provided by the NRO\footnote{$\langle$https://www.nro.nao.ac.jp/\textasciitilde nro45mrt/html/prop/eff/eff2015.html$\rangle$.} were used to convert the data taken in the second observation season into the $T_{\mathrm{MB}}$ scale. 
Only $\eta_{\mathrm{MB}}(115\,\mathrm{GHz}) = 39\,\pm\,3\%$ was public\footnote{$\langle$https://www.nro.nao.ac.jp/\textasciitilde nro45mrt/html/prop/eff/eff2014.html$\rangle$.} in the first observation season, and $\eta_{\mathrm{MB}}$ was not measured in the third observation season because of an unforeseen telescope impairment. 
We also did not use $\eta_{\mathrm{MB}}(115\,\mathrm{GHz}) = 43\,\pm\,4\%$ and $\eta_{\mathrm{MB}}(110\,\mathrm{GHz}) = 44\,\pm\,4\%$ in the fourth observation season\footnote{$\langle$https://www.nro.nao.ac.jp/\textasciitilde nro45mrt/html/prop/eff/eff\_latest.html$\rangle$.} because the intensity of the standard objects was inconsistent with the previous values when we used these efficiencies. 
Therefore, we did not divide the data taken in the three seasons, except for the second observation season by $\eta_{\mathrm{MB}}$. 
However, we used the scaling factors that corresponded to the intensity of the standard sources to those taken in the second observation season.

\subsection{Auto-reduction scripts, COMING ART}
\label{subsection:COMING-ART}
\begin{figure*}
 \begin{center}
  \includegraphics[width=16cm]{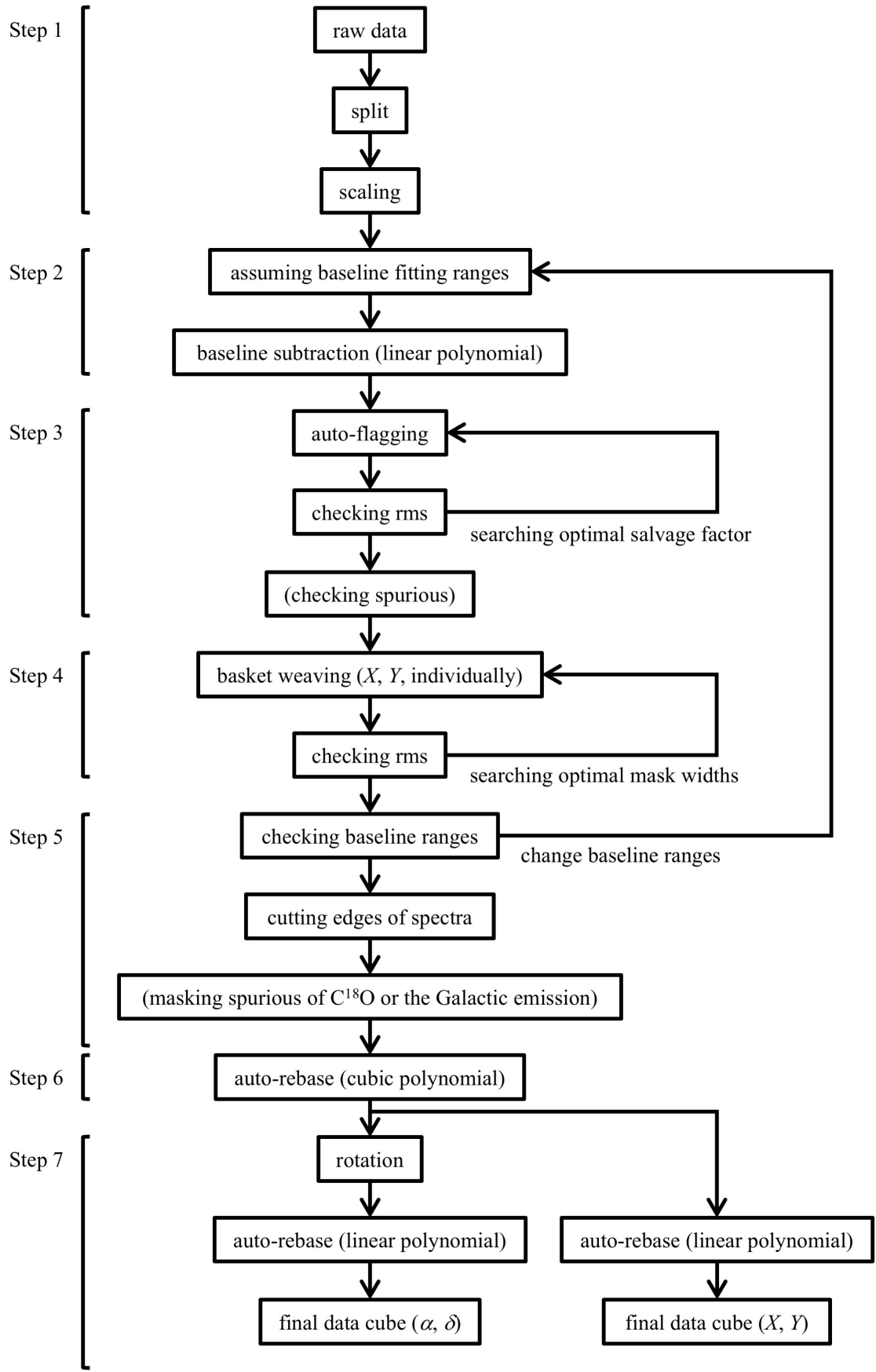}
 \end{center}
\caption{Concept of data reduction scripts, COMING ART.}
\label{fig:COMING-ART}
\end{figure*}

We developed and used the data reduction scripts COMING ART based on the Nobeyama OTF Software Tools for Analysis and Reduction (NOSTAR) developed by the NRO. 
These scripts were composed of several steps, including the evaluation of the baseline undulation, flagging of bad data, basket weaving, and automatic baseline subtraction. 
Figure \ref{fig:COMING-ART} shows an overview of the procedures in the pipeline. 
We could reduce data with very high objectivity and reproducibility using the COMING ART. 
Each step in figure \ref{fig:COMING-ART} is explained below.

\subsubsection{Step 1: data formation and intensity scaling}
\label{subsubsection:COMING-ART-Step1}
CO maps were generated as a data cube with $\timeform{6''}$ spacing and a velocity width of $10\,\mathrm{km\>s}^{-1}$. 
We selected a correlator dump time of SAM45 of 0.1\,s according to \citet{Sawada+2008}, which resulted in an OTF scan speed of approximately $\timeform{4''}$ to $\timeform{39''}$ per second depending on the map size. 
The data were split with a task {\sf Split} in the NOSTAR and calibrated with a task {\sf Scaling} in the NOSTAR using the scaling factors described in subsection \ref{subsection:calibration}. 
We produced data cubes in the abovementioned intervals according to \citet{Sawada+2008}. 
The effective spatial resolution of the OTF maps was $\timeform{17''}$ in both $115\,\mathrm{GHz}$ and $110\,\mathrm{GHz}$, which was larger than the size of the observation beam of $\timeform{14''}$. 
The data cubes were then subject to the reduction steps described below.

\subsubsection{Step 2: baseline subtraction}
\label{subsubsection:COMING-ART-Step2}
A linear baseline was fitted for the two velocity ranges with $350\,\mathrm{km\>s}^{-1}$ width on both sides of the emission range using the task {\sf Baseline} in the NOSTAR. 
The velocity range of emission was based on the previous \atom{C}{}{12}\atom{O}{}{} or H\,\emissiontype{I} data rounded to a multiple of $10\,\mathrm{km\>s}^{-1}$. 
The two baseline ranges were immediately set adjacent to the emission range.

\subsubsection{Step 3: discarding heavily undulated spectra (``{\sf auto-flag}'')}
\label{subsubsection:COMING-ART-Step3}
\begin{figure*}
 \begin{center}
  \includegraphics[width=16cm]{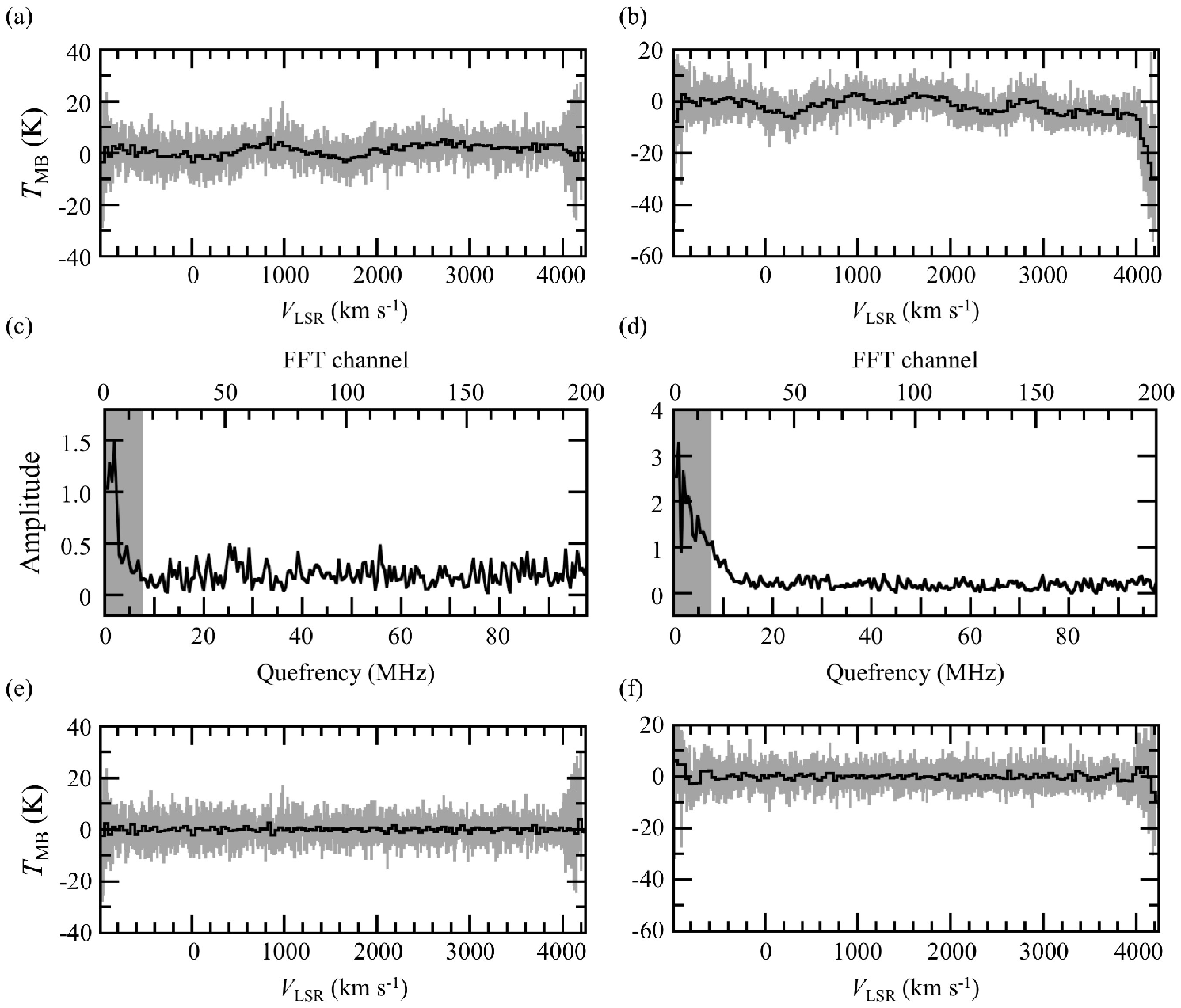}
 \end{center}
\caption{Examples of the undulating spectra of NGC\,337. 
(a), (b) Gray lines are original spectra and black lines are 32-ch smoothed spectra. 
(c), (d) The Fourier transforms of (a) and (b), respectively. 
Data below 200 channels are plotted. 
Gray hatched regions indicate below 15 FFT channels, which corresponds to $7.3\,\mathrm{MHz}$ in quefrency. 
(e), (f) Spectra masked lower than $7.3\,\mathrm{MHz}$ in quefrency of (a) and (b), respectively. 
Gray lines are masked spectra and black lines are 32-ch smoothed spectra.}
\label{fig:undulated_FFT}
\end{figure*}

This step begins with evaluating the rms noise of an ideal non-undulated spectrum using a fast Fourier transform (FFT) analysis and removing lower-frequency components. 
The baseline undulation over a wide frequency range was non-trivial because distinguishing it from the wide and weak emission line of a galaxy was very difficult. 
Even a loose undulation increased the rms noise and prevented the evaluation of the degree of undulation; thus, we performed an FFT analysis and calculated the rms noise of the spectrum whose frequency components lower than $7.3\,\mathrm{MHz}$ were removed, $\sigma_{\mathrm{non-smoothed, FFT}}$. 
The $7.3\,\mathrm{MHz}$ frequency was empirically adopted from the analysis of the observed data of several galaxies. 
Figures \ref{fig:undulated_FFT}a and b show examples of the undulated spectra. 
Figures \ref{fig:undulated_FFT}c and d depict the Fourier transforms of these undulated spectra. 
We obtained undulation-corrected spectra (figures \ref{fig:undulated_FFT}e and f) when we masked the frequency components lower than $7.3\,\mathrm{MHz}$ shown in the gray hatch in these panels.

\begin{figure}
 \begin{center}
  \includegraphics[width=8cm]{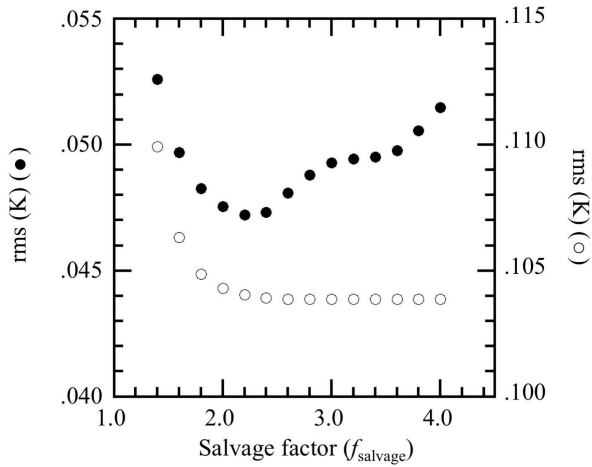}
 \end{center}
\caption{Behavior of the rms with the salvage factor ($f_{\mathrm{salvage}}$). 
The filled circles show an example of a case with a local minimum (NGC\,660), while the open circles depicts a case with a declining rms (NGC\,628).}
\label{fig:salvage_factor-rms}
\end{figure}

Second, we flagged and removed the poor-quality data after checking the rms noise behavior considering a {\it salvage factor}. 
When a spectrum is smoothed by summing up the channels, the rms noise should inversely decrease proportional to a square root of the number of the summed-up channels if the spectrum no longer suffers from undulation. 
Almost all the spectra will be identified as poor-quality data if we strictly apply this criterion; thus, we introduced a salvage factor $f_{\mathrm{salvage}}$, which was larger than unity and lower than $\sim\,4$, in the case of 
\begin{equation}
\frac{\sigma_{\mathrm{non-smoothed, FFT}}}{\sqrt{n}} \times f_{\mathrm{salvage}} < \sigma_{n\mathrm{ch-smoothed}}, 
\label{eq:salvage_factor}
\end{equation}
\noindent
where $\sigma_{n\mathrm{ch-smoothed}}$ is the rms noise for the $n-$channel smoothed spectrum. 
We will then discard that specific spectrum. 
We smoothed a spectrum of 4096 channels for every 32 channels and evaluated the baseline undulation by comparing the rms of the smoothed spectrum with that of the non-smoothed one. 
In this process, the rms was calculated for all channels, including the channels containing the CO emission, except for the NGC\,3034. 
The emission of NGC\,3034 was particularly strong; therefore, the rms was calculated for the channels over the default baseline ranges. 
The resultant rms noise had a local minimum or monotonously decreased with the increasing $f_{\mathrm{salvage}}$ (figure \ref{fig:salvage_factor-rms}) because removing the spectra with the undulated baselines resulted in a lower rms noise at first, but decreasing the number of spectra in the integration then increased the noise levels. 
We fixed $n$ as 32 and assigned 1.4 to 4.0 in steps of 0.2 for $f_{\mathrm{salvage}}$. 
We then searched for the minimum value of the resultant rms noise, and adopted the salvage factor when the resultant rms was minimum. 
This flagging method was called {\sf auto-flag}. 
After flagging the poor-quality data, the resulting data cube was constructed with a task {\sf Make Map} in the NOSTAR.

\subsubsection{Step 4: basket weaving}
\label{subsubsection:COMING-ART-Step4}
Basket weaving is an effective method of decreasing the impact of the scanning effect for scan observations (\cite{Emerson+1988}). 
The width is masked for a Fourier transformed map to reduce the scanning noise seen in the original map. 
Although the same mask width is adopted for the $X$- and $Y$-directions in a task {\sf Basket-Weave} in the NOSTAR, the procedure is not necessarily effective for a map with a large aspect ratio, such as an edge-on galaxy. 
Basket weaving was performed herein by independently assigning a mask width along the major ($X$-direction) and minor ($Y$-direction) axes of a given map, then adopting mask widths to minimize the resultant rms noise.

\subsubsection{Step 5: checking the baseline ranges and flagging of spurious channels}
\label{subsubsection:COMING-ART-Step5}
For the data whose signal range overflowed into the initial baseline range, the baseline range was revised, and the abovementioned procedures were iterated (figure \ref{fig:COMING-ART}); otherwise, the data were fine-tuned. 
Checking was made by the eye for a profile map and a spatially integrated total spectrum in \atom{C}{}{12}\atom{O}{}{} of each galaxy. 
After the {\it signal range} confirmation, we adopted the final {\it baseline ranges}, which were both sides of the signal range with a width of $200\,\mathrm{km\>s}^{-1}$. 
We then cut both outsides of the baseline range. 
We masked the channels to zero intensity when spiky spurious features were seen in the spectra. 
This procedure was applied to some data of \atom{C}{}{}\atom{O}{}{18}. 
The channels smeared by the Galactic emission in NGC\,1569 and UGCA\,86 were also masked.

\subsubsection{Step 6: baseline subtraction with a cubic polynomial (``{\sf auto-rebase}'')}
\label{subsubsection:COMING-ART-Step6}
We fitted the baseline again to optimize the zero level in each spectrum. 
We then applied a cubic polynomial as a baseline. 
We evaluated the possibility that a cubic polynomial fitting for the baseline extinguishes or reduces the galaxy emission. 
We calculated the rms noises in the {\it baseline ranges} and the {\it emission range} of a model Gaussian profile emission with sinusoidal undulation and various signal-to-noise ($S/N$) ratios. 
This simple evaluation indicated that the cubic polynomial fitting more effectively reduces the rms noise compared to a linear fitting without deleting the real emission.

We determined the emission ranges in the following manner: identifying which part of the spectrum indicates emission, in the case the velocity field of a galaxy is unknown, is not necessarily trivial. 
We created a data cube smoothed by $3 \times 3\>\mathrm{pixels}$ in space. 
Within this smoothed cube, we identified the channels with $S/N > 2.5$ and designated them as the {\it emission channels}. 
All remaining channels, including the $200\,\mathrm{km\>s}^{-1}$ range outside of the signal range, were identified as the {\it baseline channels} used to fit a final baseline. 
This technique is called {\sf auto-rebase}. 
The emission for the three \atom{C}{}{}\atom{O}{}{} lines was searched for in this manner, but the searching emission range of \atom{C}{}{13}\atom{O}{}{} and \atom{C}{}{}\atom{O}{}{18} was restricted to the emission channels of \atom{C}{}{12}\atom{O}{}{} because both lines were weak in comparison.

\subsubsection{Step 7: coordinate transformation and ``{\sf auto-rebase}''}
\label{subsubsection:COMING-ART-Step7}
The data cube was rotated to align with the equatorial coordinates. 
We performed {\sf auto-rebase} with a linear baseline subtraction again for the original and rotated data. 
The spatial resolution of a map along the $X$- and $Y$-directions was slightly higher than that of a map in the equatorial coordinates because the pixel interval of the equatorial map was coarser. 
Thus, we prepared both cubes, although we used the equatorial coordinate cubes for the subsequent analysis. 
The process after scaling was automated as a single \textsc{python} script.

We calculated the rms noise for each pixel over the whole baseline channels and made an rms noise map. 
The average of the rms noise map was recorded as the typical rms noise in the header of the resulting FITS data cube. 
We also made FITS cubes where the baseline channels were represented as unity, and the emission channels were represented as zero. 
The final data cubes are available at the Japanese Virtual Observatory (JVO) website\footnote{$\langle$https://jvo.nao.ac.jp/portal/nobeyama/coming.do$\rangle$.}.

\section{Data analysis}
\label{section:analysis}

\subsection{CO}
\label{subsection:CO}
The integrated intensity maps were constructed by summing up the emission channels for each pixel 
\begin{eqnarray}
I(\alpha, \delta) &=& \int{T_{\mathrm{MB}}(\alpha, \delta, v) dv} \nonumber \\
 &=& \sum_{k \in \mathrm{emission\>channels}}{T_{\mathrm{MB}}(\alpha, \delta, k) \Delta v_{\mathrm{ch}}}, 
\end{eqnarray}
for each line of \atom{C}{}{12}\atom{O}{}{}, \atom{C}{}{13}\atom{O}{}{}, or \atom{C}{}{}\atom{O}{}{18}, where $k$ is the channel number, and $\Delta v_{\mathrm{ch}}$ is the velocity width of a channel ($10\,\mathrm{km\>s}^{-1}$). 
The error of the integrated intensity was calculated as: 
\begin{equation}
\sigma_{\mathrm{integ}} = \sigma_{\mathrm{rms}} \times \sqrt{\Delta V_{\mathrm{signal}} \times \Delta v_{\mathrm{ch}}}, 
\label{eq:rms_integint}
\end{equation}
where $\sigma_{\mathrm{rms}}$ is the rms noise calculated over the baseline channels, and $\Delta V_{\mathrm{signal}}$ is the total velocity width of the emission channels. 
Note that $\Delta V_{\mathrm{signal}}$ became zero for the emission-free pixels when we applied the {\sf auto-rebase} to the data; thus, $\sigma_{\mathrm{integ}}$ was also zero following equation (\ref{eq:rms_integint}). 
This clearly underestimated the error of the integrated intensity of each pixel; therefore, we calculated the average width of the emission channels for the pixels with more than one emission channel and adopted this average as a typical emission width and applied it to equation (\ref{eq:rms_integint}) for the emission-free pixels. 
In the case of \atom{C}{}{}\atom{O}{}{18}, even this typical emission width became zero because the emission was hardly detected. 
We applied a typical line width of \atom{C}{}{13}\atom{O}{}{} instead of that of \atom{C}{}{}\atom{O}{}{18} in this case.

The total molecular gas mass of a galaxy $M_{\mathrm{mol}}$ was calculated by summing up the pixels within an infrared-defined radius described in subsection \ref{subsection:WISE3.4} and applying the standard conversion factor (\cite{Bolatto+2013}): 
\begin{equation}
X_{\atom{C}{}{12}\atom{O}{}{}} \equiv \frac{\mathcal{N}(\atom{H}{}{}_{2})}{I_{\atom{C}{}{12}\atom{O}{}{}}} = 2 \times 10^{20}\,[\mathrm{cm}^{-2}\>(\mathrm{K\>km\>s}^{-1})^{-1}].
\label{eq:Xco}
\end{equation}
This conversion factor had an error of 30\,\% (\cite{Bolatto+2013}), but we did not include this error in our error analysis. 
We also did not consider the calibration uncertainty of $<10\,\%$ when summing up the calibration fluctuation described in subsection \ref{subsection:SystemSetting} and the fluctuation in $\eta_{\mathrm{MB}}$ described in subsection \ref{subsection:calibration}. 
Thus, $M_{\mathrm{mol}}$ may have an uncertainty of $\sim\,30\,\%$, which is not explicitly included. 
We further multiplied by a factor of 1.36 to include the contributions from \atom{He}{}{}.

The first- and second-degree moment maps were made using the CASA routine {\sf immoments}. 
The first-degree moment 
\begin{equation}
\frac{\int{v T_{\mathrm{MB}}(\alpha, \delta, v) dv}}{\int{T_{\mathrm{MB}}(\alpha, \delta, v) dv}} \equiv \bar{v}
\end{equation}
provides an intensity-weighted mean velocity, while the second-degree moment 
\begin{equation}
\sqrt{\frac{\int{\left(v - \bar{v} \right)^{2} T_{\mathrm{MB}}(\alpha, \delta, v) dv}}{\int{T_{\mathrm{MB}}(\alpha, \delta, v) dv}}} \equiv \Delta v
\end{equation}
provides the velocity dispersion. 
We smoothed and masked the data following the method outlined in \citet{Miyamoto+2018} to avoid the invalid moment values caused by noise. 
First, the data were convolved with 1.5 times of the beam size, and then the intensity-weighted moments and the rms noise were calculated for the spatially smoothed cubes. 
Second, the pixels whose $S/N$ of the integrated intensity was lower than 4 were masked using these smoothed data.

\subsection{WISE $3.4\,\micron$}
\label{subsection:WISE3.4}
We analyzed the archival $3.4\,\micron$ band data of the WISE (\cite{Wright+2010}) All-Sky Survey to measure the stellar mass of our sample galaxies. 
According to \citet{Wen+2013}, a $3.4\,\micron$ luminosity provides a reasonable index of the stellar mass of galaxies. 
We downloaded the data corresponding to the coordinates of the galactic center listed in table \ref{tab:Galpars2} from the NASA/IPAC Infrared Science Archive\footnote{$\langle$http://irsa.ipac.caltech.edu/Missions/wise.html$\rangle$.}. 
The downloaded images covered more than two times of $D_{25}$. 
A pixel scale of the images was $\timeform{1''.375}$. 
The background level was estimated from a histogram of the pixel values within a ring between radii $2 \times R_{25}$ and $3 \times R_{25}$ ($R_{25} \equiv D_{25} / 2$) and subtracted following the documents from the WISE project\footnote{$\langle$http://wise2.ipac.caltech.edu/docs/release/allsky/expsup/sec1\_4c.html$\rangle$.}. 
Panel (a) of figures \ref{fig:maps_NGC0891}, \ref{fig:maps_NGC3627} and supplementary figures 1 -- 134 in the supplementary section of the online version depict the background-subtracted images.

\begin{figure*}
 \begin{center}
  \includegraphics[width=16cm]{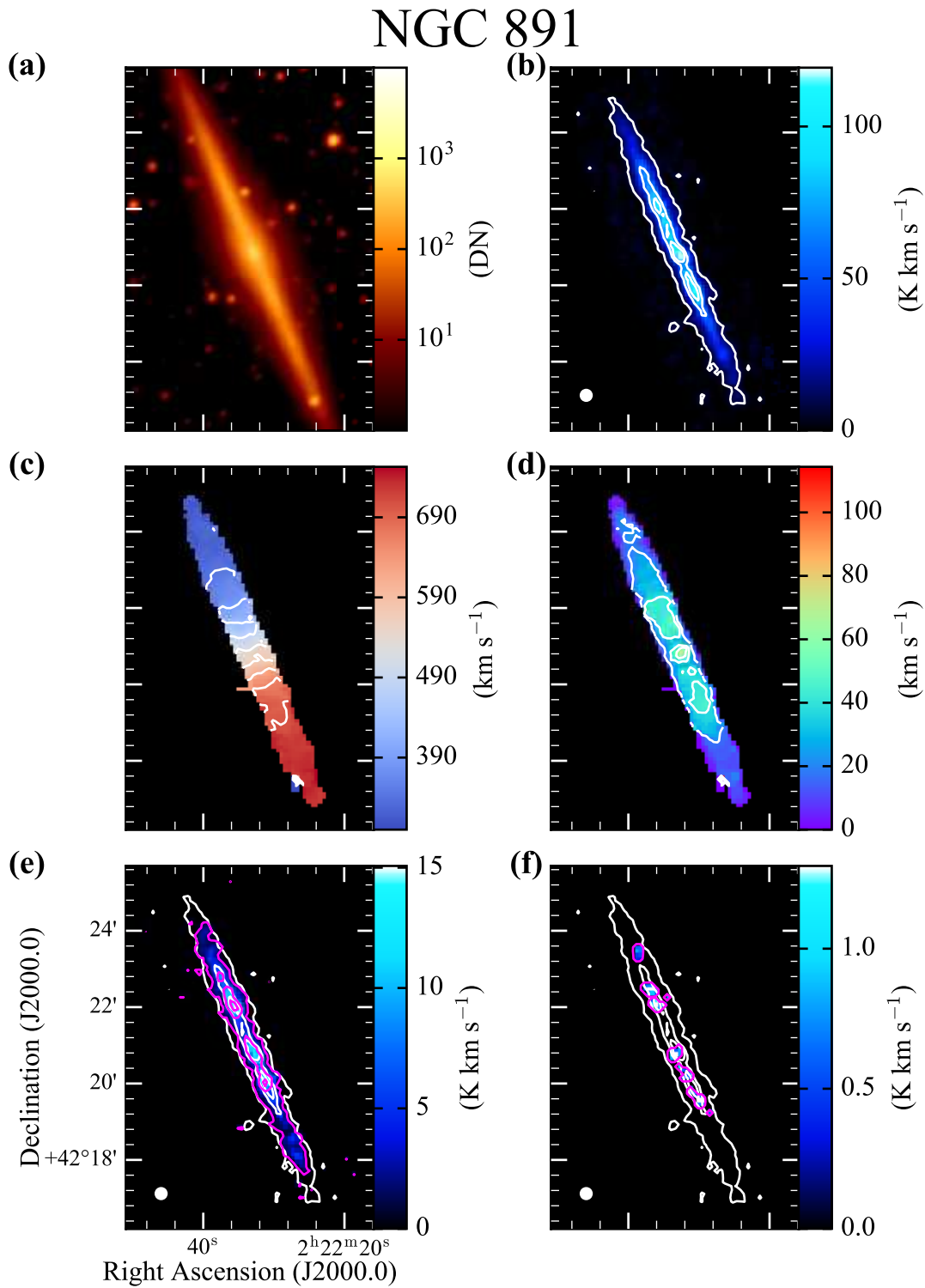}
 \end{center}
\caption{NGC\,891. (a) Background-subtracted WISE $3.4\,\micron$ image. 
(b) Integrated intensity of \atom{C}{}{12}\atom{O}{}{} $(J=1-0)$. 
The contours are plotted at 5\,\%, 40\,\%, and 60\,\% of the maximum intensity of $152.45\,\mathrm{K\>km\>s}^{-1}$. 
(c) First-degree moment map of \atom{C}{}{12}\atom{O}{}{}. 
The contours are in $45\,\mathrm{km\>s}^{-1}$ steps. 
(d) Second-degree moment map of \atom{C}{}{12}\atom{O}{}{} with the contours in steps of $20\,\mathrm{km\>s}^{-1}$. 
(e) Integrated intensity of \atom{C}{}{13}\atom{O}{}{} $(J=1-0)$. 
The white contours are the same as the panel (b). 
The magenta contours are plotted at 5\,\%, 45\,\% and 85\,\% of the maximum intensity of $16.53\,\mathrm{K\>km\>s}^{-1}$ in \atom{C}{}{13}\atom{O}{}{}. 
(f) Integrated intensity of \atom{C}{}{}\atom{O}{}{18} $(J=1-0)$. 
White contours are the same as the panel (b). 
The magenta contours are plotted at 5\,\% of the maximum intensity of $2.26\,\mathrm{K\>km\>s}^{-1}$ in \atom{C}{}{}\atom{O}{}{18}. 
The OTF beam size is indicated as a white filled circle in the bottom-left corner in panel (b), (e), and (f).}
\label{fig:maps_NGC0891}
\end{figure*}

\begin{figure*}
 \begin{center}
  \includegraphics[width=16cm]{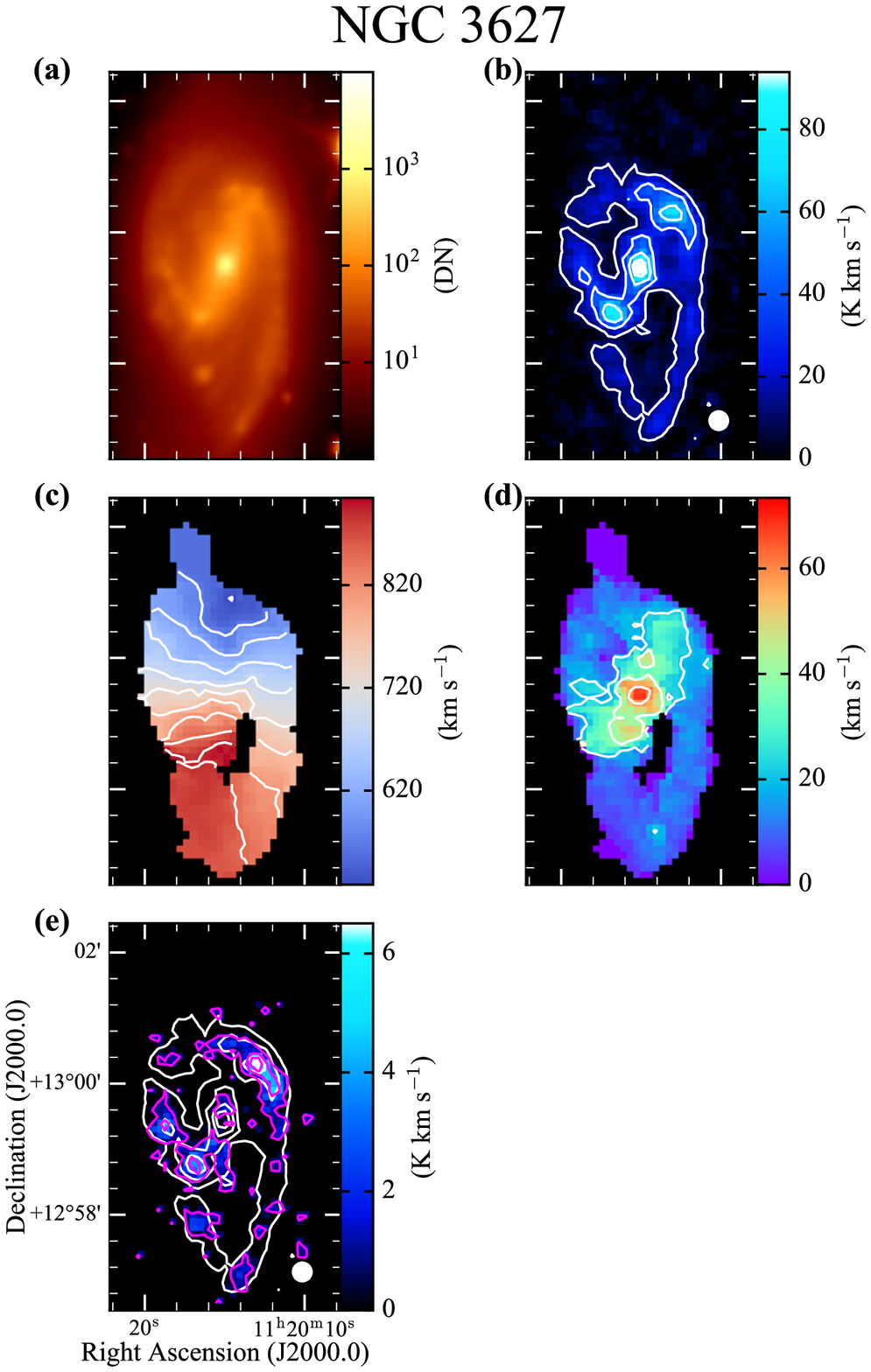}
 \end{center}
\caption{Same as figure \ref{fig:maps_NGC0891}, but for NGC\,3627. 
The contours are plotted at 10\,\%, 30\,\%, 55\,\%, and 80\,\% of the maximum intensity of $109.87\,\mathrm{K\>km\>s}^{-1}$ in (b) and (e) ({\it white}), in steps of $40\,\mathrm{km\>s}^{-1}$ in (c), in steps of $20\,\mathrm{km\>s}^{-1}$ in (d), and at 10\,\%, 45\,\%, and 80\,\% of the maximum intensity of $7.88\,\mathrm{K\>km\>s}^{-1}$ in (e) ({\it magenta}).}
\label{fig:maps_NGC3627}
\end{figure*}

Many Galactic stars were imaged in the WISE $3.4\,\micron$ images; hence, we roughly masked these stars to measure a more precise stellar mass. 
We identified the Galactic stars by simply assuming that they corresponded to bright, point-like sources. 
Although the typical resolution of the Atlas image of $3.4\,\micron$ is $\approx \timeform{8''.3}$, we fitted a Gaussian with a 1.3 times larger FWHM than the resolution to capture the extended halo around the brightest Galactic stars. 
We checked all the images in the logarithmic intensity scale and confirmed that the Galactic stars were removed. 
Some images showed strong spider diffraction patterns caused by very bright stars. 
These images were identified and automatically masked. 
The pixels on the stars and the spider diffraction patterns were substituted with an average value around the stars or the spider diffraction patterns.

This rough identification caused some misidentification of the Galactic stars, especially for the galaxies just behind the Galactic plane or galaxies with bright foreground stars. 
We could not mask a foreground Galactic cluster in UGCA\,86. 
Many foreground stars resulted in a large uncertainty in the infrared flux in IC\,10 and IC\,356. 
One or two bright stars with spider diffraction patterns also resulted in an uncertainty in NGC\,1569 and NGC\,2276. 
In the case of NGC\,5792, a bright star could not be completely subtracted, and the galaxy emission was overestimated. 
We did not use the stellar mass and its related values for UGCA\,86 because our measured stellar mass had large associated uncertainties. 
In the case of NGC\,1055, NGC\,3338, NGC\,3627, NGC\,5055, and NGC\,6951, a foreground bright star likely contributes to additional uncertainty. 
In NGC\,5364 and NGC\,5907, two foreground stars could not be removed because of their positions coincident with the target galactic discs. 
However, the data contamination for these seven galaxies seems only minor within an error bar and remains as such in our sample for further analysis. 
Our star masking, except for the six galaxies (i.e., IC\,10, UGCA\,86, IC\,356, NGC\,1569, NGC\,2276, and NGC\,5792) with a large uncertainty, affects the derived total stellar mass of approximately 10\,\% at most, as obtained from the comparison of the star-masked and non-masked data.

We measured the typical radial extent of $3.4\,\micron$ images and adopted it as a galaxy radius (hereafter $R_{3.4\,\micron}$). 
We then constructed a radial distribution from the star-masked images using the galaxy parameters listed in table \ref{tab:Galpars1} and interpolated in $\timeform{1''}$ spacing. 
The $R_{3.4\,\micron}$ radius is defined as the radius where the radially averaged intensity becomes smaller than the dispersion ($1\>\sigma$) of the background pixels. 
This radius typically corresponds to a bending point of the radial profiles. 
The stellar distributions for the interacting galaxies were divided into each galaxy using the local minima near the line linking both galaxy centers. 
Individual radial profiles were then calculated. 
Our samples contained ``no'' clear overlapping regions in the interacting systems, which would, otherwise, complicate this approach. 
We make special note of NGC\,660, where the extended optical emission indicates a $\mathrm{PA} = \timeform{170D}$ and $i = \timeform{67D.7}$, which was quite different from the values adopted in table \ref{tab:Galpars1}, although we adopted the former for putting a high priority on outer disc. 
The radial distribution of the four edge-on galaxies, namely NGC\,891, NGC\,3628, NGC\,4302, and NGC\,5907, which have inclinations higher than $\timeform{85D}$ were made only using the data on the major axis.

\begin{figure}
 \begin{center}
  \includegraphics[width=8cm]{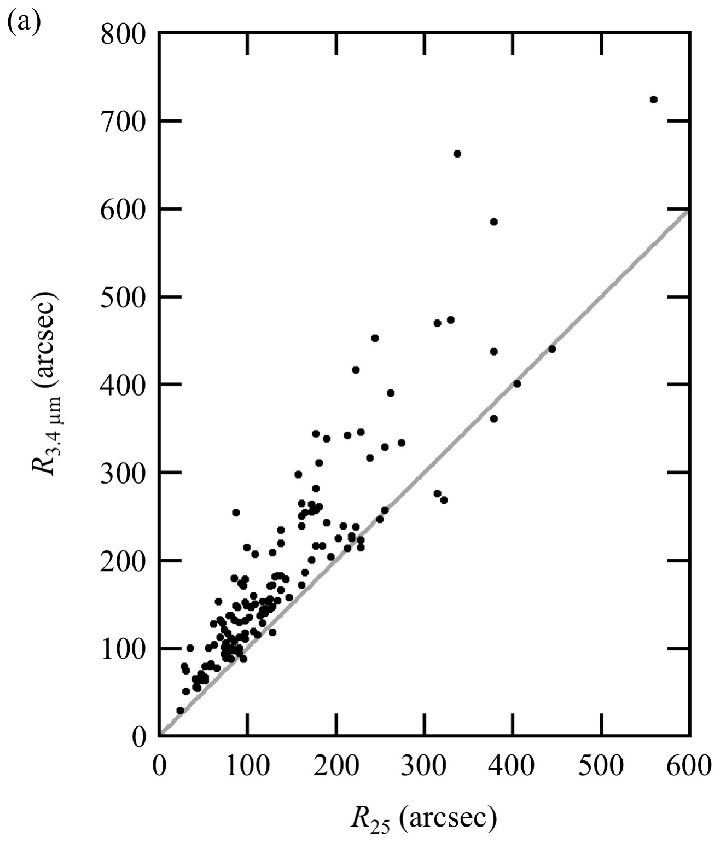}
  \includegraphics[width=8cm]{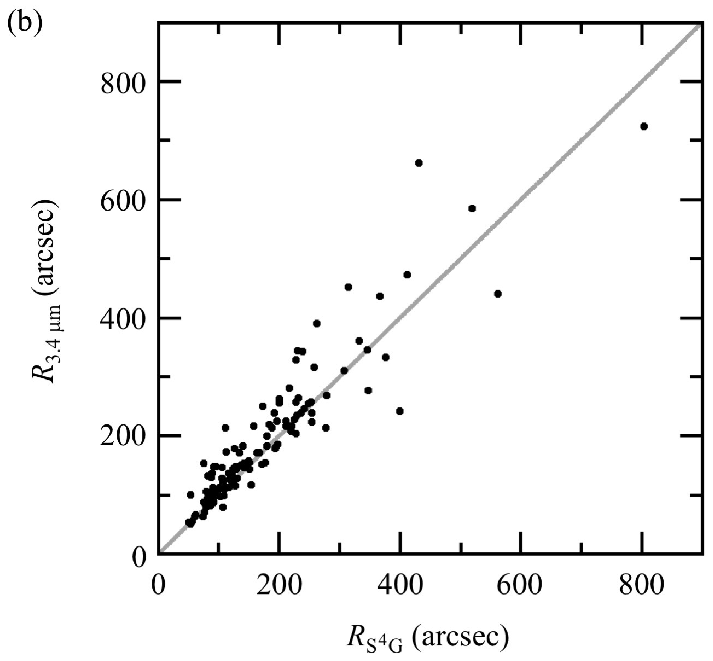}
 \end{center}
\caption{Relation between the radius derived from WISE $3.4\,\micron$ data in this work ($R_{3.4\,\micron}$) and the optical radius ($R_{25} = D_{25} / 2$). 
The solid line indicates $R_{3.4\,\micron} = R_{25}$. 
(b) Same as (a), but comparing with the semi-major axis in S$^4$G (\cite{Sheth+2010}).}
\label{fig:comparing_radii}
\end{figure}

$R_{3.4\,\micron}$ is typically more than 30\,\% larger than $R_{25}$. 
We compared both radii in figure \ref{fig:comparing_radii}a. 
Although most galaxies had a larger $R_{3.4\,\micron}$ than $R_{25}$, a few galaxies had smaller $R_{3.4\,\micron}$ than $R_{25}$ (e.g., NGC\,628). 
We also compared $R_{3.4\,\micron}$ and the semi-major axis reported in the Spitzer Survey of Stellar Structure in Galaxies (S$^4$G, \cite{Sheth+2010}) for the 115 galaxies\footnote{The number of common samples between our survey and S$^4$G is 116, but the semi-major axis of NGC\,4647 is not listed.} common between that and COMING (figure \ref{fig:comparing_radii}b).

The total stellar mass of a galaxy was calculated from the star-masked images using the formula from \citet{Wen+2013}. 
The star-masked images were converted into luminosity scale using the adopted distance listed in table \ref{tab:Galpars1} within a projected ellipse, where $R_{3.4\,\micron}$ was calculated. 
The total luminosity of a galaxy was converted into the total stellar mass according to the following equation (\cite{Wen+2013}): 
\begin{eqnarray}
\log_{10} &\left ( \frac{M_{*}}{M_{\solar}} \right )& = (0.679\,\pm\,0.002) \nonumber \\
&+& (1.033\,\pm\,0.001) \times \log_{10} \left( \frac{\nu L_{\nu} (3.4\,\micron)}{L_{\solar}} \right ). 
\label{eq:stellar-mass}
\end{eqnarray}
We did not consider herein the contribution from the active galactic nucleus (AGN) to the $3.4\,\micron$ luminosity. 
Some interacting galaxies overlapped each other, and our image separation process resulted in a comparatively poorer flux accuracy in these systems. 
The stellar mass of NGC\,660 was calculated within the ellipse projected using $\mathrm{PA} = \timeform{170D}$ and $i = \timeform{67D.7}$ (i.e., determined from CO). 
The stellar mass of the four edge-on galaxies was calculated within a rectangular region with the major axis length corresponding to $2 R_{3.4\,\micron}$. 
The pixel values along the lines parallel to the major axis were averaged. 
Its distribution in the direction perpendicular to the major axis was then used in the same way as measuring $R_{3.4\,\micron}$. 
The shorter axis was adopted to be twice this value.

\begin{figure}
 \begin{center}
  \includegraphics[width=8cm]{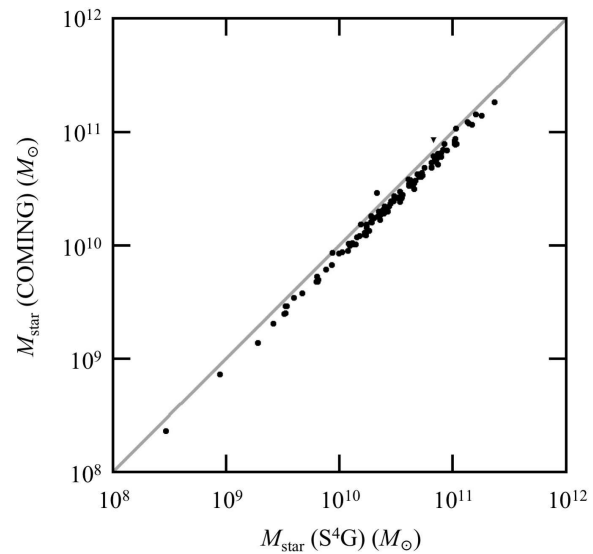}
 \end{center}
\caption{Stellar masses derived in this work and S$^4$G (\cite{Sheth+2010}). 
The solid line indicates that both masses are equal. 
The errors are comparable to or smaller than the size of each marker. 
The triangle indicates the upper limit of the stellar mass estimated in this work.}
\label{fig:comparing_stellar-mass}
\end{figure}

We compared our derived stellar masses with those measured in S$^4$G (\cite{Sheth+2010}; \cite{Munoz-Mateos+2013}; \cite{Querejeta+2015}) after substituting their adopted galaxy distance for our adopted value (figure \ref{fig:comparing_stellar-mass}). 
The comparison was done for 115 galaxies common to both samples. 
The errors were comparable to or smaller than the size of each marker, and we did not consider these errors because of the adopted models (i.e., only the contribution from the flux errors). 
The stellar mass derived in this work was typically approximately 20\,\% lower than that in the previous work, although the measured sizes were not necessarily common (figure \ref{fig:comparing_radii}b). 
This offset is mainly explained by our conservative star masking method which masks slightly larger area than the extent of stars and background subtraction. 
The offset cannot be explained by difference of the assumed IMFs, as it has a contrary effect (\cite{Zahid+2012}): Kroupa IMF (\cite{Kroupa2001}) for \citet{Wen+2013} and Chabrier IMF (\cite{Chabrier2003}) for S$^4$G (\cite{Querejeta+2015}). 
In addition, a minor offset was found between the WISE $3.4\,\micron$ band and the Spitzer $3.6\,\micron$ band at a higher magnitude (\cite{Wen+2013}). 
Table \ref{tab:results} lists the total stellar masses and $R_{3.4\,\micron}$.

\begin{longtable}{lcccc}
  \caption{Measured radius, stellar mass, \atom{C}{}{12}\atom{O}{}{} luminosity, and molecular gas mass.}\label{tab:results}
  \hline
galaxy & $R_{3.4\>\mathrm{\mu m}} (\mathrm{arcsec})$  & $\mathrm{log}_{10}\,M_{\mathrm{star}} (M_{\solar})$  & $\mathrm{log}_{10}\,L'_{\atom{C}{}{12}\atom{O}{}{}} (\mathrm{K\>km\>s}^{-1}\>\mathrm{pc}^{2})$ & $\mathrm{log}_{10}\,M_{\mathrm{mol}} (M_{\solar})$ \\
\endfirsthead
  \hline
galaxy & $R_{3.4\>\mathrm{\mu m}} (\mathrm{arcsec})$  & $\mathrm{log}_{10}\,M_{\mathrm{star}} (M_{\solar})$  & $\mathrm{log}_{10}\,L'_{\atom{C}{}{12}\atom{O}{}{}} (\mathrm{K\>km\>s}^{-1}\>\mathrm{pc}^{2})$ & $\mathrm{log}_{10}\,M_{\mathrm{mol}} (M_{\solar})$ \\
  \hline
\endhead
  \hline
\endfoot
  \hline
\multicolumn{5}{@{}l@{}}{\hbox to 0pt{\parbox{160mm}{\footnotesize
Notes. 
\par\noindent
Column (1): Galaxy name. 
Same as the column (1) in table \ref{tab:Galpars1}. 
\par\noindent
Column (2): The $3.4\>\micron$ radius ($R_{3.4\>\mathrm{\mu m}}$) in arcsec. 
\par\noindent
Column (3): Logarithmic total stellar mass in $M_{\solar}$. 
\par\noindent
Column (4): Logarithmic total integrated intensity of \atom{C}{}{12}\atom{O}{}{} in $\mathrm{K\>km\>s}^{-1}\>\mathrm{pc}^{2}$. 
\par\noindent
Column (5): Logarithmic total molecular gas mass in $M_{\solar}$. 
\par\noindent
$*$: Uncertain because of many foreground stars. 
\par\noindent
$\dag$: Slightly uncertain due to a foreground bright star and a spider diffraction pattern. 
\par\noindent
$\ddag$: Uncertain due to a foreground star cluster. 
\par\noindent
$\S$: Uncertain due to the Galactic emission. 
\par\noindent
$\parallel$: Uncertain because of several foreground stars, including a bright one and spider diffraction patterns. 
\par\noindent
$\sharp$: Slightly lower limit due to masking the channel corresponding to the Galactic emission. 
\par\noindent
$* *$: Uncertain due to two foreground stars and spider diffraction patterns. 
\par\noindent
$\dag \dag$: Slightly uncertain because of a foreground bright star. 
\par\noindent
$\ddag \ddag$: Slight upper limit caused by not masking two slightly bright stars on the galaxy. 
\par\noindent
$\S \S$: Upper limit due to a foreground bright star and spider diffraction patterns. 
}\hss}}
\endlastfoot
 \hline
\multicolumn{5}{c}{isolated galaxies}\\
 \hline
IC\,10 & 339  & 8.423\,$\pm$\,.008 $*$ & 5.915\,$\pm$\,.008 & 6.551\,$\pm$\,.008 \\
NGC\,150 & 129  & 10.425\,$\pm$\,.010 & 8.391\,$^{+.013}_{-.014}$ & 9.027\,$^{+.013}_{-.014}$ \\
NGC\,157 & 156  & 10.156\,$\pm$\,.009 & 8.412\,$\pm$\,.004 & 9.049\,$\pm$\,.004 \\
NGC\,278 & 105  & 10.336\,$^{+.009}_{-.010}$ & 8.693\,$\pm$\,.005 & 9.329\,$\pm$\,.005 \\
NGC\,337 & 99  & 9.953\,$\pm$\,.009 & 8.036\,$^{+.014}_{-.015}$ & 8.672\,$^{+.014}_{-.015}$ \\
NGC\,470 & 109  & 10.732\,$\pm$\,.010 & 8.805\,$\pm$\,.012 & 9.441\,$\pm$\,.012 \\
NGC\,520 & 155  & 9.462\,$\pm$\,.009 & 7.973\,$\pm$\,.003 & 8.609\,$\pm$\,.003 \\
NGC\,613 & 187  & 11.086\,$\pm$\,.010 & 9.395\,$\pm$\,.004 & 10.031\,$\pm$\,.004 \\
NGC\,628 & 277  & 10.247\,$^{+.009}_{-.010}$ & 8.653\,$\pm$\,.003 & 9.289\,$\pm$\,.003 \\
NGC\,660 & 247  & 10.436\,$\pm$\,.010 & 9.029\,$\pm$\,.002 & 9.665\,$\pm$\,.002 \\
NGC\,701 & 102  & 10.018\,$\pm$\,.009 & 8.140\,$^{+.014}_{-.015}$ & 8.776\,$^{+.014}_{-.015}$ \\
NGC\,891 & 401  & 10.638\,$\pm$\,.010 & 9.192\,$\pm$\,.001 & 9.828\,$\pm$\,.001 \\
NGC\,1022 & 129  & 10.185\,$^{+.009}_{-.010}$ & 8.444\,$\pm$\,.008 & 9.080\,$\pm$\,.008 \\
NGC\,1055 & 215  & 10.844\,$\pm$\,.010 $\dag$ & 9.327\,$\pm$\,.003 & 9.963\,$\pm$\,.003 \\
NGC\,1084 & 179  & 10.581\,$\pm$\,.010 & 8.875\,$\pm$\,.005 & 9.511\,$\pm$\,.005 \\
NGC\,1087 & 116  & 10.002\,$\pm$\,.009 & 8.139\,$^{+.010}_{-.011}$ & 8.775\,$^{+.010}_{-.011}$ \\
NGC\,1156 & 149  & 9.182\,$^{+.008}_{-.009}$ & 6.941\,$^{+.030}_{-.032}$ & 7.577\,$^{+.030}_{-.032}$ \\
NGC\,1241 & 109  & 11.142\,$\pm$\,.010 & 9.339\,$\pm$\,.009 & 9.975\,$\pm$\,.009 \\
UGC\,2765 & 140  & 10.236\,$^{+.009}_{-.010}$ & 8.317\,$\pm$\,.009 & 8.953\,$\pm$\,.009 \\
NGC\,1482 & 121  & 10.311\,$^{+.009}_{-.010}$ & 8.728\,$\pm$\,.005 & 9.364\,$\pm$\,.005 \\
UGCA\,86 & 30  & 6.729\,$\pm$\,.006 $\ddag$ & 5.906\,$\pm$\,.017 $\S$ & 6.542\,$\pm$\,.017 $\S$ \\
IC\,356 & 298  & 11.318\,$^{+.010}_{-.011}$ $*$ & 9.179\,$\pm$\,.004 & 9.815\,$\pm$\,.004 \\
NGC\,1530 & 167  & 10.424\,$\pm$\,.010 & 8.882\,$\pm$\,.005 & 9.518\,$\pm$\,.005 \\
NGC\,1569 & 207  & 8.942\,$\pm$\,.008 $\parallel$ & 6.090\,$^{+.034}_{-.037}$ $\sharp$ & 6.726\,$^{+.034}_{-.037}$ $\sharp$ \\
NGC\,2146 & 261  & 11.285\,$^{+.010}_{-.011}$ & 9.818\,$\pm$\,.002 & 10.454\,$\pm$\,.002 \\
NGC\,2273 & 118  & 10.665\,$\pm$\,.010 & 8.776\,$\pm$\,.009 & 9.412\,$\pm$\,.009 \\
NGC\,2339 & 101  & 10.790\,$\pm$\,.010 & 9.073\,$\pm$\,.006 & 9.709\,$\pm$\,.006 \\
NGC\,2268 & 132  & 10.618\,$\pm$\,.010 & 8.914\,$\pm$\,.008 & 9.550\,$\pm$\,.008 \\
NGC\,2276 & 180  & 10.866\,$\pm$\,.010 $* *$ & 9.337\,$\pm$\,.005 & 9.973\,$\pm$\,.005 \\
NGC\,2633 & 94  & 10.406\,$\pm$\,.010 & 9.041\,$\pm$\,.004 & 9.677\,$\pm$\,.004 \\
NGC\,2681 & 151  & 10.427\,$\pm$\,.010 & 8.315\,$\pm$\,.008 & 8.951\,$\pm$\,.008 \\
NGC\,2742 & 113  & 10.421\,$\pm$\,.010 & 8.528\,$\pm$\,.012 & 9.164\,$\pm$\,.012 \\
NGC\,2715 & 158  & 10.010\,$\pm$\,.009 & 8.329\,$^{+.008}_{-.009}$ & 8.965\,$^{+.008}_{-.009}$ \\
NGC\,2775 & 209  & 10.712\,$\pm$\,.010 & 8.408\,$\pm$\,.009 & 9.044\,$\pm$\,.009 \\
NGC\,2748 & 130  & 10.121\,$\pm$\,.009 & 8.301\,$^{+.010}_{-.011}$ & 8.937\,$^{+.010}_{-.011}$ \\
NGC\,2782 & 148  & 9.938\,$\pm$\,.009 & 8.238\,$^{+.009}_{-.010}$ & 8.874\,$^{+.009}_{-.010}$ \\
NGC\,2841 & 453  & 10.898\,$\pm$\,.010 & 8.660\,$\pm$\,.004 & 9.296\,$\pm$\,.004 \\
NGC\,2903 & 438  & 10.608\,$\pm$\,.010 & 9.013\,$\pm$\,.002 & 9.649\,$\pm$\,.002 \\
NGC\,2967 & 101  & 10.208\,$^{+.009}_{-.010}$ & 8.713\,$\pm$\,.006 & 9.349\,$\pm$\,.006 \\
NGC\,2976 & 282  & 9.140\,$^{+.008}_{-.009}$ & 7.241\,$\pm$\,.006 & 7.877\,$\pm$\,.006 \\
NGC\,2985 & 235  & 10.799\,$\pm$\,.010 & 8.829\,$\pm$\,.007 & 9.465\,$\pm$\,.007 \\
NGC\,3034 & 663  & 10.293\,$^{+.009}_{-.010}$ & 8.896\,$\pm$\,.001 & 9.532\,$\pm$\,.001 \\
NGC\,3079 & 317  & 10.808\,$\pm$\,.010 & 9.443\,$\pm$\,.001 & 10.079\,$\pm$\,.001 \\
NGC\,3077 & 265  & 9.309\,$\pm$\,.009 & 7.016\,$^{+.013}_{-.014}$ & 7.652\,$^{+.013}_{-.014}$ \\
NGC\,3166 & 179  & 10.781\,$\pm$\,.010 & 8.619\,$\pm$\,.007 & 9.255\,$\pm$\,.007 \\
NGC\,3169 & 182  & 10.890\,$\pm$\,.010 & 9.136\,$\pm$\,.005 & 9.772\,$\pm$\,.005 \\
NGC\,3177 & 55  & 10.110\,$\pm$\,.009 & 8.622\,$\pm$\,.005 & 9.258\,$\pm$\,.005 \\
NGC\,3147 & 144  & 11.268\,$^{+.010}_{-.011}$ & 9.660\,$\pm$\,.003 & 10.296\,$\pm$\,.003 \\
NGC\,3198 & 329  & 10.128\,$\pm$\,.009 & 8.375\,$\pm$\,.006 & 9.011\,$\pm$\,.006 \\
Mrk\,33 & 52  & 9.581\,$\pm$\,.009 & 7.409\,$^{+.049}_{-.055}$ & 8.046\,$^{+.049}_{-.055}$ \\
NGC\,3310 & 174  & 9.829\,$\pm$\,.009 & 7.971\,$\pm$\,.008 & 8.607\,$\pm$\,.008 \\
NGC\,3338 & 217  & 10.479\,$\pm$\,.010 $\dag \dag$ & 9.046\,$\pm$\,.005 & 9.682\,$\pm$\,.005 \\
NGC\,3344 & 214  & 10.077\,$\pm$\,.009 & 8.182\,$\pm$\,.008 & 8.818\,$\pm$\,.008 \\
NGC\,3351 & 239  & 10.388\,$\pm$\,.010 & 8.578\,$\pm$\,.004 & 9.214\,$\pm$\,.004 \\
NGC\,3367 & 89  & 10.497\,$\pm$\,.010 & 8.768\,$\pm$\,.007 & 9.404\,$\pm$\,.007 \\
NGC\,3359 & 226  & 10.300\,$^{+.009}_{-.010}$ & 8.924\,$\pm$\,.006 & 9.560\,$\pm$\,.006 \\
NGC\,3368 & 346  & 10.520\,$\pm$\,.010 & 8.373\,$\pm$\,.005 & 9.009\,$\pm$\,.005 \\
NGC\,3370 & 88  & 10.093\,$\pm$\,.009 & 8.375\,$\pm$\,.013 & 9.011\,$\pm$\,.013 \\
NGC\,3437 & 106  & 10.286\,$^{+.009}_{-.010}$ & 8.661\,$\pm$\,.009 & 9.297\,$\pm$\,.009 \\
NGC\,3471 & 64  & 9.934\,$\pm$\,.009 & 8.183\,$\pm$\,.012 & 8.819\,$\pm$\,.012 \\
NGC\,3521 & 474  & 11.066\,$\pm$\,.010 & 9.379\,$\pm$\,.002 & 10.015\,$\pm$\,.002 \\
NGC\,3556 & 391  & 10.245\,$^{+.009}_{-.010}$ & 8.759\,$\pm$\,.002 & 9.395\,$\pm$\,.002 \\
NGC\,3583 & 98  & 10.543\,$\pm$\,.010 & 8.771\,$^{+.007}_{-.008}$ & 9.407\,$^{+.007}_{-.008}$ \\
NGC\,3627 & 334  & 10.605\,$\pm$\,.010 $\dag \dag$ & 9.031\,$\pm$\,.002 & 9.667\,$\pm$\,.002 \\
NGC\,3628 & 441  & 10.638\,$\pm$\,.010 & 9.123\,$\pm$\,.002 & 9.759\,$\pm$\,.002 \\
NGC\,3655 & 65  & 10.683\,$\pm$\,.010 & 9.208\,$\pm$\,.005 & 9.844\,$\pm$\,.005 \\
NGC\,3672 & 145  & 10.625\,$\pm$\,.010 & 8.942\,$\pm$\,.006 & 9.578\,$\pm$\,.006 \\
NGC\,3675 & 345  & 10.895\,$\pm$\,.010 & 9.123\,$\pm$\,.003 & 9.759\,$\pm$\,.003 \\
NGC\,3686 & 111  & 10.008\,$\pm$\,.009 & 8.399\,$\pm$\,.007 & 9.035\,$\pm$\,.007 \\
NGC\,3810 & 148  & 10.328\,$^{+.009}_{-.010}$ & 8.761\,$^{+.004}_{-.005}$ & 9.397\,$^{+.004}_{-.005}$ \\
NGC\,3813 & 154  & 10.263\,$^{+.009}_{-.010}$ & 8.611\,$\pm$\,.006 & 9.247\,$\pm$\,.006 \\
NGC\,3888 & 68  & 10.529\,$\pm$\,.010 & 8.632\,$^{+.011}_{-.012}$ & 9.268\,$^{+.011}_{-.012}$ \\
NGC\,3893 & 183  & 10.295\,$^{+.009}_{-.010}$ & 8.679\,$\pm$\,.004 & 9.315\,$\pm$\,.004 \\
NGC\,3938 & 173  & 10.431\,$\pm$\,.010 & 8.829\,$\pm$\,.004 & 9.465\,$\pm$\,.004 \\
NGC\,3949 & 149  & 10.203\,$^{+.009}_{-.010}$ & 8.149\,$\pm$\,.011 & 8.785\,$\pm$\,.011 \\
UGC\,6973 & 138  & 10.544\,$\pm$\,.010 & 9.001\,$\pm$\,.004 & 9.637\,$\pm$\,.004 \\
NGC\,4027 & 113  & 10.272\,$^{+.009}_{-.010}$ & 8.700\,$^{+.006}_{-.007}$ & 9.336\,$^{+.006}_{-.007}$ \\
NGC\,4030 & 172  & 11.077\,$\pm$\,.010 & 9.506\,$\pm$\,.003 & 10.142\,$\pm$\,.003 \\
NGC\,4041 & 112  & 10.685\,$\pm$\,.010 & 9.236\,$\pm$\,.005 & 9.872\,$\pm$\,.005 \\
NGC\,4045 & 138  & 10.628\,$\pm$\,.010 & 8.891\,$\pm$\,.009 & 9.527\,$\pm$\,.009 \\
NGC\,4085 & 133  & 10.023\,$\pm$\,.009 & 8.340\,$\pm$\,.010 & 8.976\,$\pm$\,.010 \\
NGC\,4088 & 201  & 10.390\,$\pm$\,.010 & 8.910\,$\pm$\,.003 & 9.546\,$\pm$\,.003 \\
NGC\,4214 & 258  & 8.864\,$\pm$\,.008 & 7.141\,$\pm$\,.010 & 7.777\,$\pm$\,.010 \\
NGC\,4258 & 725  & 10.551\,$\pm$\,.010 & 8.732\,$\pm$\,.002 & 9.368\,$\pm$\,.002 \\
NGC\,4303 & 205  & 10.773\,$\pm$\,.010 & 9.334\,$\pm$\,.003 & 9.970\,$\pm$\,.003 \\
NGC\,4433 & 78  & 10.625\,$\pm$\,.010 & 9.103\,$\pm$\,.006 & 9.739\,$\pm$\,.006 \\
NGC\,4527 & 217  & 10.688\,$\pm$\,.010 & 9.222\,$\pm$\,.003 & 9.858\,$\pm$\,.003 \\
NGC\,4536 & 224  & 10.464\,$\pm$\,.010 & 8.725\,$\pm$\,.005 & 9.361\,$\pm$\,.005 \\
NGC\,4559 & 269  & 9.679\,$\pm$\,.009 & 8.164\,$\pm$\,.005 & 8.800\,$\pm$\,.005 \\
NGC\,4579 & 258  & 10.839\,$\pm$\,.010 & 8.831\,$\pm$\,.005 & 9.467\,$\pm$\,.005 \\
NGC\,4605 & 256  & 9.401\,$\pm$\,.009 & 7.214\,$^{+.013}_{-.014}$ & 7.850\,$^{+.013}_{-.014}$ \\
NGC\,4602 & 136  & 10.754\,$\pm$\,.010 & 9.033\,$\pm$\,.007 & 9.669\,$\pm$\,.007 \\
NGC\,4632 & 113  & 9.700\,$\pm$\,.009 & 7.949\,$^{+.011}_{-.012}$ & 8.585\,$^{+.011}_{-.012}$ \\
NGC\,4666 & 220  & 10.573\,$\pm$\,.010 & 8.952\,$\pm$\,.003 & 9.588\,$\pm$\,.003 \\
NGC\,4750 & 128  & 10.618\,$\pm$\,.010 & 8.531\,$\pm$\,.011 & 9.167\,$\pm$\,.011 \\
NGC\,4753 & 311  & 11.146\,$\pm$\,.010 & 8.709\,$\pm$\,.009 & 9.345\,$\pm$\,.009 \\
NGC\,4818 & 173  & 9.940\,$\pm$\,.009 & 8.196\,$^{+.004}_{-.005}$ & 8.832\,$^{+.004}_{-.005}$ \\
NGC\,5005 & 264  & 10.922\,$\pm$\,.010 & 9.148\,$\pm$\,.003 & 9.784\,$\pm$\,.003 \\
NGC\,5055 & 586  & 10.737\,$\pm$\,.010 $\dag \dag$ & 9.213\,$\pm$\,.002 & 9.849\,$\pm$\,.002 \\
NGC\,5248 & 217  & 10.369\,$^{+.009}_{-.010}$ & 8.958\,$\pm$\,.003 & 9.594\,$\pm$\,.003 \\
NGC\,5364 & 226  & 10.445\,$\pm$\,.010 $\ddag \ddag$ & 8.636\,$\pm$\,.008 & 9.272\,$\pm$\,.008 \\
NGC\,5480 & 80  & 10.226\,$^{+.009}_{-.010}$ & 8.871\,$\pm$\,.008 & 9.507\,$\pm$\,.008 \\
NGC\,5678 & 215  & 10.895\,$\pm$\,.010 & 9.394\,$\pm$\,.003 & 10.030\,$\pm$\,.003 \\
NGC\,5665 & 82  & 9.790\,$\pm$\,.009 & 8.035\,$\pm$\,.011 & 8.671\,$\pm$\,.011 \\
NGC\,5676 & 145  & 10.940\,$\pm$\,.010 & 9.415\,$\pm$\,.003 & 10.051\,$\pm$\,.003 \\
NGC\,5713 & 98  & 10.296\,$^{+.009}_{-.010}$ & 8.830\,$\pm$\,.004 & 9.466\,$\pm$\,.004 \\
NGC\,5792 & 240  & 10.931\,$\pm$\,.010 $\S \S$ & 8.896\,$\pm$\,.007 & 9.532\,$\pm$\,.007 \\
NGC\,5907 & 362  & 10.814\,$\pm$\,.010 $\ddag \ddag$ & 9.076\,$\pm$\,.003 & 9.712\,$\pm$\,.003 \\
NGC\,6015 & 251  & 10.257\,$^{+.009}_{-.010}$ & 8.251\,$\pm$\,.013 & 8.887\,$\pm$\,.013 \\
NGC\,6503 & 343  & 9.682\,$\pm$\,.009 & 7.655\,$\pm$\,.006 & 8.291\,$\pm$\,.006 \\
NGC\,6574 & 64  & 10.974\,$\pm$\,.010 & 9.367\,$\pm$\,.006 & 10.003\,$\pm$\,.006 \\
NGC\,6643 & 138  & 10.471\,$\pm$\,.010 & 8.874\,$\pm$\,.005 & 9.510\,$\pm$\,.005 \\
NGC\,6764 & 133  & 10.117\,$\pm$\,.009 & 8.330\,$^{+.012}_{-.013}$ & 8.966\,$^{+.012}_{-.013}$ \\
NGC\,6951 & 154  & 10.873\,$\pm$\,.010 $\dag \dag$ & 9.091\,$\pm$\,.005 & 9.727\,$\pm$\,.005 \\
NGC\,7331 & 470  & 10.935\,$\pm$\,.010 & 9.256\,$\pm$\,.002 & 9.892\,$\pm$\,.002 \\
NGC\,7448 & 88  & 10.346\,$^{+.009}_{-.010}$ & 8.674\,$\pm$\,.008 & 9.310\,$\pm$\,.008 \\
NGC\,7479 & 154  & 11.029\,$\pm$\,.010 & 9.427\,$\pm$\,.004 & 10.063\,$\pm$\,.004 \\
NGC\,7541 & 147  & 10.787\,$\pm$\,.010 & 9.070\,$\pm$\,.006 & 9.706\,$\pm$\,.006 \\
NGC\,7625 & 64  & 10.188\,$^{+.009}_{-.010}$ & 8.701\,$\pm$\,.005 & 9.337\,$\pm$\,.005 \\
NGC\,7721 & 120  & 10.411\,$\pm$\,.010 & 8.521\,$\pm$\,.011 & 9.157\,$\pm$\,.011 \\
NGC\,7798 & 56  & 10.301\,$^{+.009}_{-.010}$ & 8.628\,$\pm$\,.008 & 9.264\,$\pm$\,.008 \\
 \hline
\multicolumn{5}{c}{interacting galaxies}\\
 \hline
NGC\,772 / NGC\,770 &   & 11.186\,$\pm$\,.010 & 9.680\,$\pm$\,.004 & 10.316\,$\pm$\,.004 \\
\ \ NGC\,770 & 101  & 9.988\,$\pm$\,.009 & 8.715\,$^{+.012}_{-.013}$ & 9.351\,$^{+.012}_{-.013}$ \\
\ \ NGC\,772 & 228  & 11.157\,$\pm$\,.010 & 9.631\,$\pm$\,.004 & 10.267\,$\pm$\,.004 \\
NGC\,2207 / IC\,2163 &   & 11.123\,$\pm$\,.007 & 9.364\,$\pm$\,.006 & 10.000\,$\pm$\,.006 \\
\ \ NGC\,2207 & 119  & 10.888\,$\pm$\,.010 & 9.093\,$\pm$\,.009 & 9.729\,$\pm$\,.009 \\
\ \ IC\,2163 & 95  & 10.745\,$\pm$\,.010 & 9.030\,$\pm$\,.009 & 9.666\,$\pm$\,.009 \\
Arp\,283 &   & 10.449\,$^{+.008}_{-.009}$ & 8.780\,$\pm$\,.009 & 9.416\,$\pm$\,.009 \\
\ \ NGC\,2798 & 94  & 10.392\,$\pm$\,.010 & 8.732\,$\pm$\,.009 & 9.368\,$\pm$\,.009 \\
\ \ NGC\,2799 & 101  & 9.542\,$\pm$\,.009 & 7.801\,$^{+.035}_{-.038}$ & 8.437\,$^{+.035}_{-.038}$ \\
Arp\,245 &   & 10.716\,$\pm$\,.008 & 8.973\,$\pm$\,.008 & 9.609\,$\pm$\,.008 \\
\ \ NGC\,2992 & 160  & 10.614\,$\pm$\,.010 & 8.742\,$\pm$\,.011 & 9.378\,$\pm$\,.011 \\
\ \ NGC\,2993 & 66  & 10.036\,$\pm$\,.009 & 8.589\,$\pm$\,.011 & 9.225\,$\pm$\,.011 \\
Arp\,094 &   & 10.748\,$\pm$\,.007 & 9.201\,$\pm$\,.004 & 9.837\,$\pm$\,.004 \\
\ \ NGC\,3226 & 172  & 10.256\,$^{+.009}_{-.010}$ & 8.774\,$^{+.007}_{-.008}$ & 9.410\,$^{+.007}_{-.008}$ \\
\ \ NGC\,3227 & 240  & 10.579\,$\pm$\,.010 & 8.997\,$\pm$\,.005 & 9.633\,$\pm$\,.005 \\
NGC\,4298 / NGC\,4302 &   & 10.537\,$\pm$\,.007 & 9.043\,$\pm$\,.004 & 9.679\,$\pm$\,.004 \\
\ \ NGC\,4298 & 153  & 10.091\,$\pm$\,.009 & 8.773\,$\pm$\,.005 & 9.409\,$\pm$\,.005 \\
\ \ NGC\,4302 & 255  & 10.344\,$^{+.009}_{-.010}$ & 8.709\,$\pm$\,.005 & 9.345\,$\pm$\,.005 \\
NGC\,4383 / UGC\,7504 &   & 9.751\,$^{+.008}_{-.009}$ & 8.483\,$\pm$\,.006 & 9.119\,$\pm$\,.006 \\
\ \ UGC\,7504 & 75  & 8.481\,$\pm$\,.008 & 7.691\,$^{+.015}_{-.016}$ & 8.327\,$^{+.015}_{-.016}$ \\
\ \ NGC\,4383 & 83  & 9.727\,$\pm$\,.009 & 8.407\,$\pm$\,.006 & 9.043\,$\pm$\,.006 \\
Arp\,269 &   & 9.438\,$\pm$\,.008 & 7.335\,$\pm$\,.007 & 7.971\,$\pm$\,.007 \\
\ \ NGC\,4485 & 113  & 8.363\,$\pm$\,.008 & 6.633\,$^{+.017}_{-.018}$ & 7.269\,$^{+.017}_{-.018}$ \\
\ \ NGC\,4490 & 243  & 9.400\,$\pm$\,.009 & 7.238\,$\pm$\,.007 & 7.874\,$\pm$\,.007 \\
VV\,219 &   & 10.575\,$\pm$\,.007 & 9.124\,$\pm$\,.003 & 9.760\,$\pm$\,.003 \\
\ \ NGC\,4567 & 147  & 10.084\,$\pm$\,.009 & 8.616\,$\pm$\,.006 & 9.252\,$\pm$\,.006 \\
\ \ NGC\,4568 & 183  & 10.406\,$\pm$\,.010 & 8.963\,$\pm$\,.003 & 9.599\,$\pm$\,.003 \\
Arp\,116 &   & 11.237\,$\pm$\,.009 & 8.993\,$\pm$\,.004 & 9.629\,$\pm$\,.004 \\
\ \ NGC\,4647 & 255  & 10.419\,$\pm$\,.010 & 8.808\,$\pm$\,.005 & 9.444\,$\pm$\,.005 \\
\ \ NGC\,4649 & 417  & 11.165\,$\pm$\,.010 & 8.531\,$\pm$\,.009 & 9.167\,$\pm$\,.009 \\
Arp\,271 &   & 10.985\,$^{+.007}_{-.008}$ & 9.557\,$\pm$\,.004 & 10.193\,$\pm$\,.004 \\
\ \ NGC\,5426 & 100  & 10.501\,$\pm$\,.010 & 9.047\,$\pm$\,.008 & 9.683\,$\pm$\,.008 \\
\ \ NGC\,5427 & 108  & 10.812\,$\pm$\,.010 & 9.397\,$\pm$\,.005 & 10.033\,$\pm$\,.005 \\
Arp\,090 &   & 10.744\,$\pm$\,.007 & 8.421\,$^{+.020}_{-.021}$ & 9.057\,$^{+.020}_{-.021}$ \\
\ \ NGC\,5929 & 80  & 10.292\,$^{+.009}_{-.010}$ & 7.900\,$^{+.039}_{-.043}$ & 8.536\,$^{+.039}_{-.043}$ \\
\ \ NGC\,5930 & 71  & 10.554\,$\pm$\,.010 & 8.266\,$^{+.022}_{-.024}$ & 8.902\,$^{+.022}_{-.024}$ \\
Arp\,284 &   & 10.472\,$\pm$\,.009 & 9.021\,$\pm$\,.009 & 9.657\,$\pm$\,.009 \\
\ \ NGC\,7714 & 80  & 10.428\,$\pm$\,.010 & 8.874\,$^{+.010}_{-.011}$ & 9.510\,$^{+.010}_{-.011}$ \\
\ \ NGC\,7715 & 118  & 9.462\,$\pm$\,.009 & 8.478\,$^{+.015}_{-.016}$ & 9.114\,$^{+.015}_{-.016}$ \\
\end{longtable}

\section{Results and discussion}
\label{section:results}
\subsection{Spectra, integrated intensity maps, and moment maps}
\label{subsection:spectra_maps}
\begin{figure}
 \begin{center}
  \includegraphics[width=8cm]{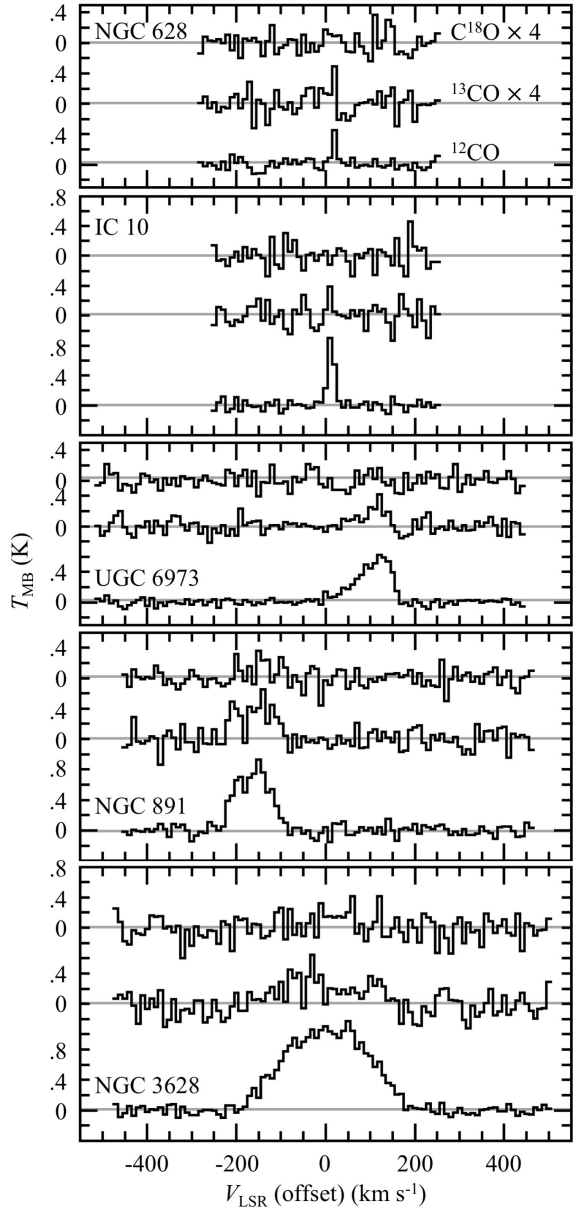}
 \end{center}
\caption{Examples of the \atom{C}{}{}\atom{O}{}{} spectra for five galaxies. 
The galaxy name is shown in each panel. 
The three transitions of \atom{C}{}{}\atom{O}{}{18} $\times 4$ (top), \atom{C}{}{13}\atom{O}{}{} $\times 4$ (middle), and \atom{C}{}{12}\atom{O}{}{} (bottom) are shown in each panel.}
\label{fig:COspectra}
\end{figure}

Figure \ref{fig:COspectra} presents a selection of \atom{C}{}{12}\atom{O}{}{} spectra with various velocity widths and also depicts the spectra of \atom{C}{}{13}\atom{O}{}{} and \atom{C}{}{}\atom{O}{}{18} at the same position in the same galaxy. 
The \atom{C}{}{13}\atom{O}{}{} emission was typically several to 10 times weaker than the \atom{C}{}{12}\atom{O}{}{} emission; thus, the $S / N$ ratio of the \atom{C}{}{13}\atom{O}{}{} spectra was low despite the lower $T_{\mathrm{sys}}$ values of \atom{C}{}{13}\atom{O}{}{} compared to that of \atom{C}{}{12}\atom{O}{}{}. 
The \atom{C}{}{}\atom{O}{}{18} emission was much weaker than the \atom{C}{}{12}\atom{O}{}{} emission. 
For example, the peak intensity of a \atom{C}{}{}\atom{O}{}{18} spectrum of NGC\,891 in figure \ref{fig:COspectra} was $\sim\,2.9\>\sigma$, which was only a marginal detection. 
Only 14 positions in 11 galaxies showed the integrated intensity of \atom{C}{}{}\atom{O}{}{18} exceeding $\sim\,4\>\sigma$ and \atom{C}{}{12}\atom{O}{}{} exceeding $\sim\,3\>\sigma$. 
However, comparing these spectra with the \atom{C}{}{12}\atom{O}{}{} spectra at the same positions, two of the spectra appeared to be merely noise, while four cannot be clearly considered as detection.

Panel (b) of figures \ref{fig:maps_NGC0891} and \ref{fig:maps_NGC3627} and supplementary figures 1 -- 134 (supplementary section of the online version) show the integrated intensity maps of \atom{C}{}{12}\atom{O}{}{}. 
The $S / N$ ratio (or noise level) was not uniform within an integrated intensity map because of the different velocity widths of the emission channels. 
Thus, we drew contours at some percentiles of the maximum intensity in each map. 
We set the lowest contour levels to match the emission extent of the first-degree moment maps by eye because the first-degree moment maps were produced by smoothing and masking, and were likely to trace the real emission extent. 
Some galaxies showed a molecular gas distribution similar to the stellar distribution tracing spiral arms and/or a bar, while some showed quite different patterns (e.g., ring-like features), as reported in previous works (e.g., \cite{Young+1995}). 
The radial distribution of molecular gas will be discussed in a forthcoming paper (Y.~Miyamoto et~al., in preparation). 
As described in subsection \ref{subsection:SystemSetting}, the UGCA\,86 and NGC\,1569 spectra suffered from Galactic emission on the off-source positions. 
The velocity of the Galactic emission slightly overlapped in the case of NGC\,1569; thus, the integrated intensity suffered from a lower limit. 
The Galactic emission overlapped in the case of UGC\,A86; therefore, the integrated intensity included the contribution from the Galactic emission.

Panel (e) of figures \ref{fig:maps_NGC0891} and \ref{fig:maps_NGC3627} and supplementary figures 3 -- 130 (supplementary section of the online version) show the integrated intensity maps of \atom{C}{}{13}\atom{O}{}{} of select galaxies. 
The \atom{C}{}{12}\atom{O}{}{} integrated intensity in each map was overlaid as white contours [the same as panel (b)]. 
Magenta contours are the level representing the significance of the \atom{C}{}{13}\atom{O}{}{} integrated intensity detection. 
The percentile level of the lowest contour in most cases is the same as the lowest one of \atom{C}{}{12}\atom{O}{}{}, except in cases where the \atom{C}{}{13}\atom{O}{}{} map is noisy. 
The sensitivity of our observations in \atom{C}{}{13}\atom{O}{}{} was insufficient; thus, the integrated intensity maps for this line suffered from considerable noise contamination. 
We only presented herein the galaxies with more than $3\>\mathrm{pixels}$ higher than $4\>\sigma$ in the \atom{C}{}{13}\atom{O}{}{} integrated intensity and higher than $3\>\sigma$ in the \atom{C}{}{12}\atom{O}{}{} integrated intensity.

\begin{figure*}
 \begin{center}
  \includegraphics[width=16cm]{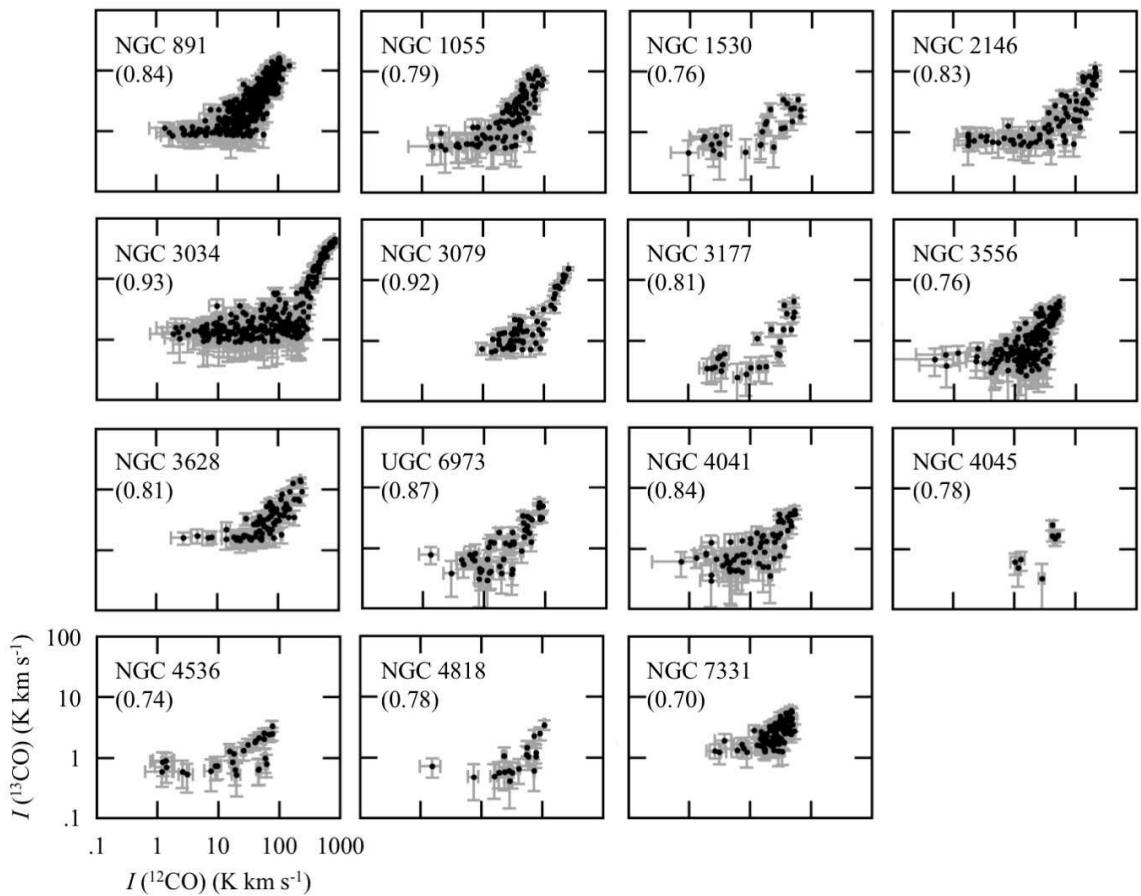}
 \end{center}
\caption{Correlation plots between the \atom{C}{}{13}\atom{O}{}{} and \atom{C}{}{12}\atom{O}{}{} integrated intensities for galaxies whose Pearson's product moment correlation coefficient between both integrated intensities [$\rho_{I(\atom{C}{}{12}\atom{O}{}{}), I(\atom{C}{}{13}\atom{O}{}{})}$] is higher than or equal to 0.7. 
The correlation coefficient is shown in parentheses below the galaxy name. 
The data in the pixels surrounded by the magenta contours in the \atom{C}{}{13}\atom{O}{}{} integrated intensity maps are plotted. 
The error bars are shown in gray lines.}
\label{fig:corr1312_well}
\end{figure*}

\begin{figure*}
 \begin{center}
  \includegraphics[width=16cm]{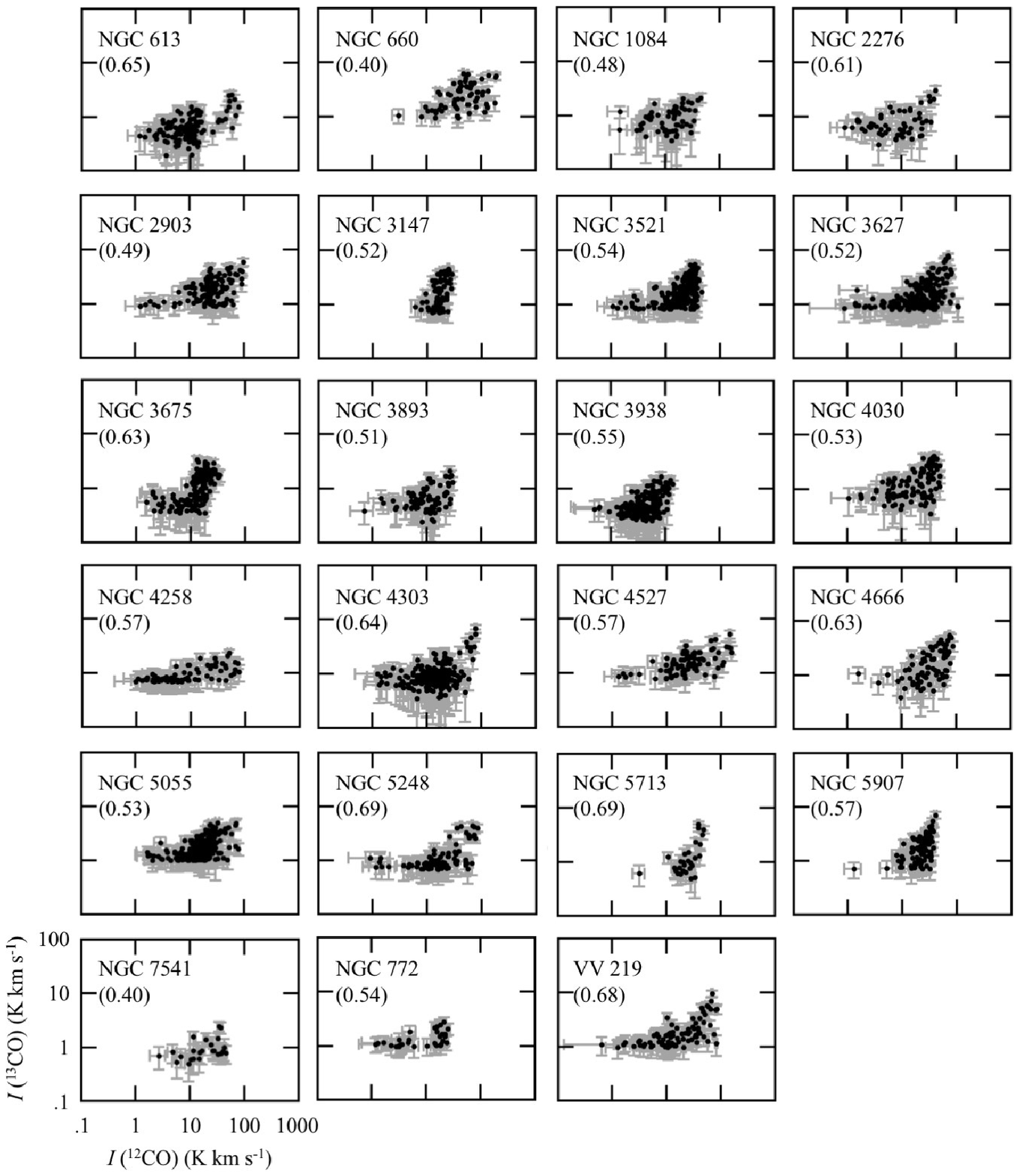}
 \end{center}
\caption{Same as figure \ref{fig:corr1312_well}, but for galaxies with $0.4 \leq \left | \rho_{I(\atom{C}{}{12}\atom{O}{}{}), I(\atom{C}{}{13}\atom{O}{}{})} \right | < 0.7$.}
\label{fig:corr1312_somewhat}
\end{figure*}

\begin{figure*}
 \begin{center}
  \includegraphics[width=16cm]{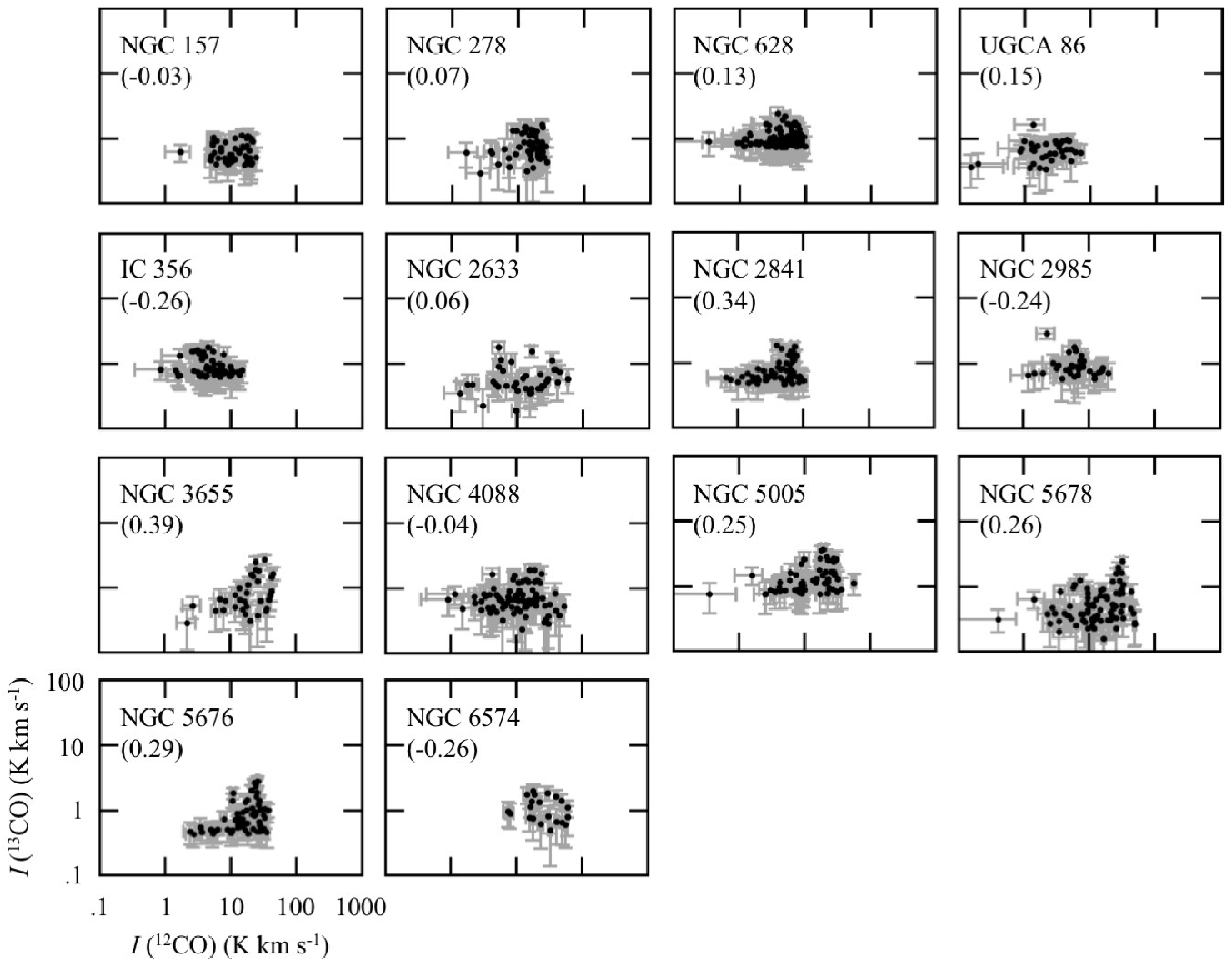}
 \end{center}
\caption{Same as figure \ref{fig:corr1312_well}, but for galaxies with $\left | \rho_{I(\atom{C}{}{12}\atom{O}{}{}), I(\atom{C}{}{13}\atom{O}{}{})} \right | < 0.4$.}
\label{fig:corr1312_different}
\end{figure*}

Figures \ref{fig:corr1312_well} -- \ref{fig:corr1312_different} depict the correlation plots between the \atom{C}{}{13}\atom{O}{}{} and \atom{C}{}{12}\atom{O}{}{} integrated intensities according to Pearson's product moment correlation coefficient between the integrated intensities [$\rho_{I(\atom{C}{}{12}\atom{O}{}{}), I(\atom{C}{}{13}\atom{O}{}{})}$]. 
Figure \ref{fig:corr1312_well} shows the galaxies with $\left | \rho_{I(\atom{C}{}{12}\atom{O}{}{}), I(\atom{C}{}{13}\atom{O}{}{})} \right | \geq 0.7$, while figures \ref{fig:corr1312_somewhat} and \ref{fig:corr1312_different} depict those with $0.4 \leq \left | \rho_{I(\atom{C}{}{12}\atom{O}{}{}), I(\atom{C}{}{13}\atom{O}{}{})} \right | < 0.7$ and $\left | \rho_{I(\atom{C}{}{12}\atom{O}{}{}), I(\atom{C}{}{13}\atom{O}{}{})} \right | < 0.4$, respectively. 
The data plotted in each panel are the pixels surrounded by the magenta contours in panel (e) of figures \ref{fig:maps_NGC0891} and \ref{fig:maps_NGC3627} and supplementary figures 3 -- 130 (supplementary section of the online version). 
The correlation coefficient is an indicator of the resemblance, although the value does not necessarily reflect the apparent resemblance between the two maps.

A comparison of the integrated intensity maps in both lines and correlation plots indicated that approximately 20 of the \atom{C}{}{13}\atom{O}{}{} integrated intensity maps (galaxies shown in figures \ref{fig:corr1312_different} and \ref{fig:corr1312_somewhat}) were noticeably different to or at least dissimilar to those of \atom{C}{}{12}\atom{O}{}{}, although the $S / N$ ratios of the \atom{C}{}{13}\atom{O}{}{} integrated intensity maps were poorer. 
About half of the lower $\rho_{I(\atom{C}{}{12}\atom{O}{}{}), I(\atom{C}{}{13}\atom{O}{}{})}$ shown in figure \ref{fig:corr1312_different} (i.e., NGC\,157, NGC\,278, NGC\,2633, NGC\,3655, NGC\,5005, NGC\,5678, NGC\,5676, and NGC\,6574) and two galaxies in figure \ref{fig:corr1312_somewhat} (i.e., NGC\,4030 and NGC\,4258) exhibited a depletion of the \atom{C}{}{13}\atom{O}{}{} emission in the central region. 
These galaxies had an H\,\emissiontype{II} region-like AGN, which suggests a \atom{C}{}{13}\atom{O}{}{} depletion in the central region (\cite{Taniguchi+1998}; \cite{Taniguchi+1999}). 
In NGC\,3627, the \atom{C}{}{12}\atom{O}{}{} integrated intensity map showed the brightest peak at the center of the galaxy and two secondary peaks at each bar end, while the \atom{C}{}{13}\atom{O}{}{} integrated intensity map showed the brightest peak at the northern bar end as shown in \citet{Watanabe+2011}. 
In NGC\,660 and NGC\,4666, the \atom{C}{}{12}\atom{O}{}{} integrated intensity maps illustrated a single peak distribution located at the center of the galaxies, while the \atom{C}{}{13}\atom{O}{}{} maps presented another peak and the central one. 
In the case of NGC\,660, the off-center peak was brighter than the central peak in \atom{C}{}{13}\atom{O}{}{}. 
The peaks at the two bar ends in the \atom{C}{}{13}\atom{O}{}{} integrated intensity map were found in NGC\,4527. 
The enhancement of the \atom{C}{}{13}\atom{O}{}{} emission at the bar ends was caused by the higher molecular gas density (\cite{Watanabe+2011}; \cite{Yajima+2019}).

The fact that the \atom{C}{}{12}\atom{O}{}{} and \atom{C}{}{13}\atom{O}{}{} maps significantly differed in some galaxies suggests that the physical properties of the molecular gas and/or abundances are not constant within a galaxy. 
We can also obtain the spectra with high significance in both \atom{C}{}{}\atom{O}{}{} lines when we stack the data with the same morphological characteristics (e.g., spiral arms or the bar) by aligning their spectra using the velocity field (\cite{Schruba+2011}; \cite{Morokuma-Matsui+2015}). 
Such stacked \atom{C}{}{13}\atom{O}{}{} spectra were made and compared with \atom{C}{}{12}\atom{O}{}{} in \citet{Muraoka+2016} and \citet{Yajima+2019}.

The integrated intensity maps of \atom{C}{}{}\atom{O}{}{18} were only presented for NGC\,891 (figure \ref{fig:maps_NGC0891}f or supplementary figure 12f), NGC\,1055 (supplementary figure 14f), and NGC\,3034 (supplementary figure 42f) in the supplementary section of the online version. 
These galaxies were those with more than $2\>\mathrm{pixels}$ higher than $4\>\sigma$ in \atom{C}{}{}\atom{O}{}{18} integrated intensity and higher than $3\>\sigma$ in \atom{C}{}{12}\atom{O}{}{} integrated intensity. 
All these galaxies were edge-on; thus, the column density of the molecular gas can be obtained. 
The $S / N$ ratios of the \atom{C}{}{}\atom{O}{}{18} integrated intensity maps were much lower than those of \atom{C}{}{13}\atom{O}{}{} because the \atom{C}{}{}\atom{O}{}{18} emission was comparatively weaker. 
The stacking of spectra via velocity alignment could lead to the \atom{C}{}{}\atom{O}{}{18} detection in some galaxies.

Panels (c) and (d) of figures \ref{fig:maps_NGC0891} and \ref{fig:maps_NGC3627} and supplementary figures 1 -- 132 (supplementary section of the online version) show the first- and second-degree moment maps in \atom{C}{}{12}\atom{O}{}{}. 
These maps were made through masking (subsection \ref{subsection:CO}); thus, the emission patterns in these maps were more smoothed and extended compared to the integrated intensity maps. 
However, even after this step, some spiky noise remained, and had a smearing effect on the moment maps. 
The discrepancy between the emission extents in the moment maps and the integrated intensity map was remarkable in the case where the emission in each pixel in the integrated intensity map was just below the $4\>\sigma$ threshold. 
One of the noticeable cases was that of IC\,356. 
On the contrary, such masking created zero emission pixels for the first- and second-degree moment maps of some galaxies. 
These moment maps were not presented in the figures for the 19 galaxies, including the two paired galaxies.

Most of the first-degree moment maps presented a circular rotation pattern, while most of the second-degree moment maps presented the highest velocity dispersion in the central region of the galaxy. 
Such velocity fields made it possible to judge the receding side of the galaxy major axis. 
Table \ref{tab:Galpars1} shows the results. 
The velocity field of NGC\,2967 in supplementary figure 39 predicted a very different PA from the one listed in table \ref{tab:Galpars1}, which was derived from its outer disc (\cite{Salo+2015}). 
A more detailed analysis also showed that the PA of the inner disc of NGC\,2967 was different from the abovementioned adopted value (\cite{Salak+2019}). 
The rotation curve of the molecular gas, PA, and {\it i} measured by fitting molecular gas distribution will be presented in Y.~Miyamoto, et~al. (in preparation). 
The molecular gas dynamics was discussed in \citet{Salak+2019} along with the kinematically determined PA and {\it i}.

\subsection{Mass}
\label{subsection:Mass}
The total CO luminosity ($L'_{\atom{C}{}{12}\atom{O}{}{}}$ in units $\mathrm{K\>km\>s}^{-1}\>\mathrm{pc}^{2}$; \cite{Solomon+1992}) and the corresponding molecular mass ($M_{\mathrm{mol}}$) within the observed region were calculated. 
The CO emission was essentially confined within the mapped region (subsection \ref{subsection:SystemSetting}); therefore, these luminosities and masses were effectively the total CO luminosity and the gas mass within $R_{3.4\,\micron}$. 
Table \ref{tab:results} lists both quantities. 
We did not use the total molecular gas mass in UGCA\,86 hereafter because the estimated value had a large uncertainty, as described in subsection \ref{subsection:spectra_maps}.

\begin{figure}
 \begin{center}
  \includegraphics[width=8cm]{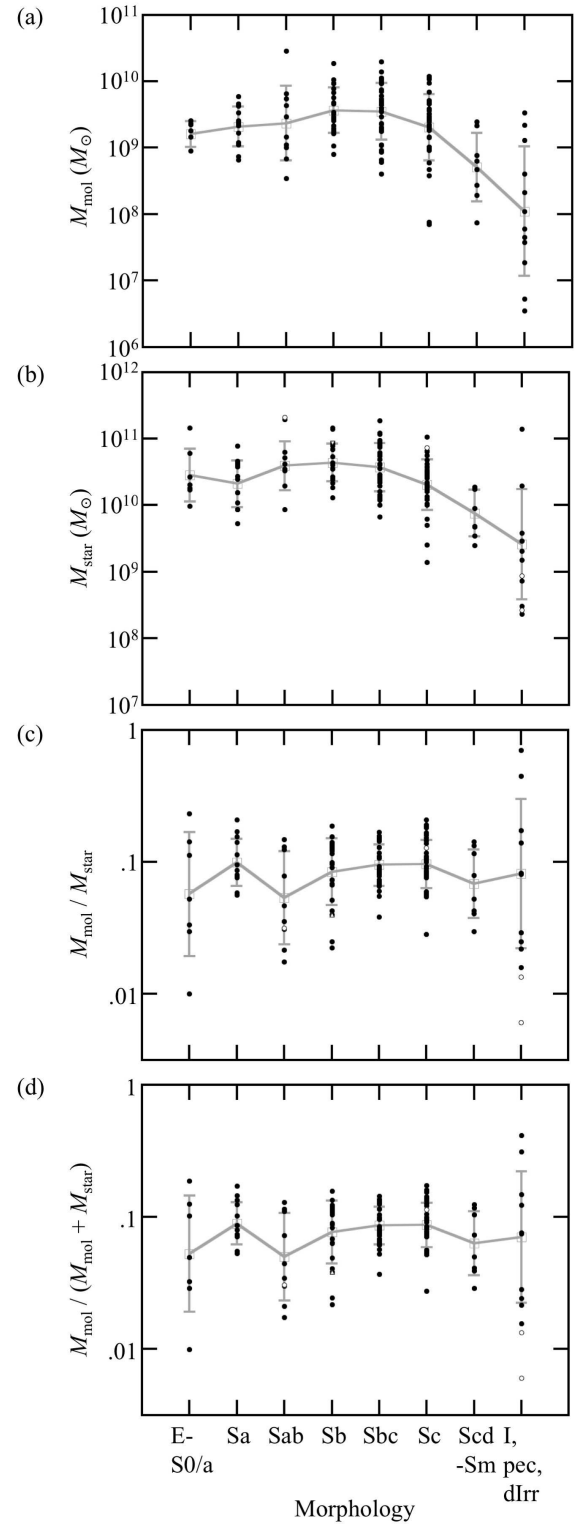}
 \end{center}
\caption{Distribution of the total molecular gas mass $M_{\mathrm{mol}}$ (a), total stellar mass $M_{\mathrm{star}}$ (b), fraction of the molecular gas mass to the stellar mass $M_{\mathrm{mol}} / M_{\mathrm{star}}$ (c), and fraction of the molecular gas mass to the total mass $M_{\mathrm{mol}} / (M_{\mathrm{mol}} + M_{\mathrm{star}})$ (d) along the morphological type. 
The open circles indicate the data with a large uncertainty (see text). 
The open downward triangle in (b) indicates an upper limit on $M_{\mathrm{star}}$, while the open upward triangle in (c) and (d) indicates a lower limit on $M_{\mathrm{mol}} / M_{\mathrm{star}}$ and $M_{\mathrm{mol}} / (M_{\mathrm{mol}} + M_{\mathrm{star}})$. 
The open squares linked with the lines are the averaged value with the standard deviation for each morphological type.}
\label{fig:Mmol_Mstar_fgas_morphology}
\end{figure}

The $M_{\mathrm{mol}}$ distribution along the Hubble sequence indicates that galaxies with Sb and Sbc types have the tendency to host a larger molecular gas content, although the scatter is considerable (figure \ref{fig:Mmol_Mstar_fgas_morphology}a). 
The geometric mean value of $M_{\mathrm{mol}}$ highlighted this trend, showing this peak and noticeably lower values in the irregular and peculiar galaxies. 
This trend was consistent with that in the previous works (\cite{Komugi+2008}). 
The distribution of the total stellar mass ($M_{\mathrm{star}}$) along the Hubble sequence was similar to that of the total molecular gas (figure \ref{fig:Mmol_Mstar_fgas_morphology}b).

The fraction of the total molecular gas mass to the total stellar mass, $M_{\mathrm{mol}} / M_{\mathrm{star}}$, showed no clear trend against the Hubble type (figure \ref{fig:Mmol_Mstar_fgas_morphology}c). 
Despite our sample having a strong bias toward molecular gas-rich FIR bright galaxies, these results were consistent with more complete observational samples [e.g., those of \citet{Saintonge+2011} (showing no difference in the concentration index rather than the Hubble type), \citet{Boselli+2014b}, \citet{Bolatto+2017}, and \citet{Young+1991} (comparing the ratio of the molecular gas mass to the dynamical mass)]. 
We also showed herein the fraction of the total molecular gas mass to the total baryonic mass, that is, the sum of the molecular and stellar masses [$M_{\mathrm{mol}} / (M_{\mathrm{mol}} + M_{\mathrm{star}})$] along the Hubble sequence in figure \ref{fig:Mmol_Mstar_fgas_morphology}d. 
The molecular gas mass fraction was small; thus, the tendency of $M_{\mathrm{mol}} / (M_{\mathrm{mol}} + M_{\mathrm{star}})$ was very similar to that of $M_{\mathrm{mol}} / M_{\mathrm{star}}$. 
We ignored the atomic gas mass because it typically contributed much less than the molecular gas to the baryonic mass within a radius where the molecular gas was dominant (\cite{Honma+1995}). 
Thus, hereafter, we used $M_{\mathrm{mol}} / M_{\mathrm{star}}$ as a proxy for molecular gas to baryon ratio.

\begin{figure}
 \begin{center}
  \includegraphics[width=8cm]{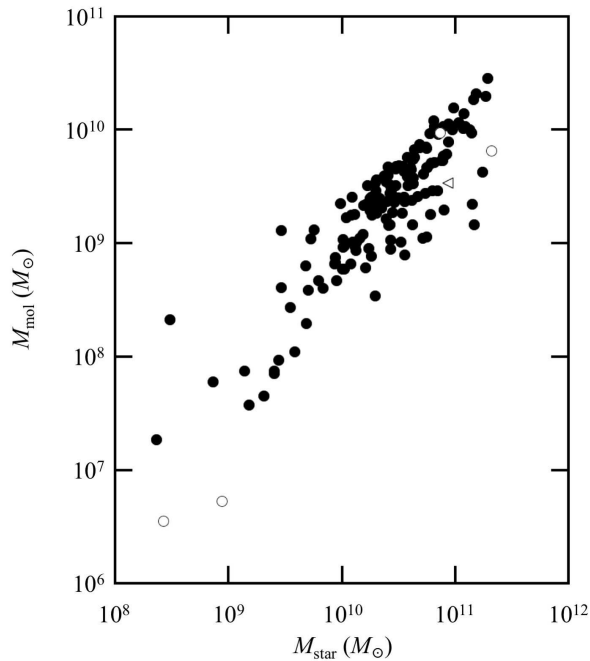}
 \end{center}
\caption{Relation between the molecular gas mass $M_{\mathrm{mol}}$ and the stellar mass $M_{\mathrm{star}}$. 
The typical errors are smaller than the size of each marker. 
The open circles indicate $M_{\mathrm{star}}$ with a considerable uncertainty. 
The open leftward triangle indicates an upper limit on $M_{\mathrm{star}}$.}
\label{fig:corr_Mmol_Mstar}
\end{figure}

$M_{\mathrm{mol}}$ correlated with $M_{\mathrm{star}}$ over three orders of magnitude (figure \ref{fig:corr_Mmol_Mstar}). 
That is, $M_{\mathrm{mol}}$ increased with $M_{\mathrm{star}}$. 
This trend was already predicted in figure \ref{fig:Mmol_Mstar_fgas_morphology}, in which $M_{\mathrm{mol}}$ and $M_{\mathrm{star}}$ behaved in a similar way, and $M_{\mathrm{mol}} / M_{\mathrm{star}}$ was constant over the Hubble types. 
Although we applied the standard conversion factor to every location in each galaxy, the factor is not constant, and could depend on the physical and chemical conditions of molecular gas (e.g., \cite{Kennicutt+2012}; \cite{Bolatto+2013}). 
Deriving variable conversion factors in various positions in individual galaxies and calculating a more precise molecular gas mass will be reported in a future study. 
The relation between the molecular gas and the star formation in these samples was presented in \citet{Muraoka+2019} and will be also reported in a future study.

\begin{figure}
 \begin{center}
  \includegraphics[width=8cm]{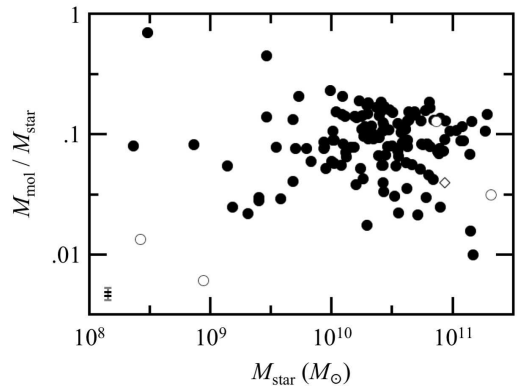}
 \end{center}
\caption{Molecular gas to stellar mass fraction $M_{\mathrm{mol}} / M_{\mathrm{star}}$ versus stellar mass $M_{\mathrm{star}}$. 
The open circles indicate $M_{\mathrm{star}}$ (and $M_{\mathrm{mol}} / M_{\mathrm{star}}$) with a considerable uncertainty. 
The open diamond indicates the upper limit on $M_{\mathrm{star}}$ (and lower limit on $M_{\mathrm{mol}} / M_{\mathrm{star}}$). 
The typical ({\it black}) and maximum ({\it gray}) errors are shown in the bottom-left corner.}
\label{fig:fgas_Mstar}
\end{figure}

\begin{figure*}
 \begin{center}
  \includegraphics[width=16cm]{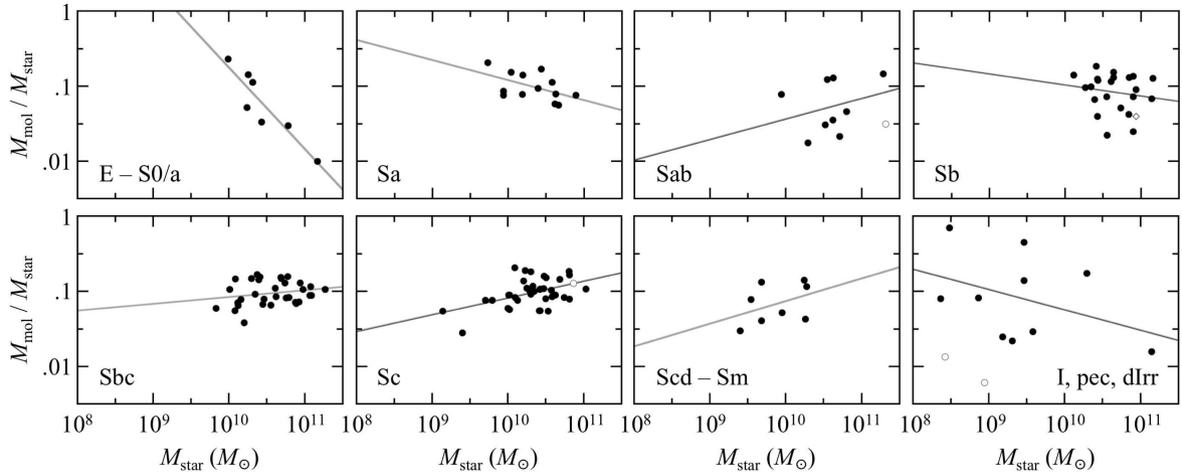}
 \end{center}
\caption{Same as figure \ref{fig:fgas_Mstar}, but as a function of the Hubble types. 
The solid line indicates the power law fitting. 
The data shown as open circles and open diamond are not used for the fitting.}
\label{fig:fgas_Mstar_HubbleTypes}
\end{figure*}

\begin{figure*}
 \begin{center}
  \includegraphics[width=16cm]{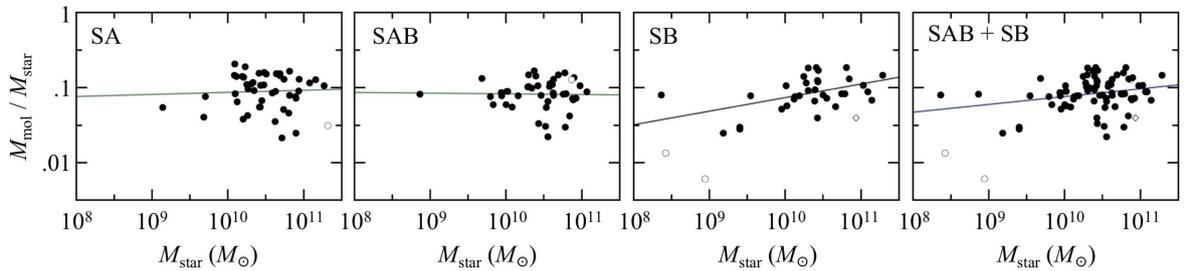}
 \end{center}
\caption{Same as figure \ref{fig:fgas_Mstar_HubbleTypes}, but as a function of the existence of a bar or not. 
The galaxies not classified into these categories are not shown.}
\label{fig:fgas_Mstar_bars}
\end{figure*}

We also showed the relation between $M_{\mathrm{mol}} / M_{\mathrm{star}}$ and $M_{\mathrm{star}}$ in figure \ref{fig:fgas_Mstar}. 
The errors were comparable to or smaller than the size of each marker. 
The relation between both quantities was expected to provide a constraint on constructing the galaxy formation and evolution models (\cite{Morokuma-Matsui+2015b}; \cite{Saintonge+2017}). 
The span was similar to the results for the local galaxies in the previous works, in which a decreasing trend after around $M_{\mathrm{star}} \sim\,10^{10}\,M_{\solar}$ (\cite{Jiang+2015}; \cite{Morokuma-Matsui+2015b}; \cite{Bolatto+2017}; \cite{Saintonge+2017}) was observed. 
We also showed herein the same plots divided by the morphologies in figures \ref{fig:fgas_Mstar_HubbleTypes} and \ref{fig:fgas_Mstar_bars}. 
A power law was fitted in each plot (solid line). 
Table \ref{tab:fgas_Mstar} lists the fitted slope and Pearson's product moment correlation coefficient. 
Although no clear tendency was seen along the Hubble sequence, $M_{\mathrm{mol}} / M_{\mathrm{star}}$ decreased with $M_{\mathrm{star}}$ in the early-type galaxies (E -- S0 and Sa), but increased in the late-type spirals (Sc and Scd -- Sm). 
It also effectively showed an uncorrelated scatter in intermediate types and irregular galaxies (figure \ref{fig:fgas_Mstar_HubbleTypes}). 
The fraction seemed to increase with $M_{\mathrm{star}}$ for the SB galaxies; however, no clear difference was seen among the SA and SAB galaxies (figure \ref{fig:fgas_Mstar_bars}). 
Thus, this tendency probably resulted from the bias of our samples because SB had predominantly less massive galaxies with irregular or peculiar morphology.

\begin{table}
  \tbl{Correlation between the total molecular gas mass fraction and the total stellar mass.}{%
  \begin{tabular}{ccccc}
      \hline
      Morphology & Samples & Power\footnotemark[$*$] & Correlation coefficient & Remarks  \\ 
      \hline
      \multicolumn{5}{c}{Hubble types}\\
      \hline
      E -- S0/a & 7 & -1.09 & -0.92 & \\
      Sa & 13 & -0.27 & -0.52 & \\
      Sab & 10 & 0.27 & 0.28 & \footnotemark[$\dag$]\\
      Sb & 23 & -0.15 & -0.16 & \footnotemark[$\ddag$] \\
      Sbc & 34 & 0.09 & 0.21 & \\
      Sc & 39 & 0.22 & 0.46 & \footnotemark[$\S$] \\
      Scd -- Sm & 8 & 0.30 & 0.40 & \\
      I, pec, dIrr & 13 & -0.27 & -0.39 & \footnotemark[$\parallel$] \\
      \hline
      \multicolumn{5}{c}{with or without a bar}\\
      \hline
      SA & 47 & 0.03 & 0.05 & \footnotemark[$\dag$] \\
      SAB & 44 & -0.01 & -0.02 & \footnotemark[$\S$] \\
      SB & 36 & 0.18 & 0.48 & \footnotemark[$\sharp$] \\
      SAB + SB & 80 & 0.10 & 0.25 & \footnotemark[$* *$] \\
      \hline
    \end{tabular}}\label{tab:fgas_Mstar}
\begin{tabnote}
\footnotemark[$*$]: Index $a$ in $M_{\mathrm{mol}} / M_{\mathrm{star}} \propto M_{\mathrm{star}}^{a}$.  \\
\footnotemark[$\dag$]: IC\,356 is included in the number of samples, but not in the statistics. \\
\footnotemark[$\ddag$]: NGC\,5792 is included in the number of samples, but not in the statistics. \\
\footnotemark[$\S$]: NGC\,2276 is included in the number of samples, but not in the statistics. \\
\footnotemark[$\parallel$]: IC\,10 and NGC\,1569 are included in the number of samples, but not in the statistics. \\
\footnotemark[$\sharp$]: IC\,10, NGC\,1569, and NGC\,5792 are included in the number of samples, but not in the statistics. \\
\footnotemark[$* *$]: IC\,10, NGC\,1569, NGC\,2276, and NGC\,5792 are included in the number of samples, but not in statistics. \\
\end{tabnote}
\end{table}

\subsection{Molecular gas content in the barred and non-barred spirals}
\label{subsection:bar-effect}
\begin{figure}
 \begin{center}
  \includegraphics[width=8cm]{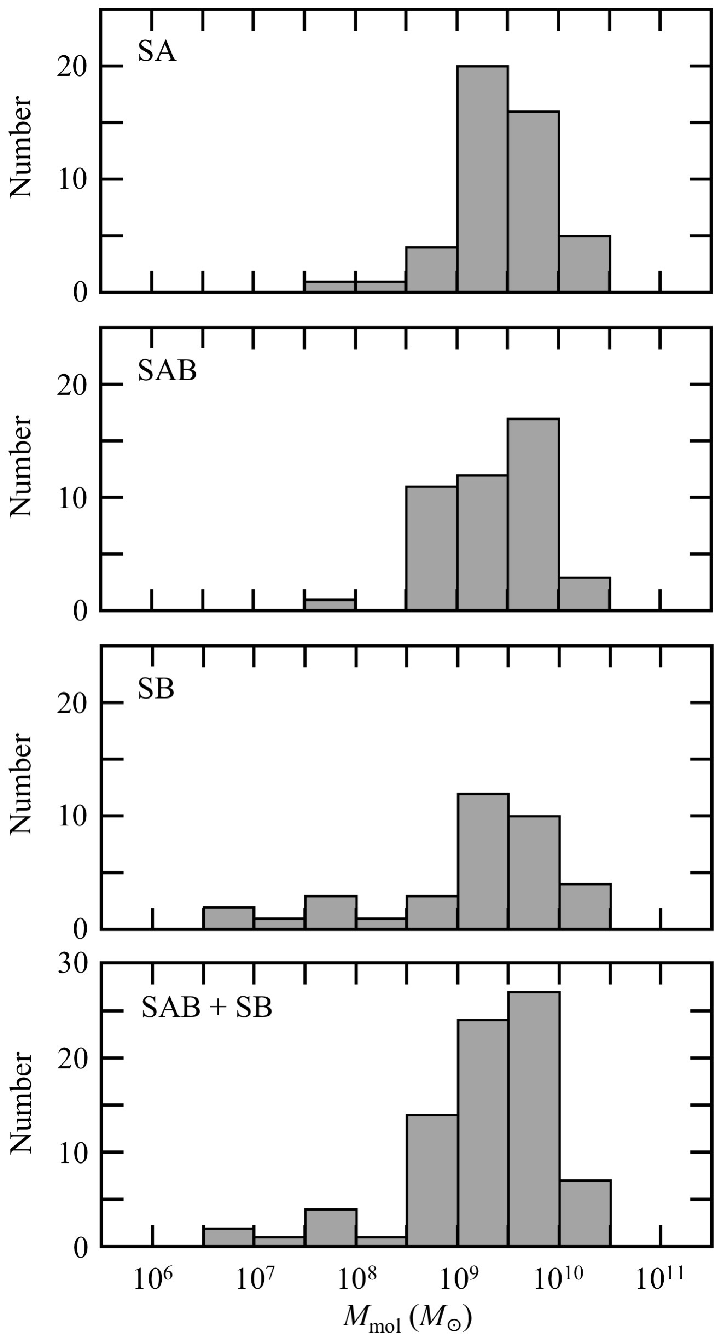}
 \end{center}
\caption{Histograms of the total molecular gas mass $M_{\mathrm{mol}}$ for bar types SA, SAB, SB, and SAB + SB.}
\label{fig:Mmol_bar}
\end{figure}

\begin{figure}
 \begin{center}
  \includegraphics[width=8cm]{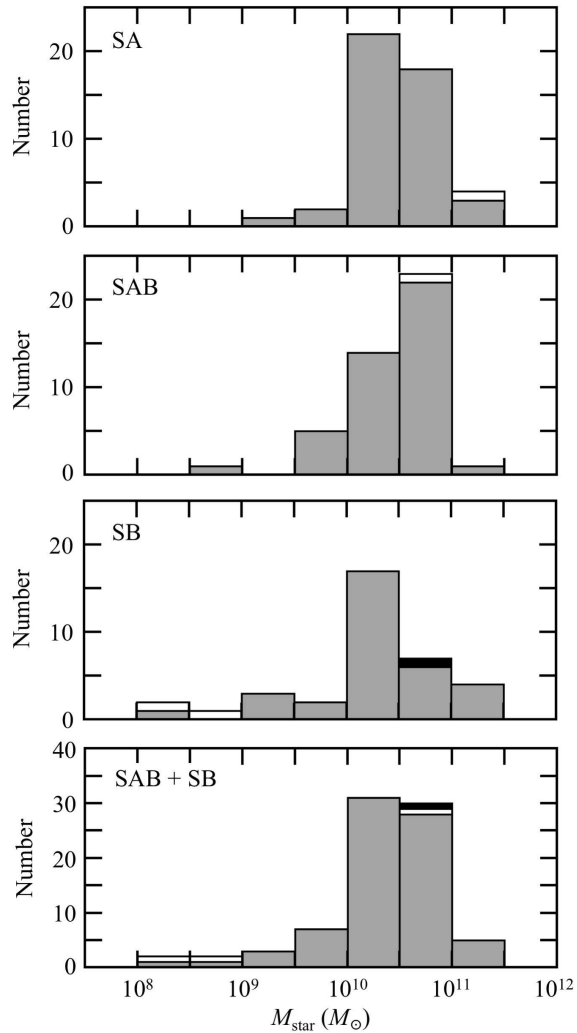}
 \end{center}
\caption{Same as figure \ref{fig:Mmol_bar}, but for the stellar mass $M_{\mathrm{star}}$. 
The open columns indicate $M_{\mathrm{star}}$ with a considerable uncertainty. 
The black filled column indicates the upper limit on $M_{\mathrm{star}}$.}
\label{fig:Mstar_bar}
\end{figure}

\begin{figure}
 \begin{center}
  \includegraphics[width=8cm]{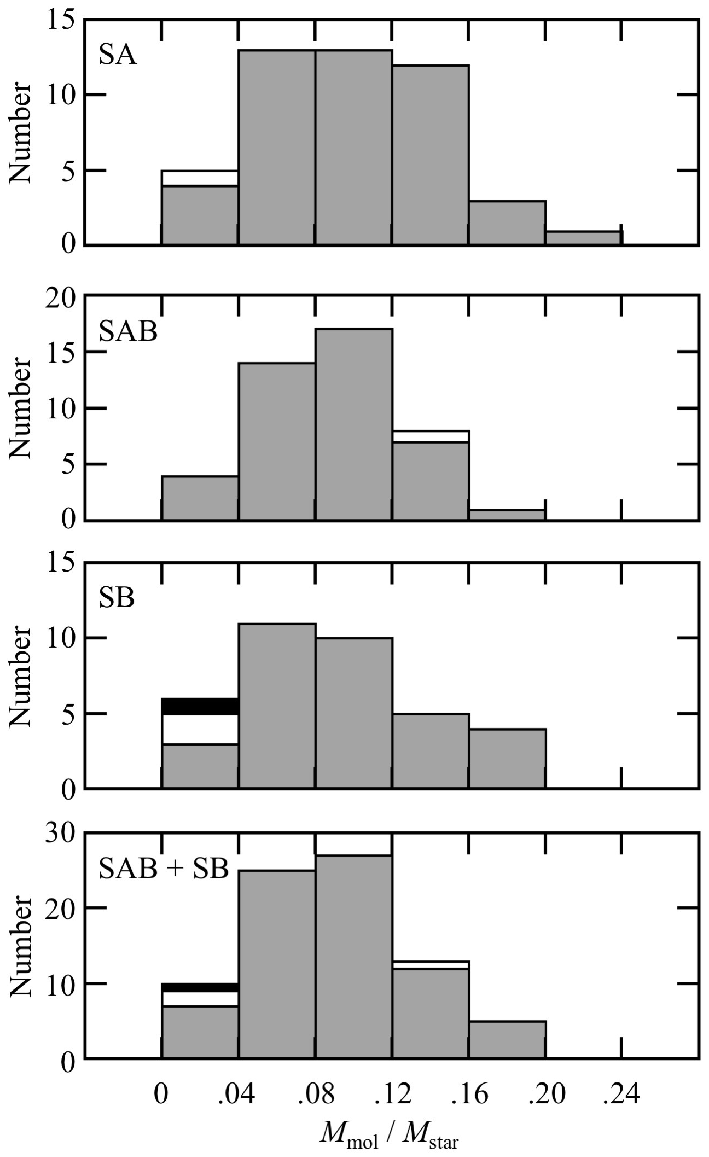}
 \end{center}
\caption{Same as figure \ref{fig:Mmol_bar}, but for $M_{\mathrm{mol}} / M_{\mathrm{star}}$. 
The open columns indicate $M_{\mathrm{mol}} / M_{\mathrm{star}}$ with a considerable uncertainty. 
The black filled column indicates the lower limit on $M_{\mathrm{mol}} / M_{\mathrm{star}}$.}
\label{fig:fgas_bar}
\end{figure}

Figures \ref{fig:Mmol_bar} -- \ref{fig:fgas_bar} show a comparison of the distributions of $M_{\mathrm{mol}}$, $M_{\mathrm{star}}$, and $M_{\mathrm{mol}} / M_{\mathrm{star}}$ over different bar morphological types. 
The $M_{\mathrm{mol}}$ and $M_{\mathrm{star}}$ histograms for the non-barred and barred spirals were very similar, although some SB galaxies had a very low $M_{\mathrm{mol}}$ or $M_{\mathrm{star}}$. 
A Kolmogorov-Smirnov (K-S) test cannot reject the null hypothesis that the $M_{\mathrm{mol}}$ or $M_{\mathrm{star}}$ distribution coincides in the barred and non-barred spirals at a significant level of 10\,\%. 
The $M_{\mathrm{mol}} / M_{\mathrm{star}}$ fraction did not correlate with the presence of bars, with each type spanning the same range of $M_{\mathrm{mol}} / M_{\mathrm{star}}$, although there seems a slight preference for higher values in the SA galaxies. 
A K-S test cannot reject the null hypothesis that the two populations are drawn from the same distribution at a significant level of 10\,\%.

\begin{figure}
 \begin{center}
  \includegraphics[width=8cm]{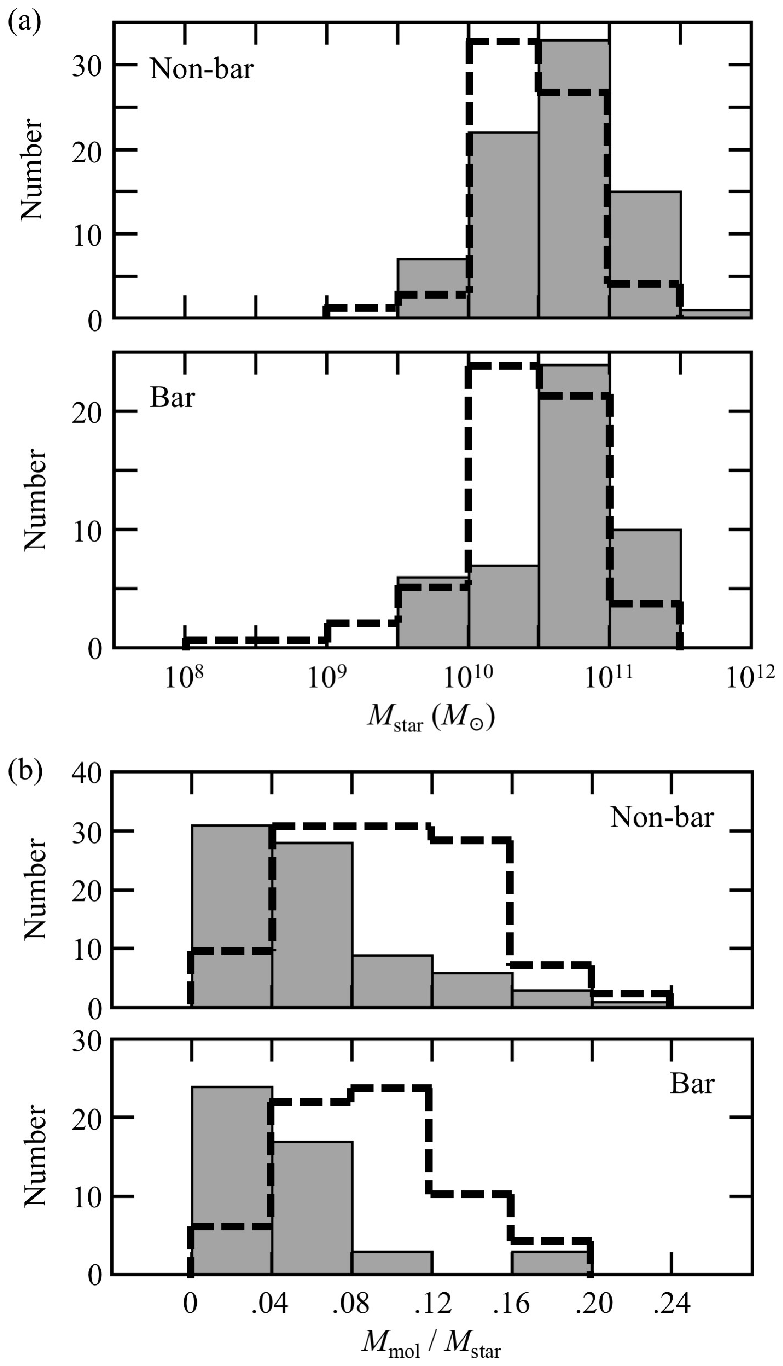}
 \end{center}
\caption{(a) Histograms of the total stellar mass $M_{\mathrm{star}}$ for the non-barred and barred galaxies in the EDGE sample (\cite{Bolatto+2017}). 
The dashed line indicates the COMING results shown in figure \ref{fig:Mstar_bar} for SA and SAB + SB. 
(b) Same as (a), but for $M_{\mathrm{mol}} / M_{\mathrm{star}}$. 
The dashed line indicates the COMING results shown in figure \ref{fig:fgas_bar} for SA and SAB + SB. 
The COMING results are normalized for the peak to correspond to that of the EDGE results, and the data with uncertainty or upper/lower limit data are not used.}
\label{fig:EDGE_histograms}
\end{figure}

In contrast to our results that $M_{\mathrm{mol}} $, $M_{\mathrm{star}}$, and $M_{\mathrm{mol}} / M_{\mathrm{star}}$ do not depend on the bar presence, previous works showed that $M_{\mathrm{mol}} / M_{\mathrm{star}}$ may be lower in the barred spirals. 
Figure \ref{fig:EDGE_histograms} shows the histograms of $M_{\mathrm{star}}$ and $M_{\mathrm{mol}} / M_{\mathrm{star}}$ for the non-barred and barred galaxies, respectively, in the EDGE sample (\cite{Bolatto+2017}). 
Data from the COMING overlaid as dashed lines suggested that the barred spiral galaxies in our sample were less massive in $M_{\mathrm{star}}$ and prone to have a higher $M_{\mathrm{mol}} / M_{\mathrm{star}}$. 
The EDGE sample consisted of rather more massive galaxies in stellar mass than the COMING samples, especially for the barred spiral galaxies. 
A K-S test indicated the hypothesis that both survey samples coming from the same population are rejected at 1\,\% significance. 
Although the K-S test did not indicate a significant difference for the non-barred galaxies between both samples, it implied that the hypothesis was rejected at 2.5\,\% significance for the barred galaxies. 
The possibility of a lower $M_{\mathrm{mol}} / M_{\mathrm{star}}$ in the barred galaxies was suggested in \citet{Bolatto+2017}, although their K-S test indicated a low significance.

\begin{figure}
 \begin{center}
  \includegraphics[width=8cm]{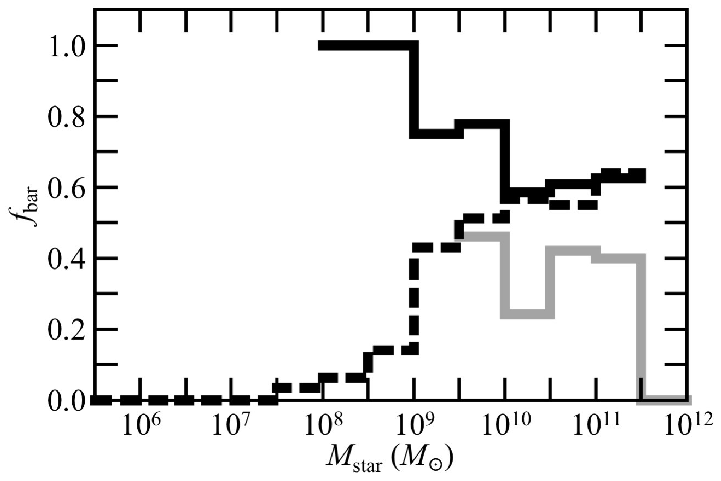}
 \end{center}
\caption{Bar fraction $f_{\mathrm{bar}}$ of the COMING sample as a function of the stellar mass $M_{\mathrm{star}}$ ({\it black solid}). 
The data with a large uncertainty or the upper limit data are not used. 
The fractions of the EDGE sample (\cite{Bolatto+2017}) ({\it gray solid}) and S$^{4}$G (\cite{Diaz-Garcia+2016}) ({\it black dashed}) are shown.}
\label{fig:fbar_Mstar}
\end{figure}

Our sample selection was biased toward the FIR bright galaxies, that is, gas-rich galaxies, which may result in this discrepancy. 
Figure \ref{fig:fbar_Mstar} shows the fraction of the barred galaxies in our sample (SAB$\,+\,$SB) ($f_{\mathrm{bar}}$) for a given $M_{\mathrm{star}}$ as well as that of EDGE (\cite{Bolatto+2017}) and S$^{4}$G (\cite{Diaz-Garcia+2016}). 
The dependency of $f_{\mathrm{bar}}$ on $M_{\mathrm{star}}$ was different from the previous works. 
\citet{Diaz-Garcia+2016} indicated that the bar fraction increased with the stellar mass in the S$^{4}$G samples. 
\citet{Algorry+2017} studied the population of the barred galaxies in the EAGLE cosmological simulation and found that the gas mass fraction decreased with the bar strength parameter. 
In other words, barred spiral galaxies were prone to having a lower gas mass fraction. 
\citet{Spinoso+2017} predicted that gas is driven inward and consumed by star formation in the barred galaxies in cosmological simulations. 
Observational results also concurred with those predictions (\cite{Cheung+2013}; \cite{Chown+2019}). 
Our sample consisted of gas-rich galaxies; barred spiral galaxies with high $M_{\mathrm{star}}$ and low $f_{\mathrm{gas}}$ were not included; and those with lower $M_{\mathrm{star}}$ and higher $M_{\mathrm{mol}} / M_{\mathrm{star}}$ were selectively included in our sample, thereby causing similar $M_{\mathrm{mol}}$, $M_{\mathrm{star}}$, and $M_{\mathrm{mol}} / M_{\mathrm{star}}$ distributions between the barred and non-barred spiral galaxies seen in our sample.

\section{Summary and conclusion}
\label{section:summary}
In this study, we simultaneously conducted COMING, an OTF-mapping survey project, targeting 147 nearby galaxies in \atom{C}{}{12}\atom{O}{}{}, \atom{C}{}{13}\atom{O}{}{}, and \atom{C}{}{}\atom{O}{}{18} $J=1-0$ lines using the NRO 45 m radio telescope. 
The spatial resolution was \timeform{17''}. 
The velocity resolution was set to $10\,\mathrm{km\>s}^{-1}$. 
The sensitivity was typically $T_{\mathrm{MB}} = 70\,\mathrm{mK}$. 
Each mapping region covered 70\,\% of the optical diameter $D_{25}$, which was expected to encompass all CO emissions, as indicated by the previous observations. 
The targets were selected from nearby galaxies based on the FIR flux. 
In other words, they were strongly biased to the FIR brightness. 
Some interacting galaxies were selected despite not satisfying the selection criteria because the counterpart of the paired galaxies satisfied the criteria. 
We only completed observations for 147 out of the 238 galaxies in the selected sample because of the observing limitations. 
The fractions of the barred and non-barred galaxies in the sample were similar, but we concentrated on those with the Hubble-type Sb--Sc. 
The galaxies with smaller disc lengths were a dominant component of the observed sample. 
The FITS cubes with \timeform{6''} spacing in the three lines are publicly available at the JVO archive.

We developed an observation ranking system and automatic data reduction tools to optimize resource observation and ensure that reduction is fully reproducible and highly efficient. 
We quantified the relative positions between the direction of the telescope and the available targets and made assessments based on the target elevation and size while considering the system noise and other factors to maximize the quantity of the observed galaxies in the allotted time. 
This system was introduced in the last two observation seasons and resulted in the reduction of the total observing time to complete a map typically by 37\,\%. 
We also developed the tools {\sf auto-flag} and {\sf auto-rebase}. 
The former quantified the undulation of the spectral baselines and removed heavily undulated spectra in the OTF data. 
Meanwhile, {\sf auto-rebase} determined the zero levels in the spectra with ambiguous features. 
These tools are a part of the COMING ART \textsc{python} package designed to optimize the objectivity and reproducibility of the reduction process.

The radii and the total stellar mass of the observed galaxies were measured from the WISE $3.4\,\micron$ archival data. 
The radius, $R_{3.4\,\micron}$, was 30\,\% larger than the optical radius for most targets. 
The total stellar mass, $M_{\mathrm{star}}$, was measured within $R_{3.4\,\micron}$ following the method of \citet{Wen+2013}. 
Our derived $M_{\mathrm{star}}$ was approximately 20\,\% less massive than the mass derived in S$^4$G (\cite{Sheth+2010}) for 115 overlapping galaxies between our project and S$^4$G. 
The offset may be caused by our conservative star masking method which masks slightly larger area than the extent of stars and background subtraction, and possibly a minor offset between the WISE $3.4\,\micron$ band and the Spitzer $3.6\,\micron$ band flux at a higher magnitude (\cite{Wen+2013}).

The \atom{C}{}{12}\atom{O}{}{} emission was detected in most of the observed galaxies, while the \atom{C}{}{13}\atom{O}{}{} emission was detected in approximately a third of the galaxies with a low $S / N$ ratio. 
The \atom{C}{}{}\atom{O}{}{18} emission was detected in only $\lesssim\,10$ targets. 
Some integrated intensity maps in \atom{C}{}{12}\atom{O}{}{} indicated a similar distribution to the stellar distribution, while some displayed quite different morphologies. 
Some galaxies also showed differences in the locations of the \atom{C}{}{12}\atom{O}{}{} and \atom{C}{}{13}\atom{O}{}{} emission, although the $S / N$ ratio of \atom{C}{}{13}\atom{O}{}{} was rather low in most cases. 
This result implied that the molecular gas properties in the galactic discs were not uniform. 
The \atom{C}{}{}\atom{O}{}{18} maps were only successfully created in three of our targets, namely NGC\,891, NGC\,1055 and NGC\,3034. 
Additionally, the first- and second-degree moment maps in \atom{C}{}{}\atom{O}{}{} were created for the majority of our sample.

The total molecular gas mass, $M_{\mathrm{mol}}$, measured using the standard conversion factor correlated well with $M_{\mathrm{star}}$ over three orders of magnitude. 
Moreover, the ratio $M_{\mathrm{mol}} / M_{\mathrm{star}}$ did not depend on the Hubble type. 
Galaxies with Sb and Sbc types have the tendency to host a larger $M_{\mathrm{mol}}$ or $M_{\mathrm{star}}$, although the scatter is considerable. 
The trend of $M_{\mathrm{mol}} / M_{\mathrm{star}}$ relation was consistent with that in the previous works, although the scatter was considerable. 
However, $M_{\mathrm{mol}} / M_{\mathrm{star}}$ seemed to decrease with $M_{\mathrm{star}}$ in early-type galaxies and vice versa in late-type galaxies.

$M_{\mathrm{mol}}$ and $M_{\mathrm{mol}} / M_{\mathrm{star}}$ did not show dependence on the presence of bars, but this was likely the result of our sample selection criteria. 
We saw no differences in the mass distribution of $M_{\mathrm{mol}}$ and $M_{\mathrm{star}}$ between the barred and non-barred spirals; however, we found that the SB types had a lower mass tail in the $M_{\mathrm{mol}}$ and $M_{\mathrm{star}}$ distribution. 
No difference of $M_{\mathrm{mol}} / M_{\mathrm{star}}$ was seen between the barred and non-barred spirals, although the previous works suggested the possibility of a comparatively lower $M_{\mathrm{mol}} / M_{\mathrm{star}}$ in the barred spirals (\cite{Bolatto+2017}). 
The fraction of the barred spirals as a function of $M_{\mathrm{star}}$ also showed a trend opposite to that in the previous works. 
This difference was caused by our sample selection, in which we prioritized observing the FIR bright (thus, molecular gas-rich) galaxies. 
The barred spirals are effective at funneling molecular gas toward their central regions, resulting in a molecular gas deficit in their discs. 
Thus, the molecular gas-rich barred spirals had a lower $M_{\mathrm{star}}$.

This survey was conducted toward approximately 150 galaxies, observing the \atom{C}{}{12}\atom{O}{}{} distribution over their entire disc area and revealing the large-scale distribution of their entire molecular gas budget. 
These data act as a benchmark for comparison with the targeted observations of more distant galaxies, interferometric observations of our sample galaxies, observations in higher-$J$ lines, and those with very high spatial resolutions of specific regions of external galaxies. 
Our sample was composed of nearby galaxies, and, as such, a wealth of archival data across multiple wavelengths was already available. 
Thus, COMING acts as an excellent resource for studies of the interstellar medium and star formation in galaxies on kpc scales.

\section*{Supplementary data}

The following supplementary data is available at PASJ online.

\noindent 
Supplementary figures 1--134. \\
\noindent  
  WISE $3.4\,\micron$ image, 
integrated  intensity map, 
first-degree moment map and second-degree moment map of \atom{C}{}{12}\atom{O}{}{} $(J=1-0)$ for all the observed galaxies [panels (c) and (d) were not presented in 17 figures], 
integrated intensity maps of \atom{C}{}{13}\atom{O}{}{} $(J=1-0)$, 
 and \atom{C}{}{}\atom{O}{}{18} $(J=1-0)$ for some galaxies are shown in each figure. 



\begin{ack}
The authors thank the referee for very many and invaluable comments and suggestions, which significantly improved the manuscript. 
We would also like to gratefully acknowledge Alex Pettitt for his helping to improve the paper greatly. 
We are grateful to the NRO staff for the operation of the 45 m telescope and their continuous efforts to improve the performance of the instruments. 
This work is based on one of the legacy programs of the Nobeyama 45 m radio telescope, which is operated by Nobeyama Radio Observatory, a branch of National Astronomical Observatory of Japan. 
This research has made use of the NASA/IPAC Extragalactic Database (NED) which is operated by the Jet Propulsion Laboratory, California Institute of Technology, under contract with the National Aeronautics and Space Administration. 
This publication makes use of data products from the Wide-field Infrared Survey Explorer, which is a joint project of the University of California, Los Angeles, and the Jet Propulsion Laboratory/California Institute of Technology, funded by the National Aeronautics and Space Administration. 
HK is supported by JSPS KAKENHI Grant Number 18K13593.
\end{ack}

\end{document}